\documentclass[article]{JHEP3}

\usepackage{graphicx}
\usepackage{amsmath,epsfig}
\usepackage{amssymb,amsfonts}
\usepackage{latexsym}
\usepackage{epstopdf}  %for macs
\usepackage{float}
\usepackage{subfigure}

\relax
\renewcommand{\theequation}{\arabic{section}.\arabic{equation}}

\def\be{\begin{equation}}
\def\ee{\end{equation}}

\allowdisplaybreaks[2]

\newcommand{\bear}{\begin{align}}
\newcommand{\eear}{\end{align}}
\newcommand{\bea}{\begin{align}}
\newcommand{\eea}{\end{align}}
\newcommand{\nn}{\nonumber}

\def\hri#1#2{\href{http://arxiv.org/abs/#1}{[ArXiv:#1]#2}}
\def\hre#1#2{\href{http://arxiv.org/abs/#1/#2}{[ArXiv:#1/#2]}}

\newbox\pippobox

\def\II{\relax{\rm I\kern-.18em I}}

\def\l{\lambda}
\def\m{\mu}

\def\pa{\partial}

\def\tr{\ensuremath{\mathrm{Tr}}}

\def\t{\tau}

\def\h{\kappa}
\def\gf{w}
\def\Awf{{A}} %The weight factor in the metric
\def\G{G} %Shorthand for the sqrt factor

\def\cla{{\cal L}_A}
\def\nt{\tilde n}
\def\qt{\tilde Q}
\def\CL{{\cal L}}

\def\mga{\mathcal{G_A}}
\def\mgs{\mathcal{G_S}}

\title{Spectral Functions in V-QCD with Matter:\\
Masses, Susceptibilities, Diffusion and Conductivity}

\author{Ioannis Iatrakis\\
Department of Physics and Astronomy,\\
Stony Brook University,  Stony Brook, NY 11794-3800.}

\author{Ismail Zahed\\
Department of Physics and Astronomy,\\
Stony Brook University,  Stony Brook, NY 11794-3800.}

\preprint{}

\abstract{We consider a holographic model of QCD in the Veneziano limit of a large number 
of colors $N_c$ and flavors $N_f$ but fixed $x=N_f/N_c$ (V-QCD).  The model exhibits a first
order deconfined but chirally broken transition, followed by a second order  chirally restored transition 
in the $\mu-T$ plane for a range of plausible holographic parameters. We study the quasi-normal mode
spectrum, and derive the pertinent vector and axial spectral functions across the transition regions. 
The pole masses,  susceptibilities, diffusion constants and electric conductivity are also discussed.
In particular, the pole masses are found to survive the deconfining transition, to quickly dissolve in the
the chirally restored phase by developing substantial widths. 
The flavor electric conductivities arise sharply in the transition region. The flavor susceptibility
is shown to be consistent with the one derived from bulk thermodynamics.}

\begin{document}

\def\g{\gamma}
\def\go{\g_{00}}
\def\gi{\g_{ii}}

\maketitle %%%%%%%%%% THIS IS IGNORED %%%%%%%%%%%

%%%%%%%%%%%%%%%%%%%%%%%%%%%%%%%%%%%%%%%%%%%%%%%%%%%%%%%%%%%%%%%%%%
\section{Introduction}
%%%%%%%%%%%%%%%%%%%%%%%%%%%%%%%%%%%%%%%%%%%%%%%%%%%%%%%%%%%%%%%%%%
Current lattice QCD simulations of the bulk thermodynamics are limited by the sign problem in the $\mu-T$ plane~\cite{Barbour:1997ej}. Of particular
interest is the faith of confinement and the spontaneous breaking of chiral symmetry across the transition plane, and the nature
of such transitions. While some of these questions are starting to be answered quantitatively along the  $\mu=0$ line, they remain
very speculative for $\mu\neq 0$~\cite{Endrodi:2011gv}. In particular, the possible occurrence of a critical point in the $\mu-T$ plane is still very elusive.
While much is understood about the space-like structure of QCD  at $\mu=0$, such as the screening masses, hadronic wave-functions and flavor susceptibilities
both numerically~\cite{Bernard:1991ah,Petreczky:2003iz,Petreczky:2012rq}
and analytically~\cite{Hansson:1991kb,Fayyazuddin:1993ua,Hansson:1994nb,Hansson:1997fz}, 
not much can be said about the time-like structure of the QCD matter in 
the $\mu-T$ plane given the limitations of the present lattice algorithms. 
An exception is the real-time analysis of QCD in hot $2+1$ dimensions~\cite{Hansson:1997fz}.
Complex Langevin techniques may provide the 
alternative needed to address these fundamental physics questions from first principles but they are currently limited to algorithm testing~\cite{Sexty:2014zya}.

Guided by the strictures of the spontaneous breaking of chiral symmetry at low energy, a number of phenomenological
models have been suggested to model QCD with matter. Of particular interest are the instanton liquid (IL) 
model~\cite{Schafer:1996wv,Diakonov:2002fq,Nowak:1996aj} and many
variations of the Nambu-Jona-Lasinio (NJL) model~\cite{Nowak:1996aj} . Both of which yield an effective Lagrangian of relativistic quarks interacting through chirally symmetric four-point interactions. The main drawback of these models is the lack of color confinement. As a result,
these models overlook the confining interactions that may affect the nature and properties of the transition in the $\mu-T$
plane. Extensions of the NJL models~\cite{Ratti:2006wg,Ratti:2007jf}  and IL models~\cite{Diakonov:2008sg} to include  the effects of confinement by the introduction of the Polyakov line as an effective field that couples to the constituent quarks have shown quantitative changes in the nature of
the transition, as well as excitation spectra.

In the past decade, holographic QCD has proven to be a useful model for addressing QCD for large number of colors
$N_c$ and t$^\prime$ Hooft coupling $\lambda=g^2N_c$ based on the gauge/gravity duality observed in string theory. The duality states that in the double limit $N_c\gg\lambda\gg 1$ certain supersymmetric gauge theories are equivalent to 
a one-dimensional higher gravity theory coupled to some bulk fields that are dual to gauge invariant operators of the boundary quantum field theory.
Although the correspondence is established for type IIB superstring theory in AdS$_5\times$S$_5$ and $\mathcal{N}=4$ super Yang-Mills theory \cite{Maldacena:1997re}, 
it is usually assumed that it holds for a more general class of strongly-coupled quantum field theories.
%to hold for string theory in a curved background. 
In the double limit $N_c\gg \lambda\gg 1$ the string theory reduces to a weakly coupled classical supergravity. 
Some of the most famous models for a top-down holographic dual of QCD with a small number of flavors or $N_f \ll N_c$, include the probe $D3/D7$ model and the Witten-Sakai-Sugimoto model (WSS). In  the probe $D3/D7$ model chiral symmetry breaking is successfully described but it can only be an abelian $U(1)$ symmetry~\cite{d3d7}.   In the WSS model chiral symmetry is non-abelian and its breaking is realized through the geometry of the flavor branes inside the $D4$ background~\cite{wss}. 
A different approach is  the bottom-up construction of gravitational models that mimic the IR behavior of QCD. The initial effort in this direction is the hard wall model which successfully describes some basic features of the  mesonic sector of QCD~\cite{hardwall}.

A more comprehensive bottom-up dual model of 4d Yang-Mills is the improved holographic model pursued by Kiritsis and others (ihQCD). The action of ihQCD is the Einstein-dilaton action, with the bulk dilaton to interpolate between strong and weak-coupling, \cite{ihqcd}. The model compares favorably to lattice results at zero and finite temperature~\cite{ihqcd2}. Flavor degrees of freedom can be described in this context by the low energy fields of $N_f$ brane-antibrane pairs \cite{sen,ckp}.  Those include a  bulk tachyon field that describes the spontaneous breaking of chiral symmetry. The back-reaction of the flavor onto color is considered by taking the Veneziano limit ($N_{c,f} \rightarrow \infty$, $x=N_f/N_c = \rm{fixed}$), which was introduced in~\cite{veneziano}. A holographic model (V-QCD) 
describing this limit of the theory was constructed as a fusion of  the ihQCD action and the flavor brane-antibrane effective action~\cite{jk}. For a wide class of dilaton and tachyon potentials the model produces a phase diagram which resembles the structure of the QCD phase diagram in the Veneziano limit~\cite{bankszaks}.   WSS has also been studied in the limit of backreacting flavor ~\cite{Burrington:2007qd}. The V-QCD holographic model was also studied at finite temperature and chemical potential in~\cite{alte,altemu}, reviling a rich structure in the phase diagram. For zero chemicall potential the phase diagram has been constructed in the $T-x$ plane for different choices of the dilaton and tachyon potentials. In case of finite chemical potential the study was restricted to $x=1$ and the class of dilaton and tachyon potentials which were used is the phenomenologically most relevant. The chemical potential was introduced in an ihQCD-like model in the probe limit in \cite{Stoffers:2010sp}.

The spectrum of V-QCD includes mesons and glueballs which are discrete and gapped in the confined phase,  \cite{aijk}.
The present work considers the deconfined phase of the model both at zero and finite chemical potential. The vacuum state 
and the phase diagram were studied in detail in  \cite{altemu}, where the charged black hole solutions with scalar hair were constructed numerically, \cite{alnum}.
Here, we would like to probe the time-like structure of V-QCD in order to unravel its transport properties
and excitation spectra both of which are of much interest  for comparison with future lattice simulations as well as collider
experiments. In particular, we calculate the quasi-normal modes of the flavored vector and axial-vector mesons, then we compute the corresponding retarded correlators 
and the transport coefficients following standard holographic prescriptions, \cite{transquasi}, \cite{Iqbal:2008by}. The conductivity is be expressed in terms of the background fields  calculated on the horizon. Vector meson spectrum at finite temperature and chemical potential has been studied in several holographic models in the probe limit. In the $D3/D7$ model vector mesons have been studied in non-zero baryon and/or isospin chemical potential, \cite{Erdmengermes} and \cite{mas}, for WSS see \cite{Kim:2006gp}- \cite{DiNunno:2014bxa}. We also calculate the diffusion constant of the longitudinal mode of vector mesons following, \cite{Iqbal:2008by}, \cite{Kovtun:2003wp}. Einstein relation for diffusion constant and charge susceptibility is also verified in the present model. The validity of Einstein relation has also been shown in the probe $D3/D7$ model at finite chemical potential in \cite{mas2}.

In section \ref{vqcdgen}, we detail the key features of the V-QCD model. In section \ref{muvac}, we characterize some aspects of its
phase diagram in the $\mu-T$ plane with a first order de-confining transition followed by a second order chiral transition. In section
\ref{flucteq}, we present the quadratic action of the different towers of meson excitations, as well as their equations of motion. 
The pertinent vector and axial-vector spectral functions are detailed in section \ref{specfun}.  The numerical analysis of the spectral
functions and their quasi-normal low-lying modes are given in section \ref{numres}. In section \ref{condifsus}, we explicitly present the bulk flavor conductivities, we derive the
vector diffusion constant and the susceptibility. We also compare to the thermodynamic calculation of the susceptibility of \cite{altemu}.
 Our conclusions and future recommendations follow in section \ref{concl}. In the Appendices, we
summarize the Shr\"odinger algorithm used for the quasi-normal mode analysis, and detail the expressions for the mode analysis. We present the details  on the expansion of the DBI action for all different excitations.

%%%%%%%%%%%%%%%%%%%%%%%%%%%%%%%%%%%%%%%%%%%%%%%%%%%%%%%%%%%%%%%%%%
\section{V-QCD}
\label{vqcdgen}
%%%%%%%%%%%%%%%%%%%%%%%%%%%%%%%%%%%%%%%%%%%%%%%%%%%%%%%%%%%%%%%%%%
Current holographic approaches to the QCD problem with light quarks are carried in the context of the probe approximation
whereby the light quarks are inserted as spectator D-branes in the limit where $N_f\ll N_c$. The Veneziano limit consists of
addressing the double limit $N_{f,c}\rightarrow \infty$ but fixed $\lambda=g^2N_c$ and $x=N_f/N_c$ in the context of  holographic models or V-QCD.  In the vacuum V-QCD exhibits chiral symmetry breaking for $x\leq 4$, restores chiral symmetry for  $4\leq x\leq 5.5$, \cite{jk}. Beyond the Banks-Casher point or $x>5.5$, \cite{bankszaks}, the theory becomes QED-like.

The bulk action describing the system
\be
 S = S_g + S_f + S_a \,.
\ee
$S_g$ is the gluon action which is the same as the ihQCD action, \cite{ihqcd}. $S_f$ is the brane-antibrane action which was introduced in \cite{sen}, and was proposed as a low energy effective action for the holographic meson sector in \cite{ckp}. The last part, $S_a$, is the action of the CP-odd sector which contains the coupling of the flavor-singlet mesons to the axion, which comes from the closed string sector, \cite{ckp,aijk}. The full form of the the first two parts of the action will be presented below. Since, we currently do not study the flavor-singlet CP-odd excitations and the vacuum state is CP-even the last term is not discussed further. To study the excitation spectrum we expand the above equations to quadratic order around the vacuum solution.

%%%%%%%%%%%%%%%%%%%%%%%%%%%%%%%%%%%%%%%%%%%%%%%%%%%%%%%%%%
\subsection{The glue sector} 
\label{sec:VQCDglue}
To describe the IR physics of large $N_c$ pure 4-dimensional Yang-Mills holographically we use a two-derivative effective action with bulk fields that correspond to the lowest dimension operators of the field theory. This are the metric, the dilaton and the axion field. The metric is dual to the energy-momentum operator of the boundary theory, the dilaton is dual to ${\mathbb Tr} F^2$  and the axion to ${\mathbb Tr} F \wedge F$. As it is mentioned above, the axion part of the action will not be studied here. The following gluonic action was proposed in \cite{ihqcd} 
\be
S_g= M^3 N_c^2 \int d^5x \ \sqrt{-g}\left( R- {4 \over 3}{
(\partial \lambda)^2 \over \lambda^2} + V_g(\lambda) \right) \, .
\label{glueact}
\ee
$M^3=1/(16\pi G_5N_c^2)$ is the 5-dimensional Plank mass and $\l=e^\phi$ is the exponential of the bulk dilaton with the corresponding potential $V_g$. $\lambda$  is interpreted as the holographic t' Hooft coupling. The Ansatz for the finite temperature vacuum solution is

\be
ds^2=e^{2 \Awf(r)} \left( -f(r)\,dt^2+dx_{3}^2+{dr^2 \over f(r)} \right) \, \, , \, \,\, \lambda=\lambda(r) ,
\label{bame}
\ee
where the conformal factor $e^{2 A}$ is identified as the energy scale in field theory and $f(r)$ is the black hole factor. The boundary of the bulk space-time is taken to be at $r=0$ and the field-asymptotics are modified from the usual AdS asymptotics such that they reproduce the perturbative running of t' Hooft coupling.  Hence in the UV, 

\be
 A \sim \ln \left( { \ell \over r} \right)+{4 \over 9 \log (\Lambda r)}+\ldots \,\, ,\,\,\,  \lambda \sim  { 1 \over  \log (\Lambda r)}+\ldots  \,, \,\,  r \to 0 \,,
\label{uvasal}
\ee
, where $\ell_{uv}$ is the AdS radius and $\Lambda$ is the UV scale of the theory.
The dilaton potential is such that it reproduces the perturbative running of t' Hooft coupling in the UV. Even if QCD in the UV is not expected to have a dual description in terms of a gravity theory, since it is weakly coupled, the UV asymptotics of our model provide the correct UV-boundary conditions for the IR description of the system. The requirement of confinement, gapped and discrete glueball spectrum as well as linear Regge trajectories fix the IR behavior of the potential to $V(\lambda) \sim \lambda^{4/3} \sqrt{\ln \lambda}$. Considering a simple interpolation of the asymptotic forms of the potential in the two regions we have 
\be
V_g(\lambda)={12\over \ell_0^2}\biggl[1+{88\lambda\over27}+{4619\lambda^2
\over 729}{\sqrt{1+\ln(1+\lambda)}\over(1+\lambda)^{2/3}}\biggr]\, ,
%\quad
% V_f (\lambda, \tau) = x_f V_{f0} (\lambda) e^{-\frac{3}{2} \tau^2/\ell^2},
\label{VfSB}
\ee
where $\ell_0$ is the AdS radius in case of no backreacting flavors, see Eq. (\ref{adsrad}). 
%By requiring that at high temperature (\ref{glueact}-\ref{bame}) reproduce the Stephan-Boltzmann law for a gluon gas it follows that $(M \elll)^3=1/(45\pi^2)$.

%%%%%%%%%%%%%%%%%%%%%%%%%%%%%%%%%%%%%%%%%%%%%%%%%%%%%%%%%%
\subsection{The flavor sector} \label{sec:VQCDflavor}

The flavor sector is described by generalizing Sen's action which was introduced on flat spacetime in \cite{sen}. This form of the action is used as an effective holographic action of the low energy meson sector of QCD, \cite{ckp},
\be
S_f= - \frac{1}{2} M^3 N_c  {\mathbb Tr} \int d^4x\, dr\,
\left(V_f(\l,T^\dagger T)\sqrt{-\det {\bf A}_L}+V_f(\l, TT^\dagger)\sqrt{-\det {\bf A}_R}\right)\,,
\label{generalact}
\ee
where ${\bf A}_{L/R}$ are
\begin{align}
{\bf A}_{L\,MN} &=g_{MN} + \gf(\l,T) F^{(L)}_{MN}
+ {\h(\l,T) \over 2 } \left[(D_M T)^\dagger (D_N T)+
(D_N T)^\dagger (D_M T)\right] \,,\nonumber\\
{\bf A}_{R\,MN} &=g_{MN} + \gf(\l,T) F^{(R)}_{MN}
+ {\h(\l,T) \over 2 } \left[(D_M T) (D_N T)^\dagger+
(D_N T) (D_M T)^\dagger\right] \, , \nn \\
D_M T &= \partial_M T + i  T A_M^L- i A_M^R T\,,
\label{Senaction}
\end{align}
where $x=N_f/N_c$ and $g_{MN}$ is the induced metric on the brane-antibrane pair. We also defined the covariant derivative appearing in ${\bf A}_{L/R}$. The fields  $A_{L}$, $A_{R}$ as well as $T$ are $N_f \times N_f$ matrices in the flavor space and are dual to the lowest dimension operators of the mesonic sector of the field theory. ${\bf A}_{L/R}$ are dual to the left and right flavor currents, respectively. And the bifundametnal scalar T is dual to the quark mass operator. The Plank mass which appears as and overall factor in front of both $S_g$ and $S_f$. It is fixed by requiring the pressure of the system  to approach the large temperature and zero chemical potential limit of free non-interacting fermions and bosons. This fixes $(M \ell)^3 = (1+7 x/4)/ 45 \pi^2$,  \cite{alte}, and the AdS radius in the presence of backreacting flavors is 

\be
\ell^3 = \ell_0^3 \left( 1+{7 x \over 4} \right) \, .
\label{adsrad}
\ee
 The boundary theory is taken to have fermions of the same mass, so the tachyon field is proportional to the unit matrix in flavor space, $T =\tau(r) \mathbb{I}_{N_f}$,
  where $\tau(r)$ is real. In addition, we consider finite baryon density in the vacuum by introducing a nonzero time component in the singlet-flavor vector combination of the left and right gauge fields.

The tachyon potential near the boundary has an analytic expansion that matches the UV running of t' Hooft coupling and the anomalous dimension of the quark mass operator, \cite{jk}. The tachyon field in the boundary is 

\be
\tau(r)= m_q r(-\log \Lambda r)^{- \gamma}+\langle {\bar q} q \rangle r^3 (-\log \Lambda r)^{\gamma} \, ,
\label{tauuv}
\ee
where the power  $\gamma$ is matched to the coefficients of the anomalous dimension of ${\bar q} q$.  In the IR, the tachyon field diverges and the tachyon potential vanishes as it is argued in \cite{sen}. As it is shown in \cite{ckp} brane - antibrane condensation in confining backgrounds leads to chiral symmetry breaking. The form of the tachyon potential that we use is
\be
V_f(\l,TT^\dagger)=V_{f0}(\l) e^{- a(\l) T T^\dagger} \, .
\label{tachpot}
\ee
This is the string theory tachyon potential where the constants have been allowed to depend on the dilaton $\lambda$.
%\be
% V_f(\l,T)=V_{f0}(\l) e^{- a(\l)\tau^2} \, .
%\label{vf}
%\ee
The rest of the potentials $\kappa(\lambda)$ and $w(\lambda)$ are taken to be independent of $T$ and  have an analytic expansion close to the boundary in terms of $\lambda$. Chiral symmetry breaking, thermodynamics and the meson spectra constrain their IR asymptotics.
The function $w(\l,\t)$ is taken

\be
w(\l,\tau)=\cla^2\kappa(\l) \, ,
\label{wfun}
\ee
where $\cla$ is a dimension full quantity which is found to be $\cla \sim \ell_0$, \cite{altemu}. The potentials $V_{f0}(\lambda)$,  $\kappa (\lambda)$  and $a(\l)$ are given by

\begin{eqnarray}
\label{Vf0SB}
%\hspace{-1cm} 
 V_{f0}& =& {12\over \ell^2}\biggl[{\ell^2\over\ell_0^2}
-1+{8\over27}\biggl(11{\ell^2\over\ell_0^2}-11+2x_f\biggr)\lambda
\nn \\ && % \hspace{-1cm}
  +{1\over729}\biggl(4619{\ell^2\over \ell_0^2}-4619+1714x - 92x^2\biggr)\lambda^2\biggr] \, , \nn \\
%\equiv W_0+W_1\l+W_2\l^2.
 \kappa(\l) &=& {[1+\ln(1+\l)]^{1/2}\over[1+\frac{3}{4}(\frac{115-16x }{27}+{1\over 2})\l]^{4/3}} \quad a(\l)=\frac{3}{2 \, \ell^2} \, .
\end{eqnarray}
It should be noticed that the potentials in IR are remarkably close to the non-critical string theory values on flat space-time, except form the logarithmic corrections, see \cite{aijk}.

%%%%%%%%%%%%%%%%%%%%%%%%%%%%%%%%%%%%%%%%%%%%%%%%%%%%%%%%%%%%%%%%%%%%%%%%%%%%%%%%%%%%%
\section{The finite density vacuum}
 \label{muvac}%Meson Spectra Equations}
%%%%%%%%%%%%%%%%%%%%%%%%%%%%%%%%%%%%%%%%%%%%%%%%%%%%%%%%%%%%%%%%%%%%%%%%%%%%%%%%%%%%%

The boundary theory is taken to have $N_f/N_c=1$, massless quarks and finite baryon density.
Consequently, in the flavor sector, only the tachyon and the flavor singlet vector field  are non-zero in the vacuum. Such vacua were studied in \cite{altemu} and the phase diagram of the model was extensively discussed. 

We briefly review how to include the background chemical potential in the model along the lines of \cite{altemu}.
The vacuum action action reads

{\small
\be
S_f= - M^3 N_c N_f \int d^4x\, dr\,
\left(V_f(\l,\t) e^{5 A(r)} \sqrt{1 + e^{-2 A(r)} \, f(r)\, \kappa(\l,\t) \, \tau^{\prime}(r)^2-e^{-4  A(r)} \,   \cla^4 \, \kappa(\l,\t)^2 \, A_0^{\prime}(r)} \right)\,,
\label{vacact}
\ee}
The equation of motion of $A_0$ can be integrated once with constant of integration $\hat n$ and solve for $A_0^{\prime}$.
%\be
%{\partial L_f \over \partial A_0^{\prime}}={- V_f(\l,\t) \, \CL_A^4\, \kappa(\l,\t)^2 \, e^{A(r)} \,  A_0^{\prime}(r) \over
%\sqrt{1 + e^{-2 A(r)} \, f(r)\, \kappa(\l,\t) \, \tau^{\prime}(r)-e^{-4  A(r)} \,   \cla^4 \, \kappa(\l,\t)^2 \, A_0^{\prime}(r)} }=\hat n.
%\label{A0eom}
%\ee
%From this one solves (we choose the $-$ sign to get $+$ sign in the chemical potential, see
%\eqref{A0cond})
\be
\cla^2 \,A_0^{\prime}(r)= -{e^{2 \, A(r)} G(r)\over \kappa(\l, \t) \, \qt(r)} \sqrt{\qt(r)^2-1},
\label{A0value}
\ee
%\biggl(1+e^{-2 A(r)}\, f(r) \, \kappa(\l,\t) \, \partial_r \tau^2 \biggr)
where we have also introduced the dimensionless factor
\be
\qt(r)=\sqrt{{\nt^2 \over \kappa(\l,\t)^2 e^{6 \Awf} V_f(\l,\t)^2}+1} \, ,
\label{qdef}
\ee
which depends on the charge density $\tilde n$,  defined by scaling out $\CL_A$ by writing
\be
\tilde n={\hat n \over\CL_A^2}. 
\label{tilden}
\ee
 $\CL_A^2$ sets the unit of charge in $n$. We have also introduced the factor
\be 
 \G(r) = \sqrt{1 + e^{-2A(r)}\h(\l,\tau) f(r) (\partial_r \t(r))^2} \, ,
\label{Gdef}
\ee
%so that the ``effective'' metric factor, which often appears due to the nonzero tachyon background, reads
%\be
% e^{2A(r)} +\h(\l,\tau) \, f(r) \, (\partial_r \t(r))^2  =  e^{2A(r)} \G(r)^2 \, .
%\ee
Whenever the bulk fields have been determined from Einstein's equations and from
the $\l$ and $\tau$ equations, $A_0(r)$ can be computed by integrating \eqref{A0value}:
\be
A_0(r)=\mu+\int_0^{r} dr \, \dot A_0(r) \, .
\label{A0int}
\ee
The usual regularity condition for the gauge field on the horizon $A_0(r_h)=0$ fixes the chemical potential
\be
\mu=-\int_0^{r_h} dr \, \dot A_0(r),
\label{A0cond}
\ee
 To construct the phase diagram of the theory one has to solve Einstein's equations for $A(r), f(r), \lambda(r)$ and the tachyon equation for $\tau(r)$ with certain boundary conditions guaranteeing correct AdS asymptotics. Since, the equations of motion are integrated by starting from the IR one should demand that $A(r \to 0)$ and $\lambda(r \to 0)$ are such that all the background solutions (for any $T$ and $\mu$) have the same UV scale $\Lambda_{UV}$. Moreover, the black hole factor should always be normalized $f(r \to 0) \to 1$ and $\tau(r)$ should be such that $m_q=0$. In general, the solutions that are found are either black hole space-times or thermal gas metrics with no horizon. By computing the pressure, the dominant vacuum is found and the phase diagram is constructed. The above process is extensively discussed in \cite{altemu}.
  In  \cite{alnum}, the  mathematica packages for the numerical solution of the background equations of motion are published and it is noticed that due to an error in the input of the numerics, the phase diagram, for the choice of potentials  (\ref{Vf0SB}), changes from the one reported in \cite{altemu}. In the current phase diagram there is no critical point and the chiral transition is second order for the whole range of temperatures.

\begin{figure}[!tb]
\begin{center}
\includegraphics[width=0.69\textwidth]{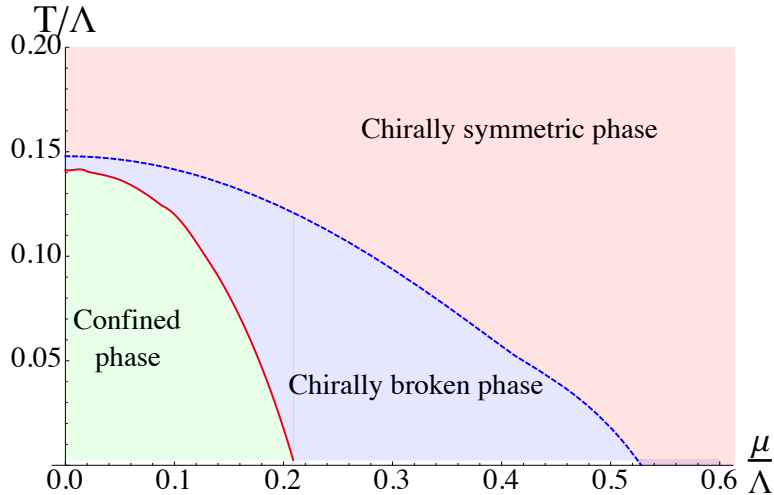}
\end{center}
\caption{The phase diagram in the $T-\mu$ plane for massless V-QCD with $N_f=N_c$. There is a first order confinement-deconfinement transition from thermal gas to black hole 
background and a second order chiral transition after which chiral symmetry is restored.}
\label{phdia}
\end{figure}

Hence, a first order confinement-deconfinement phase transition takes place and a second order chiral phase transition at higher temperature. For low $T$ and $\mu$ the theory is confined and chiral symmetry is broken (green region in Fig. \ref{phdia}). Holographically, it corresponds to a thermal gas metric of the form of (\ref{bame}) with $f(r)=0$. In this case temperature is introduced trivially by compactifing the time circle. As $T$ and $\mu$ increase the system undergoes a first order transition to a deconfined but chirally symmetric phase (blue region). This is a balck hole metric where the tachyon is non-zero.  The second order chiral transition is at higher $T$ and $\mu$ and in the rest of the phase diagram there is a deconfined chirally symmetric plasma, corresponding to $\tau=0$ solutions. The deconfinement transition at $\mu=0$ is $T_c=0.141 \Lambda$ and the chiral transition at $T_{\chi}=0.148  \Lambda$.

%%%%%%%%%%%%%%%%%%%%%%%%%%%%%%%%%%%%%%%%%%%%%%%%%%%%%%%%%%%%%%%%%%%%%%%%%%%%%%%%%%%%%
\section{The excitation equations} 
\label{flucteq}%Meson Spectra Equations}
%%%%%%%%%%%%%%%%%%%%%%%%%%%%%%%%%%%%%%%%%%%%%%%%%%%%%%%%%%%%%%%%%%%%%%%%%%%%%%%%%%%%%

In the present work, we focus to the study of the flavored mesons. %Singlet excitations couple with the dilaton and the metric from the glue action.  
These non-singlet excitations come only the DBI action $S_f$.  
We shall consider the following flavor fluctuations of the tachyon and gauge fields:
\begin{align}
T=(\tau+\mathfrak{s}^a t^a)e^{i\theta+i\,\pi^a t^a}\,.
\label{fluctdef}
\end{align}
where $t^a$ are the generators of $SU(N_f)$ group.  We use the vector and axial combinations of the left and right gauge
fields
\be
V_M = \frac{A_M^L + A_M^R}{2}\, ,\qquad
A_M = \frac{A_M^L - A_M^R}{2}\,.
\label{VAdefstext}
\ee
The associated field strengths will be $V_{MN}$, $A_{MN}$. We choose the gauge $A_r = V_r = 0$.

The fields in the bulk are Fourier transformed. Any field is written as
\be
V_\mu (t,{\bf x},r) =\int {d^4 k \over (2 \pi)^4} e^{-i \omega t+i {\bf k x}}  V_\mu (r,\omega,k)\,.
\ee
Therefore we derive the quadratic actions for the vector, axial vector, pseudoscalar and scalar sectors in the appendices \ref{vecapp}, \ref{axvecapp} and \ref{scapp}. We also present the singlet vector mesons which couple to the metric perturbation corresponding to the heat current. The mixing term is due to finite baryon density. The singlet axial-vector field has an extra term in its equation of motion compared to the non-singlet one, coming from the CP-odd action, see \cite{aijk}. The singlet scalar fluctuations include the scalar part of the metric and the dilaton and the tachyon. Those are all coupled and give rise to mixed glueball and $\sigma$ meson states. Finally, the singlet pseudoscalar fluctuations include the axion field the longitudinal part of the axial-vector and the phase of the tachyon.

%%%%%%%%%%%%%%%%%%%%%%%%%%%%%%%%%%%%%%%%%%%%%%%%%%%%%%%%%%
\subsection{Vector Mesons}
\label{app:NSvmesons}
The quadratic action for the vector mesons in the $V_r=0$ gauge is 

\be
\begin{split}
S_V &= -  {1\over2}\, M^3 N_c\,  {\mathbb Tr} \, \int d^4x\, dr
V_f(\l,\t) \, \cla^4 \,  \kappa(\l,\t)^2\, \qt \, \G(r)^{-1}\,e^{A(r)} \\
& \left[ \frac12\, {\G(r)^2 \over \qt(r)^2}\, V_{ij}V^{ij} -\frac12 \, {\G(r)^2 \over f(r)^2}\, V_{i0}V^{i0}  
 +f  \partial_r V_i \partial_r V^{i} -\qt(r)^2 \partial_r V_0 \partial_r V^{0}
\right]\, ,
\end{split}
\label{vectoracti1m}
\ee
where   $V_{ij}=\partial_i V_j-\partial_j V_i$, and the trace is over the flavor indices. The functions $G(r)$ and $\qt(r)$ are defined in (\ref{Gdef}) and (\ref{qdef}), respectively.   Without  loss of generality, we may take only $k_3=k$, non-zero. Then, the two decoupled equations for the transverse $V_i^{\bot}$ and the longitudinal gauge invariant field ${E_L}= k V_0 +\omega V_3$ are

\begin{align}
& -\partial_r \biggl(  {e^{A(r)} \, f(r) \, \qt(r) \over G(r) }   V_f(\l,\t) \, \kappa(\l,\t)^2  \, \partial_r \, V^\bot_i(r)  \biggr)  \, \nonumber \\
&+{e^{A(r)} G(r) \over \qt(r) }  V_f(\l,\t) \, \kappa(\l,\t)^2\, \biggl( {\bf k}^2 -  {\qt(r)^2  \over f(r) } \, \omega^2 \biggr)   V^\bot_i(r) \,=0 \, , 
\label{tranveceq1}
\end{align}
 
\begin{align}
&E_L(r)'' - \left( \partial_r \log  {V_f(\l,\t) \, \kappa(\l,\t)^2 e^{A(r)} \qt(r)^3 \over G(r)}  -{\omega^2 \over \omega^2 - k^2 {f \over \qt(r)}} \partial_r  \log {\qt(r)^2 \over f(r)}   \right) E_L(r)' \nn \\
&-{G(r)^2 \over f(r)^2} \left( \omega^2 - k^2 {f(r) \over \qt(r)^2} \right) E_L(r)=0 \, .
\label{longveceq1}
\end{align}
%{ G(r) \over  V_f(\l,\t) \, \kappa(\l,\t)^2 e^{A(r)} \qt(r)^3}  \biggl[ \partial_{r} \left({e^{A(r)} \qt(r)^3 \over G(r) } V_f(\l,\t) \, \kappa(\l,\t)^2 \right) {\bf k}^2 \\
%&-  {\qt(r)^4 \over f(r)^2}\, \omega^2 \, \partial_r \left( {e^{A(r)} \, f(r) \, \qt(r) \over G(r) } V_f(\l,\t) \, \kappa(\l,\t)^2  \right)  \biggr] {1 \over {\bf k}^2 -{\qt(r)^2 \over f(r) }\omega^2} \cv(r)' \\
%& +{G(r)^2 \over \qt(r)^2 f(r)} \, \left({\bf k}^2 -{\qt(r)^2 \over f(r)} \omega^2 \right)  \cv(r) =0
%\be
%\begin{split}
%&-\cv(r)'' + { G(r) \over  V_f(\l,\t) \, \kappa(\l,\t)^2 e^{A(r)} \qt(r)^3}  \biggl[ \partial_{r} \left({e^{A(r)} \qt(r)^3 \over G(r) } V_f(\l,\t) \, \kappa(\l,\t)^2 \right) {\bf k}^2 \\
%&-  {\qt(r)^4 \over f(r)^2}\, \omega^2 \, \partial_r \left( {e^{A(r)} \, f(r) \, \qt(r) \over G(r) } V_f(\l,\t) \, \kappa(\l,\t)^2  \right)  \biggr] {1 \over {\bf k}^2 -{\qt(r)^2 \over f(r) }\omega^2} \cv(r)' \\
%& +{G(r)^2 \over \qt(r)^2 f(r)} \, \left({\bf k}^2 -{\qt(r)^2 \over f(r)} \omega^2 \right)  \cv(r) =0
%\end{split}
%\ee
For ${\bf k}=0$, the equations of $V^\bot$ and $E_L$ reduce to the same equation

\be
\frac{1}{V_f(\l,\t)\, \h(\l,\t)^2\, e^{\Awf}\,\G \, \qt \, f^{-1}}
\partial_r \left( V_f(\l,\t)\, \h(\l,\t)^2 \,e^{\Awf}\,
\G^{-1}\, \qt \, f \, \partial_r \psi_V \right)
+\, \omega^2  \,\psi_V = 0 \, ,
\label{vectoreom1m}
\ee
the vector field in the bulk is written as $V_i^{\bot}=V_3=\psi_V$, where $i=1,2$ denotes the transverse coordinates.
Eq. \eqref{vectoreom1m} can be transferred to Schr\"odinger form as shown in Appendix~\ref{app:schro}.
The Schr\"odinger functions for the vector meson equation are
$C_3(r)=M(r)=0$ and
\begin{equation}
\begin{split}
C_1(r)&=V_f(\l,\t)\,  \cla^4 \, \kappa(\l,\t)^2 \, e^{\Awf(r)}\, f(r) \, \G(r)^{-1} \, \qt(r)\,, \\
C_2(r)&=V_f(\l,\t) \, \cla^4 \, \kappa(\l,\t)^2 \, e^{\Awf(r)}\, f(r)^{-1}  \,\G(r)\, \qt(r)\,.
\end{split}
\label{ABdefs1}
\end{equation}
Further defining
\be \label{XiHV}
 \Xi_V(r) = \left(C_1(r)C_2(r)\right)^{1/4} = \,  \cla^2 \, \kappa(\l,\t) \,  \sqrt{V_f(\l,\t)\, \qt(r) \, e^{\Awf(r)}}\,,\qquad H_V(r) = \frac{M(r)}{C_2(r)} = 0\,,
\ee
the Schr\"odinger potential for the flavor non-singlet vectors reads
\be
 V_V(u) = \frac{1}{\Xi_V(u)}\frac{d^2\Xi_V(u)}{du^2} + H_V(u)\,.
\ee
Here the Schr\"odinger coordinate $u$ is defined by
\be \label{udef1}
 \frac{du}{dr} = \sqrt{\frac{C_2(r)}{C_1(r)}} = {G(r) \over f(r)}\,
\ee
and the boundary condition that $u \to 0$ in the UV. $G$ is defined in~\eqref{Gdef}. The definition of the coordinate $u$ will be the same for all non-singlet meson towers, but the potential will generally change.

%%%%%%%%%%%%%%%%%%%%%%%%%%%%%%%%%%%%%%%%%%%%%%%%%%%%%%%%%%
\subsection{Axial Vector and Pseudoscalar Mesons}
Axial-vector and pseudoscalar mesons  mix at finite temperature. The coupled system equations is written in appendix \ref{axvecapp}. Here, we present the excitation equations
for  $ {\bold k}=0$, the fluctuation equation of the transverse modes $A_i^{\bot}$ reads

\begin{align}
&\frac{1}{V_f(\l,\t)\, \h(\l,\t)^2\, e^{A(r)}\,G(r) \, \qt(r) \, f(r)^{-1}}
\partial_r \left( V_f(\l,\t)\, \h(\l,\t)^2 \,e^{A(r)}\,
\G(r)^{-1}\, \qt(r) \, f \, \partial_r \psi_A(r) \right) \nn \\
&-{4 \t(r)^2\, e^{2 A(r)} \, f(r) \over \qt(r)^2 \cla^4 \kappa(\l,\t)}\, \psi_A(r)  +\omega^2 \, \psi_A(r) = 0 \, .
\label{axvectoreom1}
\end{align}
There is now an additional bulk mass term, compared to the vector excitation equation, which depends on the background tachyon field. This comes from the covariant derivative, (\ref{Senaction}). This term is responsible for the splitting of the vectors and axial vectors in the chirally broken phase.
\be \label{XiHA}
H_A(r) =  {4 \t(r)^2\, e^{2 \Awf(r)} \, f(r) \over \qt(r)^2 \cla^4 \kappa(\l,\t)} \,.
\ee
 In the case of ${\bold k}=0$, $\theta$ mixes with $A_0$

\begin{eqnarray}
&& {\mathrm i}\, \cla^4 \, \kappa(\l, \t)^2 \, e^{-2 A(r)}   \, {\qt(r)^2 \over f(r) } \, \omega \, \partial_r A_0  +2 \,  \kappa(\l,\t) \, \t(r)^2 \, \partial_r \theta=0  
\label{k0eq11} \\
&& -\partial_r \biggl( {e^{A(r)} \qt(r)^3 \over G(r) } V_f(\l,\t)\, \cla^4 \, \kappa(\l,\t)^2 \, \partial_r A_0 \biggr)\nn  \\
&& +2\, V_f(\l,\t) \, \kappa(\l,\t) \, \t(r)^2 \, {e^{3 A(r)} \, \qt(r) \, G(r) \over f(r) } \left( 2 A_0 - {\mathrm i} \, \omega \, \theta \right) =0\, .
\label{k0eq21}
\end{eqnarray}
The two equations \eqref{k0eq11} and \eqref{k0eq21} are combined to
\begin{eqnarray}
&&V_f(\l,\t) \, \kappa(\l,\t) \, \t(r)^2 \, e^{3A(r)}\, \qt(r)^{2} \, G(r)^{-1} \, f(r) \, \partial_r \left[  { 1 \over V_f(\l,\t)\, \t(r)^2\, \kappa(\l,\t)\, e^{3 A(r)} \, \qt(r)\,G(r) \, f(r)^{-1} }\partial_{r} \hat \psi_{P} \right] \nn \\
&& - 4 \tau(r)^2  {e^{2 A(r)} \, f(r) \over \cla^4 \, \kappa(\l,\t) \, \qt(r)^2}  \hat \psi_P +\omega^2 \,\hat \psi_P=0 \,,
\end{eqnarray}
where  $ \hat \psi_P=- {e^{A(r)} \qt(r)^3 \over G(r) } V_f(\l,\t)\, \cla^4 \, \kappa(\l,\t)^2 \, \partial_r A_0$ .

%The Schr\"odinger functions read
%\be
%\begin{split}
%C_1(r)&= V_f(\l,\t)^{-1}\, \t(r)^{- 2}\, \kappa(\l,\t)^{-1}\, e^{-3 \Awf} \, \qt(r)^{-1}\,G(r)^{-1} \, f(r) 
%\,,\\
%C_2(r)&=V_f(\l,\t)^{-1}  \, \kappa(\l,\t)^{-1} \, \t(r)^{-2} \, e^{-3A(r)}\, \qt(r)^{-1} \, G(r) \, f(r)^{-1}
%\,, \\
%M(r)&= C_2(r){ 4 \tau(r)^2  e^{2 A(r)} \, f(r) \over \cla^4 \, \kappa(\l,\t) \, \qt(r)^2} \,,
%\end{split}
%\ee
%and $C_3(r)=0$. Therefore,
%\be \label{XiHP}
% \Xi_P(r) = \frac{1}{\t(r)\sqrt{V_f(\l,\t)\,\h(\l,\t)\, e^{3 \Awf(r)} \, \qt(r)}}\,,\qquad H_P(r) = { 4 \tau(r)^2  e^{2 A(r)} \, f(r) \over \cla^4 \, \kappa(\l,\t) \, \qt(r)^2}\,.
%\ee

\subsection{Singlet Vector Mesons}
\label{Svmesons}
The transverse singlet vector mesons $V_i$ are mixed to leading order in $x$ with the metric excitation $h_{ti}$ due to the background charge desity. Vector fluctuations in charged Reissner-Nordstr\"om  have been studied in order to describe charge transport in finite density condensed matter systems, \cite{Hartnoll:2009sz}. Here, we derive the equations of motion of the $U(1)$ vector excitation. Without loss of generality we may consider only the $i=3$ components to be non-zero.  The coupling of the two excitations comes from the DBI action. We expand the action to quadratic order

\be
\begin{split}
S_V & =- M^3 \, N_c \,  {\mathbb Tr} \, \int d^4 x \, dr  V_f(\l,\t) \sqrt{-\mathrm{det}({\mathcal G})} \sqrt{ \mathrm{det} (\delta^M_N +{\mathcal G}^{MR}h_{RN}+\cla^2 \kappa(\lambda, \tau) {\mathcal G}^{MR}V_{RN} )} \\
 & = -{1 \over 2}M^3 \, N_c \, {\mathbb Tr} \, \int d^4 x \, dr  V_f(\l,\t) \sqrt{-\mathrm{det}({\mathcal G})} \biggl( \mathcal{G_S}^{MN} h_{MN} -\cla^2 \kappa(\l,\t) \mathcal{G_A}^{MN} V_{MN} \\
 & - { \cla^4 \kappa(\l,\t)^2 \over 2} \left[\mgs^{MS} \, V_{ST} \, \mgs^{TN} \, V_{NM} + \mga^{MS} \, V_{ST} \, \mga^{TN} V_{NM} -{1 \over 2}\, (\mga^{MN} V_{MN})^2 \right] \\
 & -{1 \over 2} \left[ \mgs^{MS} \, h_{ST} \, \mgs^{TN} \, h_{NM} + \mga^{MS} \, h_{ST} \, \mga^{TN} h_{NM} -{1 \over 2}\, (\mgs^{MN} h_{MN})^2   \right]  \\
 & -2  \cla^2 \kappa(\l,\t) \mga^{MS} h_{ST} \mgs^{TN} V_{NM}
 \biggr) \, .
\end{split}
\ee
The matrix ${\mathcal G}={\bf A}_L={\bf A}_R$, defined in Eq. (\ref{Senaction}), calculated at the background solution. It is further decomposed in its symmetric and antisymmetric parts ${\mathcal G}=\mgs+\mga$, its explicit form is presented in appendix \ref{vecapp}. We consider the case of one non-zero field component, $i=3$, $k_{\mu}=(-\omega,0,0,0)$. Then, 

%\be
%\begin{split}
%S_V & = -{1 \over 2}M^3 \, N_c \, {\mathbb Tr} \, \int d^4 x \, dr  V_f(\l,\t) \sqrt{-\mathrm{det}({\mathcal G})} \biggl(
%\cla^4 \kappa(\l,\t)^2 \mgs^{rr} \mgs^{33} V_{r3}^2      \\
%&+\cla^4 \kappa(\l,\t)^2 \mgs^{00} \mgs^{33} V_{03}^2- \mgs^{00} \mgs^{33} h_{03}^2  + 2 \cla^2 \kappa(\l,\t) \mga^{r0} \mgs^{33} h_{03} V_{r3}
% \biggr) \, .
%\end{split}
%\ee
\be
\begin{split}
S_V & = -{1 \over 2}M^3 \, N_c \, {\mathbb Tr} \, \int d^4 x \, dr  V_f(\l,\t) \cla^4 \kappa(\l,\t)^2 \qt G^{-1} e^A \, \biggl(
f (\pa_r V_{3})^2      \\
&- {G^2 \over f} (\pa_0 V_{3})^2+ {G^2 \over \cla^4 \kappa(\l,\t)^2 f } h_{03}^2  -2 {{\tilde n} G e^{-3 A} \over V_f(\l,\t) \cla^2 \kappa(\l,\t)^2 \qt } h_{03}  \pa_r V_{3}
 \biggr) \, .
\end{split}
\label{u12act}
\ee
Linearizing Einstein equations, the  fluctuation equation of $h_{03}$ reads

\be
-{1\over 2} \pa_r \left( e^{-2 A(r)} h_{03}(t,r) \right) + x\, {\tilde n}\, e^{-3 A(r)}   V_3(t,r)=0
\label{gravfl}
\ee
The equation of the gauge field excitation reads

\begin{eqnarray}
&& -\partial_r \biggl(   V_f(\l,\t) \, \kappa(\l,\t)^2   {e^{A(r)} \, f(r) \, \qt(r) \over G(r) }  \, \partial_r \, V_3  \biggr) + V_f(\l,\t) \, \kappa(\l,\t)^2 {e^{A(r)} G(r) \qt(r) \over f(r) } \, \omega^2    V_3\,  \nn \\
&& - \nt \,  \pa_r \left( e^{-2A} h_{03} \right)=0 \, , 
\label{u1vecfl}
\end{eqnarray}
where we have used $\cla=1$. Combining equations (\ref{gravfl}) and (\ref{u1vecfl})  it is found

\begin{eqnarray}
&& -\partial_r \biggl(   V_f(\l,\t) \, \kappa(\l,\t)^2   {e^{A(r)} \, f(r) \, \qt(r) \over G(r) }  \, \partial_r \, V_3  \biggr) + V_f(\l,\t) \, \kappa(\l,\t)^2 {e^{A(r)} G(r) \qt(r) \over f(r) } \, \omega^2    V_3\,  \nn \\
&& - 2 x\, \nt^2 \,  e^{-3 A(r)}   V_3 =0 \, .
\label{totu1vecfl}
\end{eqnarray}
Hence, the vector excitation obtains an effective bulk-mass term which turns out to be particularly important for the computation of holographic electric conductivity, \cite{Hartnoll:2009sz}.

%%%%%%%%%%%%%%%%%%%%%%%%%%%%%%%%%%%%%%%%%%%%%%%%%%%%%%%%%%%%%%%%%%%%%%%%%%%%%%%%%%

\section{The Spectral function}
\label{specfun}

The retarded correlator is defined as

\be
G_{\mu\nu}^{ab \, R}=- i \theta(x^0-y^0) \langle [ J_{\mu}^a(x), J_{\nu}^b(y) ] \rangle
\ee
where $a,b$ are the $SU(N_f)$ indices. The correlator of the vector and transverse axial-vector current are proportional to $P_{\mu\nu}= \eta_{\mu\nu}-{k_\mu k_\nu \over k^2}$, $G_{\mu\nu}^{ab \, R}=P_{\mu\nu} \Pi^{ab}(k^2)$. In the thermal states that we consider Lorentz symmetry is broken and only rotational symmetry is left. Then, the projector is splitted in transverse and longitudinal parts with respect to the spatial momentum ${\bf k}$, $P_{\mu\nu} = P_{\mu\nu}^T + P_{\mu\nu}^L$ and the Green's function reads

\be
G_{\mu\nu}^{ab \, R}= P_{\mu\nu}^T \Pi_T^{ab}(k^2) +P_{\mu\nu}^L \Pi_L^{ab}(k^2) \,.
\ee
We are interested in the transverse part of the correlator for vector and axial-vector fluctuations. 
%
%The quadratic action and the equation of motion for the transverse vector fluctuations is rewritten here
%
%
%
%\be
%\frac{1}{V_f(\l,\t)\, \h(\l,\t)^2\, e^{\Awf}\,\G \, \qt \, f^{-1}}
%\partial_r \left( V_f(\l,\t)\, \h(\l,\t)^2 \,e^{\Awf}\,
%\G^{-1}\, \qt \, f \, \partial_r \psi_V \right)
%+\, \omega^2  \,\psi_V = 0 \, .
%\label{vectoreom2}
%\ee
%
%\be
%\begin{split}
%S_V &= -  {1\over2}\, M^3 N_c\,  {\mathbb Tr} \, \int d^4x\, dr
%V_f(\l,\t) \, \cla^4 \,  \kappa(\l,\t)^2\, \qt \, \G^{-1}\,e^{\Awf} \\
%& \left[ \frac12\, {\G^2 \over \qt^2}\, V_{ij}V^{ij} +\frac12 \, {\G^2 \over f^2}\, V_{i0}V^{i0}  
% -f  \partial_r V_i \partial_r V^{i} - \qt^2 \partial_r V_0 \partial_r V^{0}
%\right]\, .
%\end{split}
%\label{vectoracti2}
%\ee
We Fourier decompose the vector field, choose $k_\m=(-\omega,0,0,0,k)$ and use the variable $E_L=\omega V_3 + k V_0$. Using the equations of motion (\ref{tranveceq1}) and (\ref{longveceq1}), we find from (\ref{vectoracti1m}) the on-shell action

\begin{eqnarray}
S_V =   { M^3 N_c N_f \over 2} {\delta^{ab} \over 2 N_f} \int &&  {d^4k \over (2 \pi)^4} \Big[
{V_f(\l,\t) \, \cla^4 \,  \kappa(\l,\t)^2\, \qt \, f\,e^{\Awf}  \over \, \G} \nn \\ 
&& \Big(   {{\cal E}^a(-k) {\cal E}^b(k) \psi_E(r) \partial_r \psi_E(r) \over \omega^2 -k^2 {f \over \qt^2}}   + {\cal V}^a_i(-k) {\cal V}^b_i(k)  \psi_V(r) \partial_r \psi_V(r) \Big) \Big]_{r=\epsilon}\, .
 \label{vectoractionsh}
\end{eqnarray}
where $V_i (k, r) =  \psi_V(r) \, {\cal V}^a_i (k) t^a$ and $E_L (k, r) =  \psi_E(r) \, {\cal E}^a (k) t^a$. Taking the second derivative with respect to the sources ${\cal V}_i$ we find the transverse part of the retarded two-point function

\be
G_{ij}^{ab \, R}(\omega,k)= {\delta^{ab} \over 2 N_f} \left(\delta_{ij} - {k_i k_j \over {\bf k}^2 }\right) \Pi_T(\omega,k) \, .
 \ee
$\Pi_T$ is calculated from the bulk on-shell action and is normalized in order to be the same as as in the singlet case. It is found directly from (\ref{vectoractionsh})

\be 
\Pi_T= -M^{3} N_c N_f   \left. {V_f(\l,\t) \, \cla^4 \,  \kappa(\l,\t)^2\, \qt \, f\,e^{\Awf}  \over \, \G} \mathrm{Im} \left( \psi_V \partial_r \psi_V \right) \right|_{r=\epsilon} \, ,
\label{spectdensfin}
\ee
where $\psi_V$ is the solution of \eqref{vectoreom1m} with infalling boundary condition on the horizon and $\psi_{V}(\epsilon)=1$ at the boundary. The spectral density is defined as

 \be 
 \rho=- 2\, \mathrm{Im} G_{ii}^{R}= -6\, \mathrm{Im} \Pi_T\, ,
 \ee
with the spatial indices being contracted.The spectral density can be calculated using the membrane paradigm by redefining a variable proportional to the canonical momentum of the field $V_i^{\bot}$

\be
\zeta=-{V_f(\l,\t) \, \cla^4 \,  \kappa(\l,\t)^2\, \qt \, f\,e^{\Awf} \over   \G \, \omega} { \partial_r \psi_V \over \psi_V} \, .
\ee
The new variable satisfies the following first order equation

\be
\zeta'-{G \omega \over f} \left({\zeta^2 \over V_f(\l,\t) \, \cla^4 \,  \kappa(\l,\t)^2\, \qt \, e^{\Awf}} + V_f(\l,\t) \, \cla^4 \,  \kappa(\l,\t)^2\, \qt \, e^{\Awf} \left( 1-{k^2 \over \omega^2}{f \over \qt^2} \right) \right)=0 \, .
\label{1ordeq}
\ee
The incoming regularity condition on the horizon for $\psi_V$ translates to $\zeta'(r_h)=0$, hence 

\be
\zeta_h = i\, V_{f}(\l_h,\t_h) \cla^4 \,  \kappa(\l_h,\t_h)^2\, \qt_h \,e^{\Awf_h} \, .
\label{bchor}
\ee
And the spectral density in terms of $\zeta$ is

\be 
{ \rho(\omega) \over \omega}=6 M^{3} N_c N_f  \mathrm{Im }\zeta(\epsilon)\,. 
\ee
In the numerical computation of $\rho$, we use the first order equation, (\ref{1ordeq}), since the numerical errors close to the boundary are reduced significantly.

\section{Numerical results}
\label{numres}
\begin{figure}[H]
\begin{center}
\includegraphics[width=0.49\textwidth]{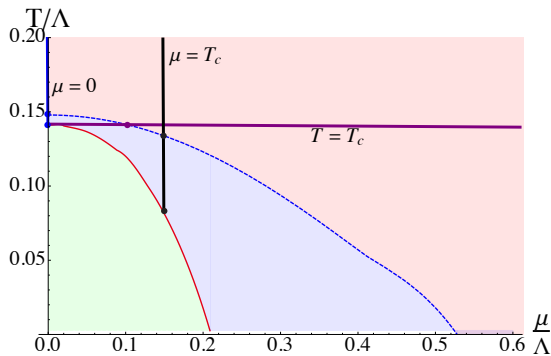}
\end{center}
\caption{The phase diagram in the $T-\mu$ plane with the three lines along which we have studied the quasinormal mode spectrum and the spectral densities of the wave functions. The dots on the lines denote the points where each line meets a transition curve. The three lines are $\mu=0$ (blue), $\mu=T_c$ (black) and $T=T_c=0.1412 \Lambda$ (purple), where $T_c$ is the deconfinement temperature for $\mu=0$.
}
\label{phdialines}
\end{figure}

In this section, the numerical results for the meson spectrum and spectral densities are presented. The temperature and chemical potential dependence of the lowest quasinormal mode of vector and axial-vector mesons for zero spatial momentum has been studied. We found the modes in three different lines of the phase diagram, see figure (\ref{phdialines}).

\subsection{Quasi-normal modes}
\begin{figure}[H]
\begin{center}
\subfigure[]{%
\includegraphics[width=0.49\textwidth]{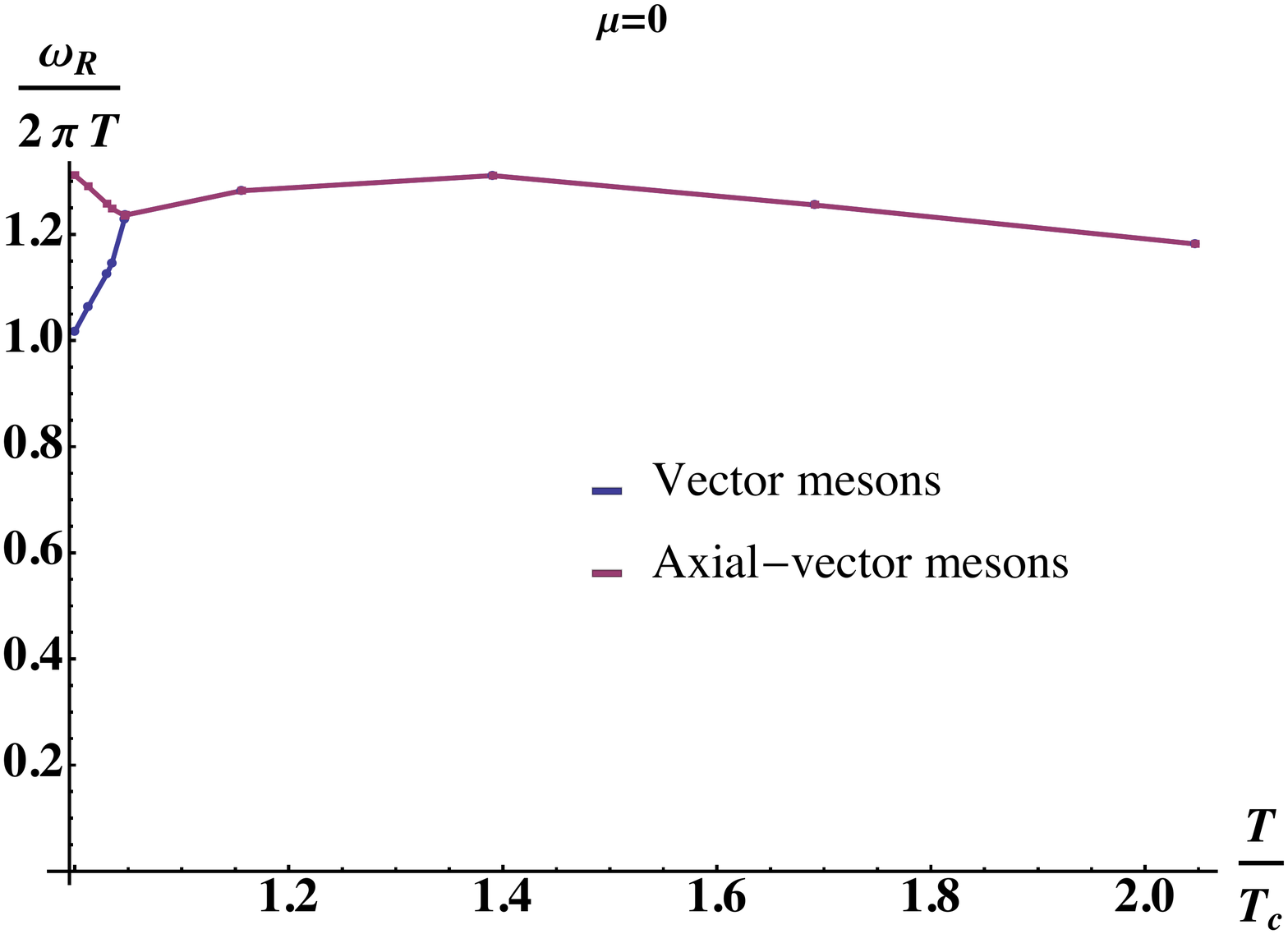}}
\subfigure[]{%
\includegraphics[width=0.49\textwidth]{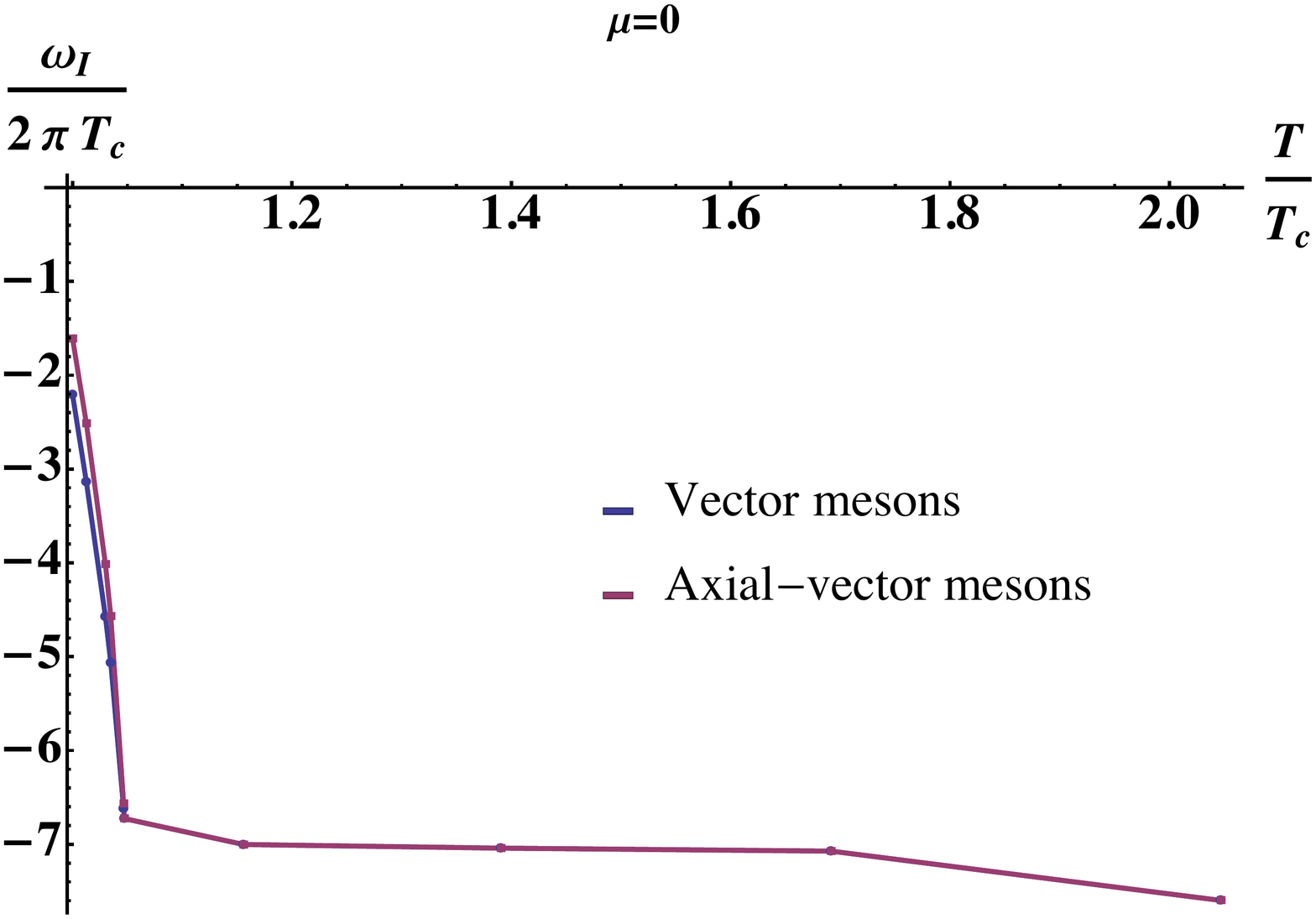}}\quad
\subfigure[]{%
\includegraphics[width=0.49\textwidth]{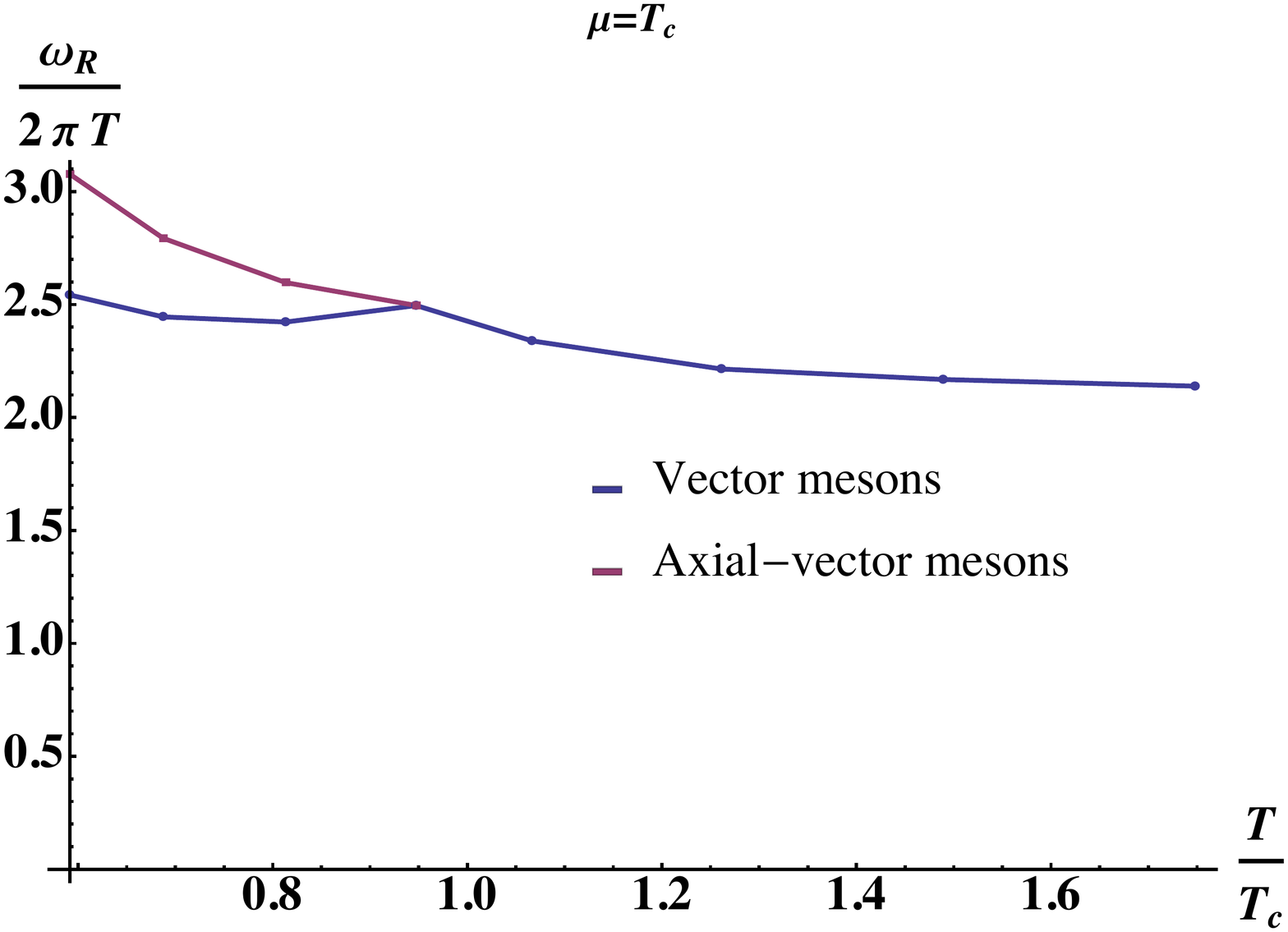}}
\subfigure[]{%
\includegraphics[width=0.49\textwidth]{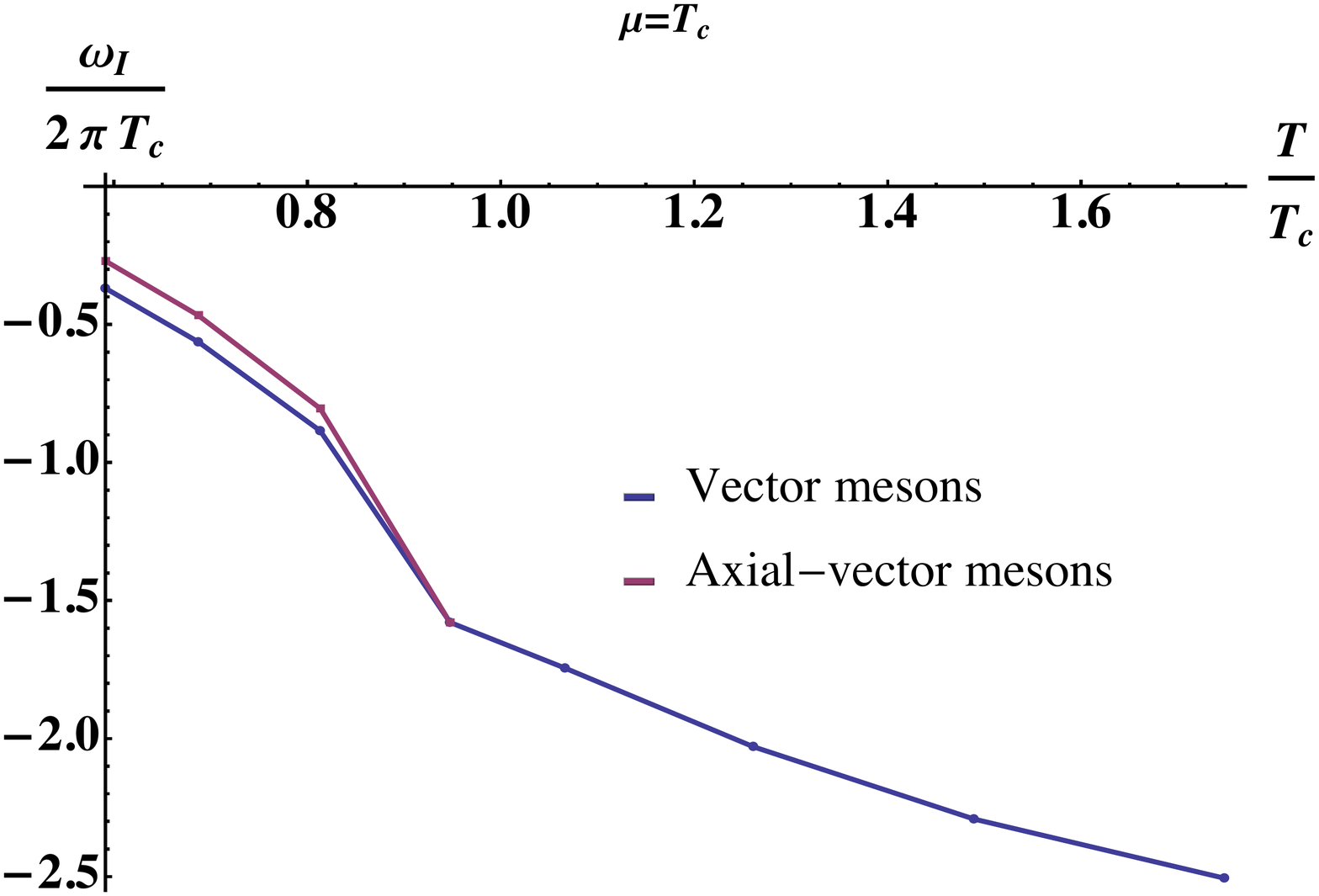}}\quad
\subfigure[]{%
\includegraphics[width=0.49\textwidth]{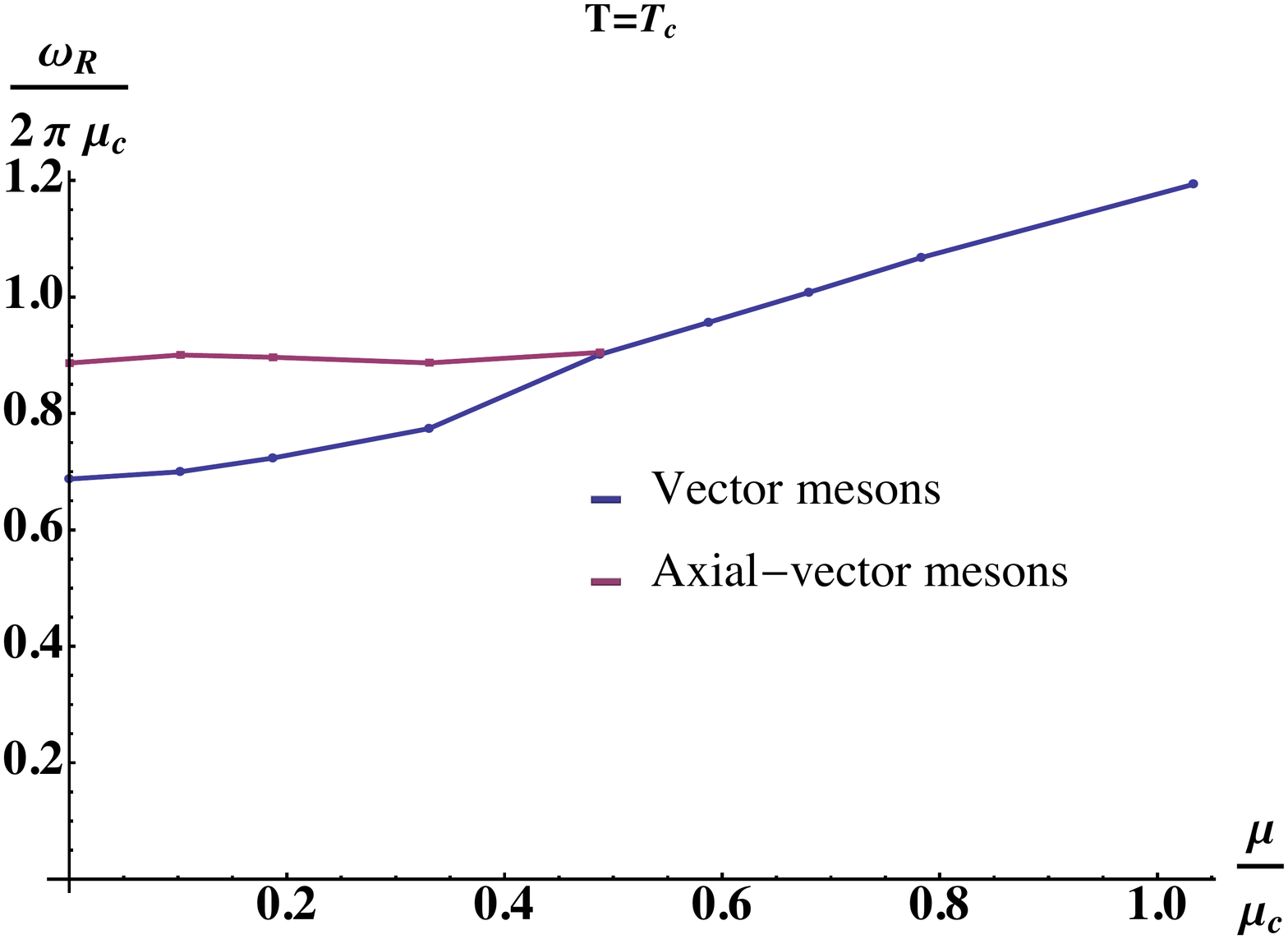}}
\subfigure[]{%
\includegraphics[width=0.49\textwidth]{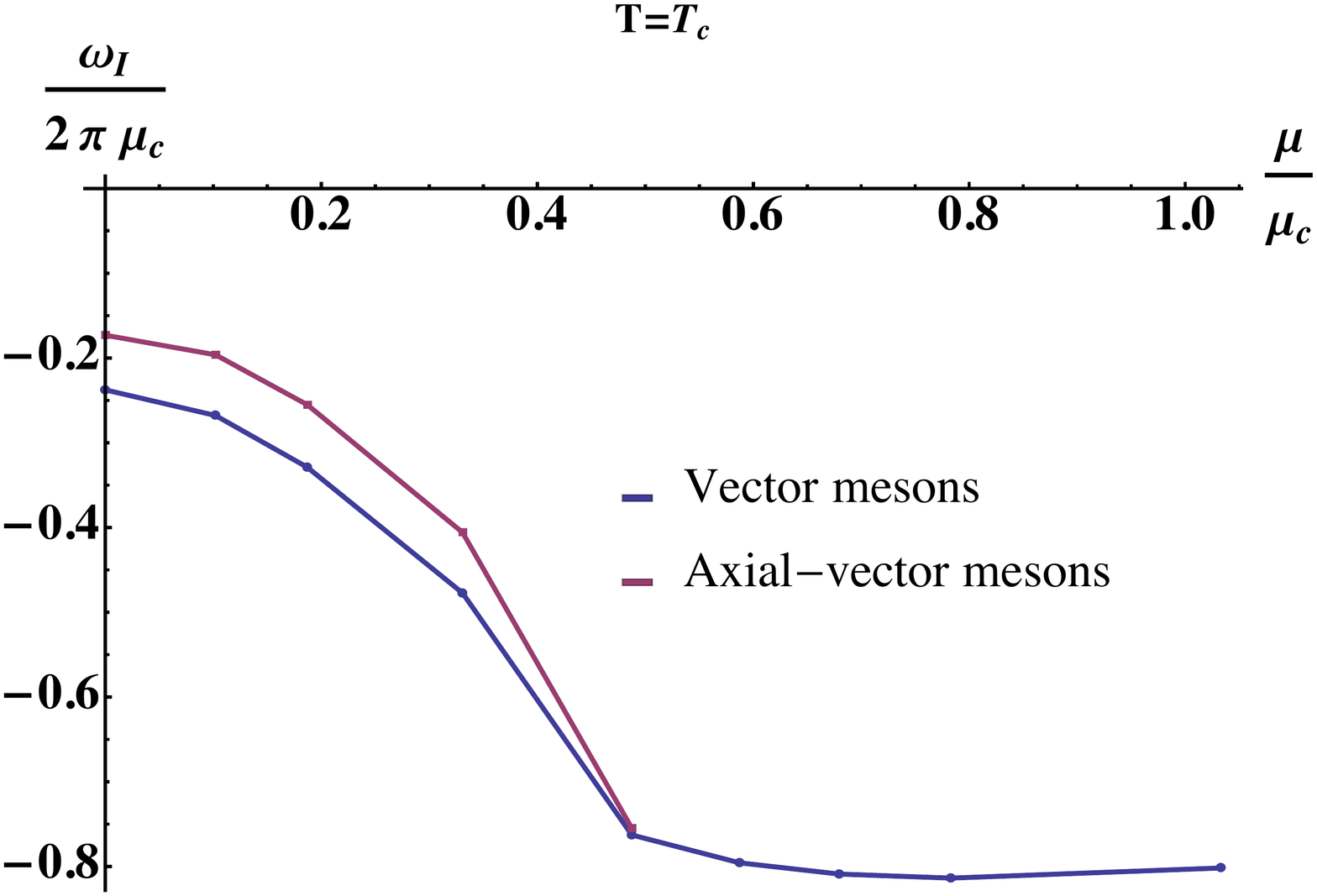}}\quad
\end{center}
\caption{(a), (b): The real ($\omega_R$) and imaginary ($\omega_I$) part respectively of the lowest quasi-normal frequency of vector and axial-vector mesons as a function of  temperature at zero chemical potential (blue line in Fig.  \protect\ref{phdialines}). 
(c), (d): The lowest quasi-normal frequency of vector and axial-vector mesons in terms of temperature at $\mu= T_c$, (black line in Fig. \protect\ref{phdialines}) . 
(e), (f): The lowest quasi-normal frequency of vector and axial-vector mesons as a function of the chemical potential for $T=T_c$ . $\omega$ and $\mu$ are measured in units of $2 \pi \mu_c$. $\mu_c$ is the de-confinement transition chemical potential at  $T=0$}
\label{spect1}
\end{figure}

We solve numerically the quadratic equation of vector and axial-vector mesons, Eq. (\ref{vectoreom1m}) and (\ref{axvectoreom1}). We require normalizability at the boundary, $\psi_{V/A}(r \to0)=0$, and impose infalling boundary condition on the horizon, $\psi_{V/A}(r\to r_h)= (r_h-r)^{- i{\omega \over 4 \pi T}} F(r)$, where $F(r)$ is an analytic function  with a regular expansion on the horizon. We then determine the discrete quasi-normal modes of the mesons. The frequencies are complex, hence it is necessary to scan the complex $\omega$-plane. We numerically shoot from the horizon to the boundary and start with the infalling solution.  It is stressed that in order to have stable numerical results for the frequencies, we expand $F(r)$ to quite high order around $r=r_h$.  We show in Figs.~\ref{spect1}a-b the $T$ and $\mu=0$ dependence of the real and imaginary parts of the lowest vector (rho) and axial-vector (a1) frequencies at zero momentum or $k^i=0$. In this limit the longitudinal and transverse modes coincide. Recall that the longitudinal channel admits also a hydro-dynamical diffusive mode related to global flavor charge conservation, see section \ref{sec:dif}. 
We note that with the restoration of chiral symmetry, the vector and axial-vector frequencies and width coincide as they should. The real part asymptotes about $2\pi T$
while the imaginary appears to change rapidly from $-2T_c$ to level at $-2\pi T_c$ for $T$ in the range 1-2 $T_c$.  For
$\mu=0.15\Lambda_{UV}$ we show in Figs.~\ref{spect1}c-d the same quasi-normal modes. The real part is now seen to asymptote about $4\pi T$ in the same range of temperatures. 

In Figs.~\ref{spect1}e-f we show the quasi-normal frequencies as a function of  $\mu/\mu_c$ along the de-confining line $T=T_c$  (purple line in Fig. \protect\ref{phdialines}). 
The quasi-normal modes presented in this section are the analogue of the time-like poles in real time correlators.
They are to be contrasted with the space-like poles of Euclidean correlators or screening masses widely studied on the lattice~\cite{Bernard:1991ah}. The screening masses are real and asymptote $2\pi T$ at high temperature. It is amusing
to note that about the same asymptotics appear to emerge for the real part of the quasi-normal modes in our analysis
in the chirally restored phase. While the Euclidean $G_E$ and retarded $G_R$ correlators are tied by analyticity 
e.g. $G_E(\omega_n,k)=G_R(i\omega_n,k)$ at discrete Matsubara frequencies $\omega_n=2\pi nT$, the extraction of the time-like poles from Euclidean correlators is still challenging. Conversely, the limit $\omega_n\rightarrow 0$ in the
retarded correlator does not allow us to extract screening masses by analytically continuing the quasi-normal mode
equations. An exception is the hydrodynamical diffusive mode (see below) whose dispersion relation $\omega\approx -iDk^2$ can be extrapolated from the Euclidean screening masses~\cite{Brandt:2014cka}.

\subsection{The Spectral functions}

We also calculate the spectral densities of the vector and axial-vector currents for different temperatures and chemical potentials. We are interested in the transverse channel of the correlator. 
We present the numerical results for $\rho_V$ which was defined in \eqref{spectdensfin}. Similar expression holds for $\rho_A$. The spectral densities are calculated both for zero and and non-zero spatial momentum, $\bf{k}$.  At $\mu=0$ the vector spectral function in Fig.~\ref{spect2}a is finite at $\omega=0$
due to a finite flavor conductivity (see below) and rises almost continuously from the de-confining but chirally broken phase at $T=T_c$ through the chirally
restored phase at higher temperature. The small wiggle at about $\omega/T_c=6$ is due to the broad rho meson noted in the quasi-normal modes. Fig.~\ref{spect2}b is the corresponding flavor axial spectral function with the broad a1at about $\omega/T_c=8$. Figs.~\ref{spect2}b-c show the vector spectral function again at $\mu=0$ but $T=T_c$ and $T=2T_c$ respectively for different momenta. Figs.~\ref{spect2}e-f show the same for the axial spectral function. In both cases, the rho and a1 quasi-normal modes are considerably diluted by a finite momentum. These spectral functions are to be contrasted
with the high temperature vector spectral functions extracted  from lattice simulations~\cite{Brandt:2012jc} (and references therein). We note that the vector
spectral function is continuously increasing in V-QCD in contrast to the extracted lattice vector spectral function which is characterized by a bump at low
$\omega/T$ in~\cite{Brandt:2012jc} (see Fig.9). The same lattice behavior can be reproduced with soft gluon condensates at high temperature~\cite{Lee:2014pwa}. The same results for $\mu=0$ and finite spatial momenta are displayed in Figs.~\ref{spect2}. The rho and a1 are now visible in Figs~\ref{spect2}a-b. The behavior of the vector and axial-vector spectral functions at the de-confining temperature $T_c$ but finite $\mu$ are shown in Figs~\ref{spect3}a-b for for zero spatial momentum. The effects of a finite spatial momentum are shown in Figs.~\ref{spect3}c-f. There are no curent lattice simulations of these
spectral functions at finite $\mu$ and $k$. It is worth emphasizing that both the vector and axial-vector spectral functions at finite $T,\mu$ and varying $\omega,k$ play a key role in the electro-magnetic emissivities of hadonic and partonic matter at collider energies~\cite{Lee:2014pwa} (and references therein). The present holographic set-up for V-QCD offers therefore valuable insights to these spectral functions especially at finite $\mu$ and $k$.

\begin{figure}[H]
\begin{center}
\subfigure[]{%
\includegraphics[width=0.49\textwidth]{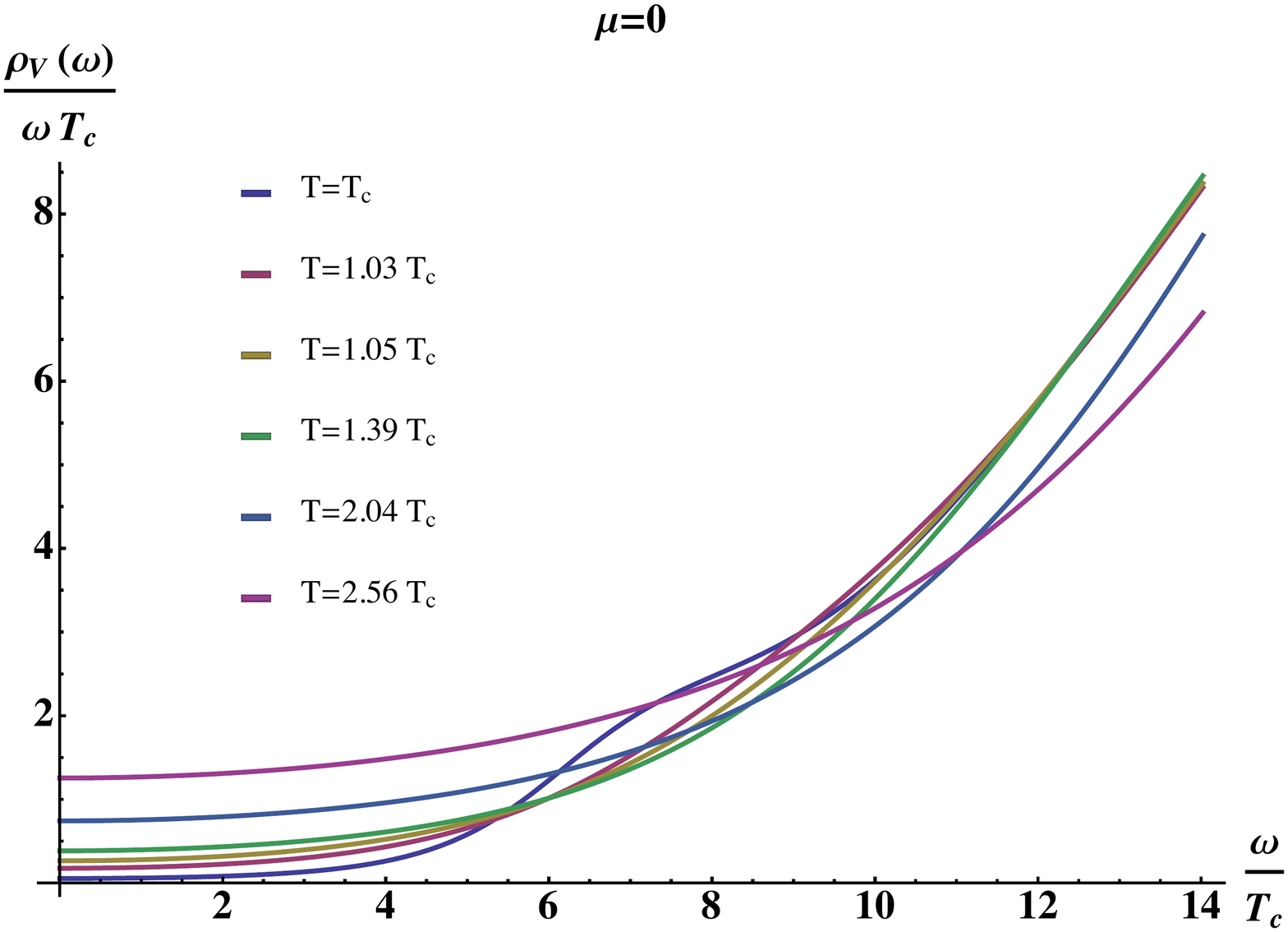}}
\subfigure[]{%
\includegraphics[width=0.49\textwidth]{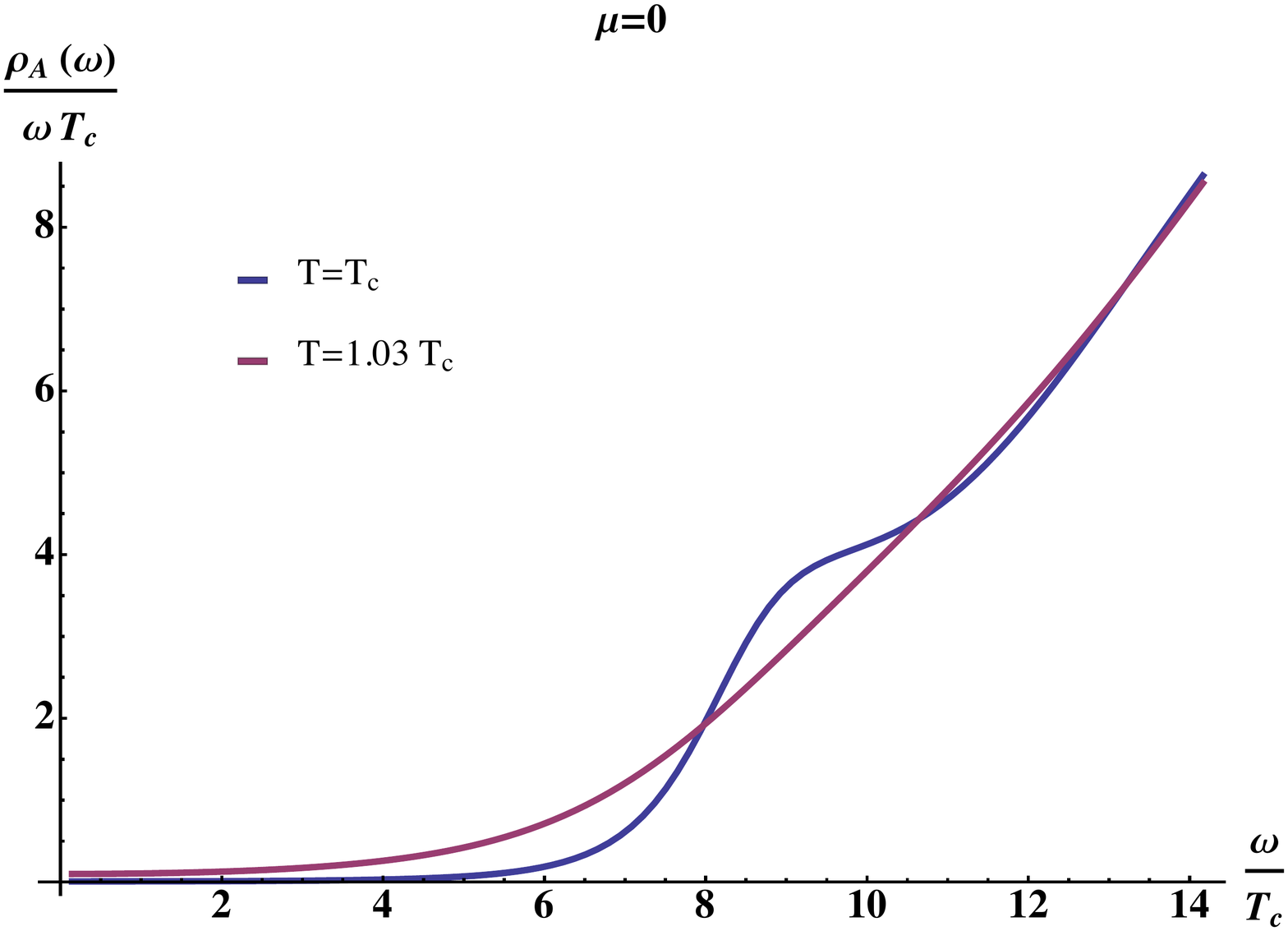}}\quad
%\end{center}
%\end{figure}
%
%\begin{figure}[H]
%\begin{center}
\subfigure[]{%
\includegraphics[width=0.49\textwidth]{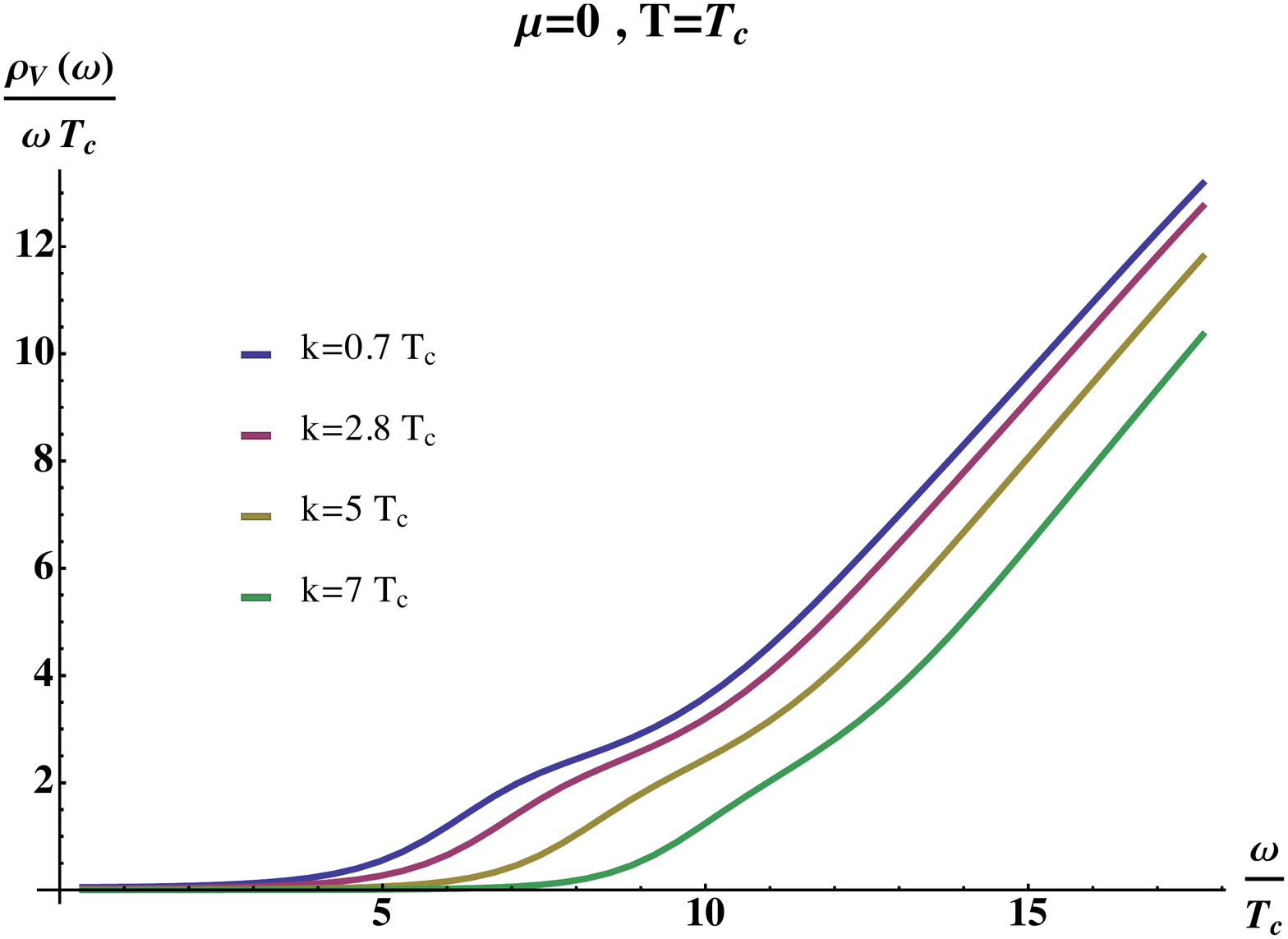}}
\subfigure[]{%
\includegraphics[width=0.49\textwidth]{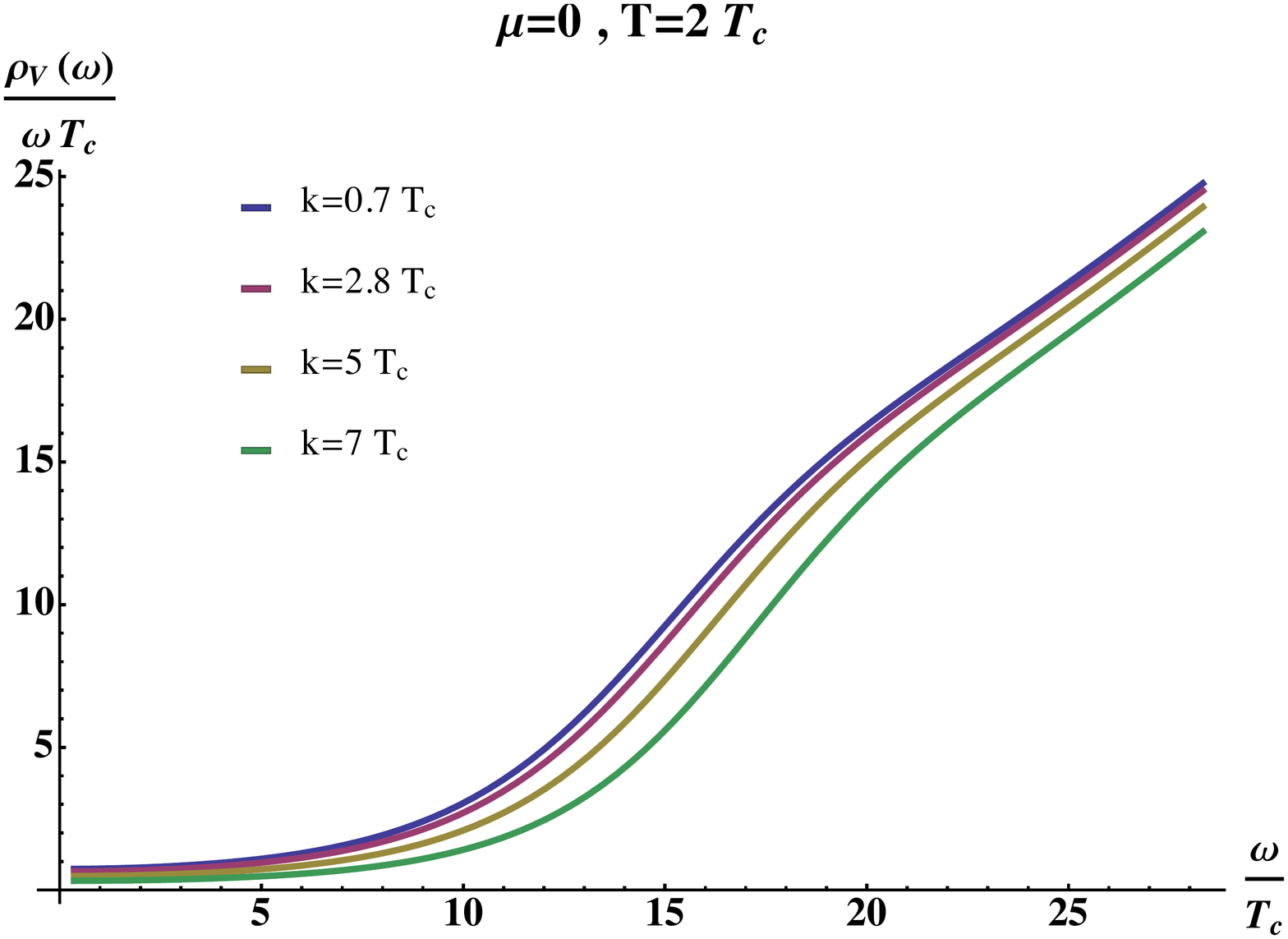}}\quad
\subfigure[]{%
\includegraphics[width=0.49\textwidth]{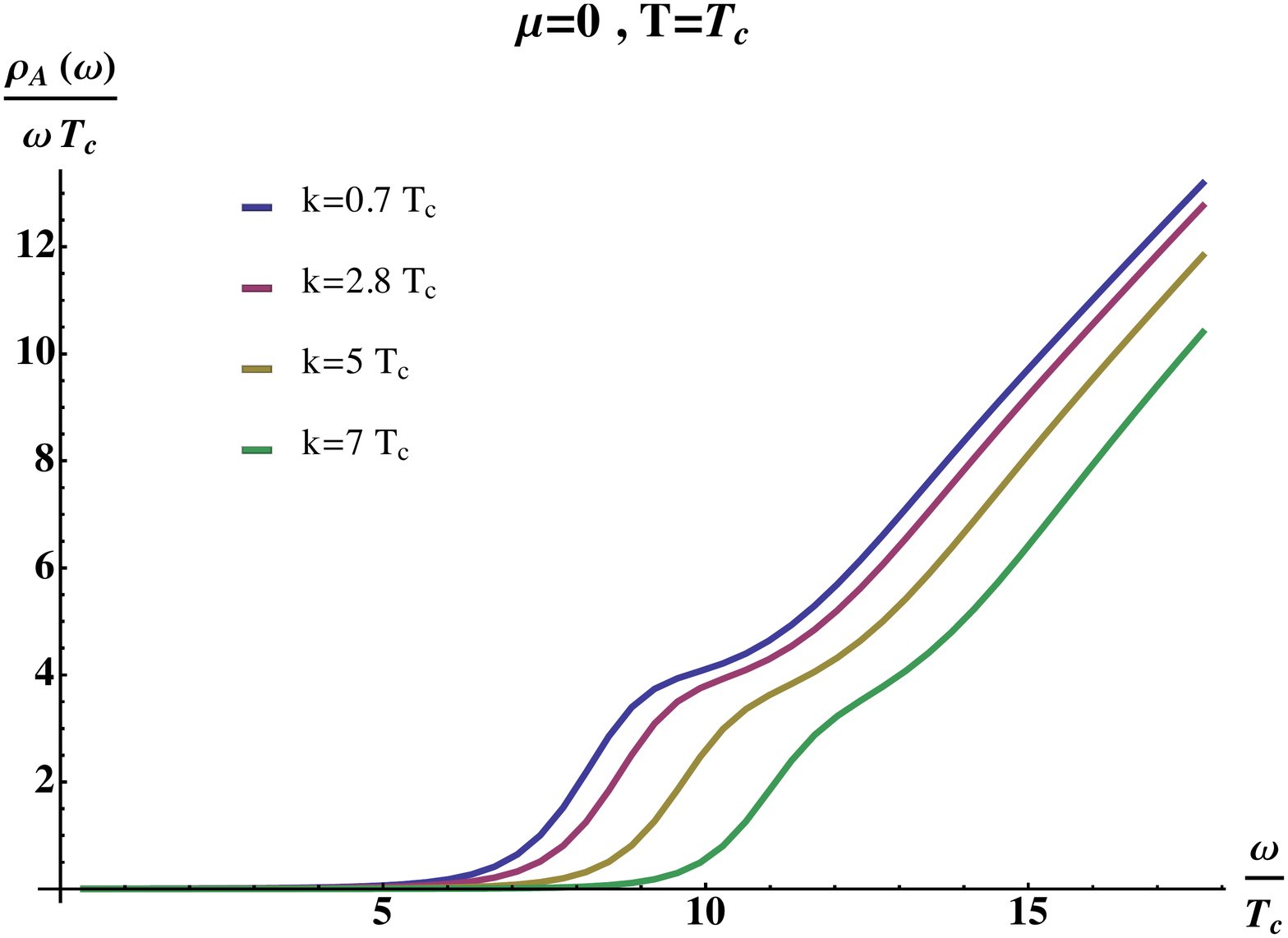}}
\subfigure[]{%
\includegraphics[width=0.49\textwidth]{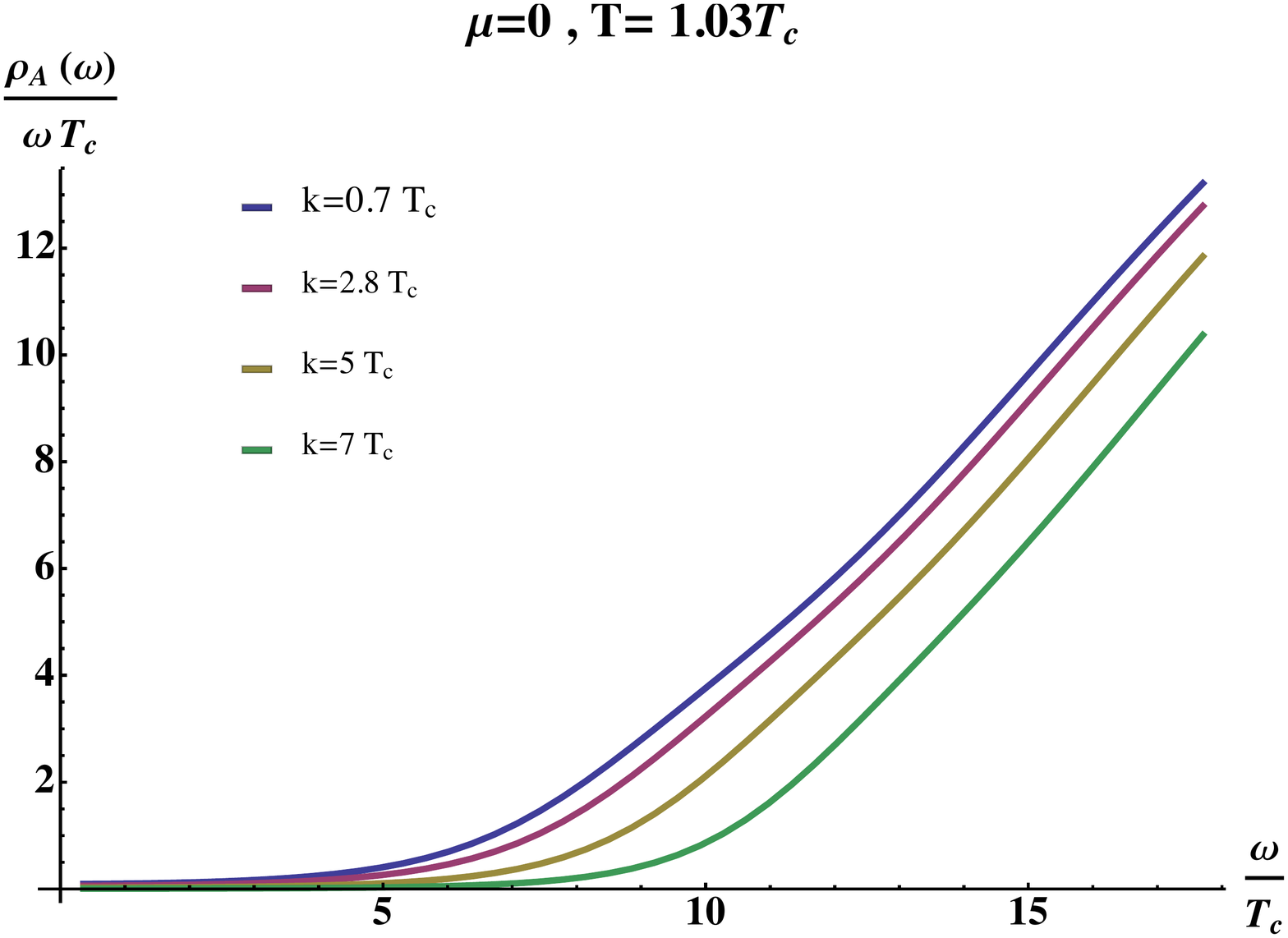}}\quad
\end{center}
\caption{(a), (b): Vector and Axial-vector spectral densities, respectively, at $\mu=0$ for different temperatures as a function of $\omega$ for ${\bf k}=0$. 
(c), (d): Vector spectral density at $\mu=0$ for different spatial momenta ${\bf k}$ as a function of $\omega$ at $T=T_c$ and $T=2T_c$, respectively. 
(e), (f):  Axial-vector spectral density at $\mu=0$ for different spatial momenta ${\bf k}$ as a function of $\omega$ at $T=T_c$ and $T=1.03T_c$, respectively.The dimensional quantities are expressed in units of $T_c$.}
\label{spect2}
\end{figure}

\begin{figure}[H]
\begin{center}
\subfigure[]{%
\includegraphics[width=0.49\textwidth]{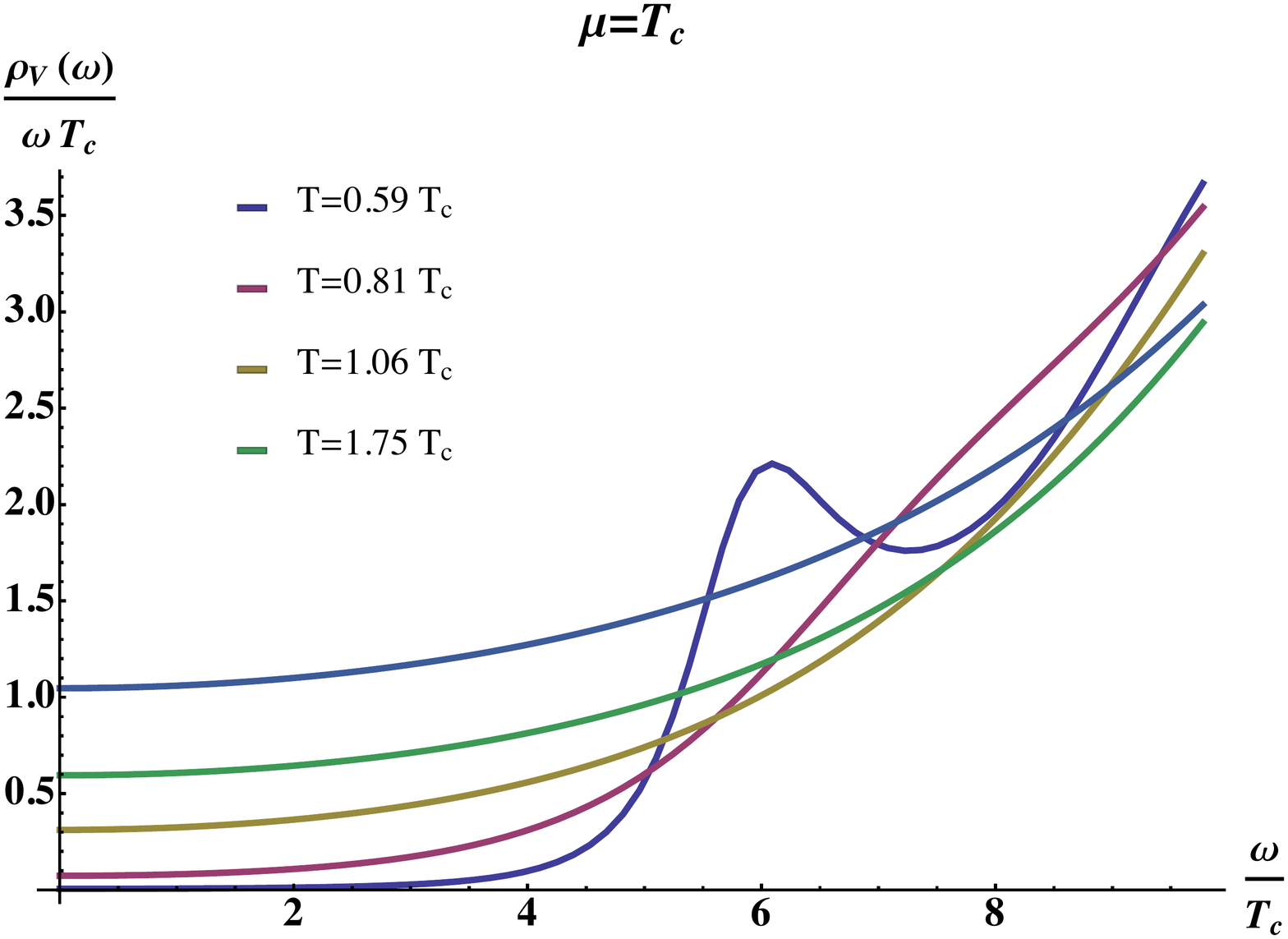}}
\subfigure[]{%
\includegraphics[width=0.49\textwidth]{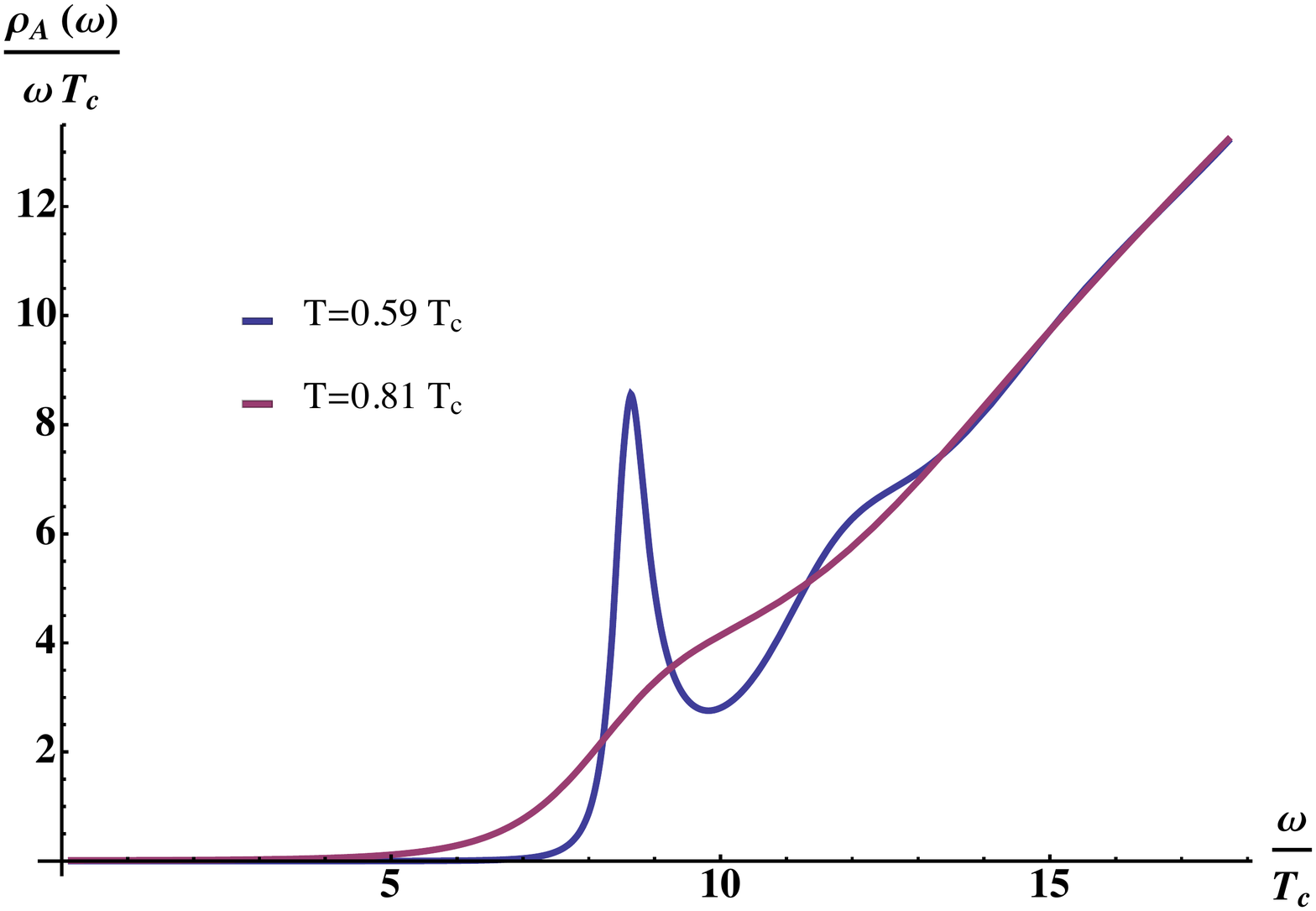}
} \quad
\subfigure[]{%
\includegraphics[width=0.49\textwidth]{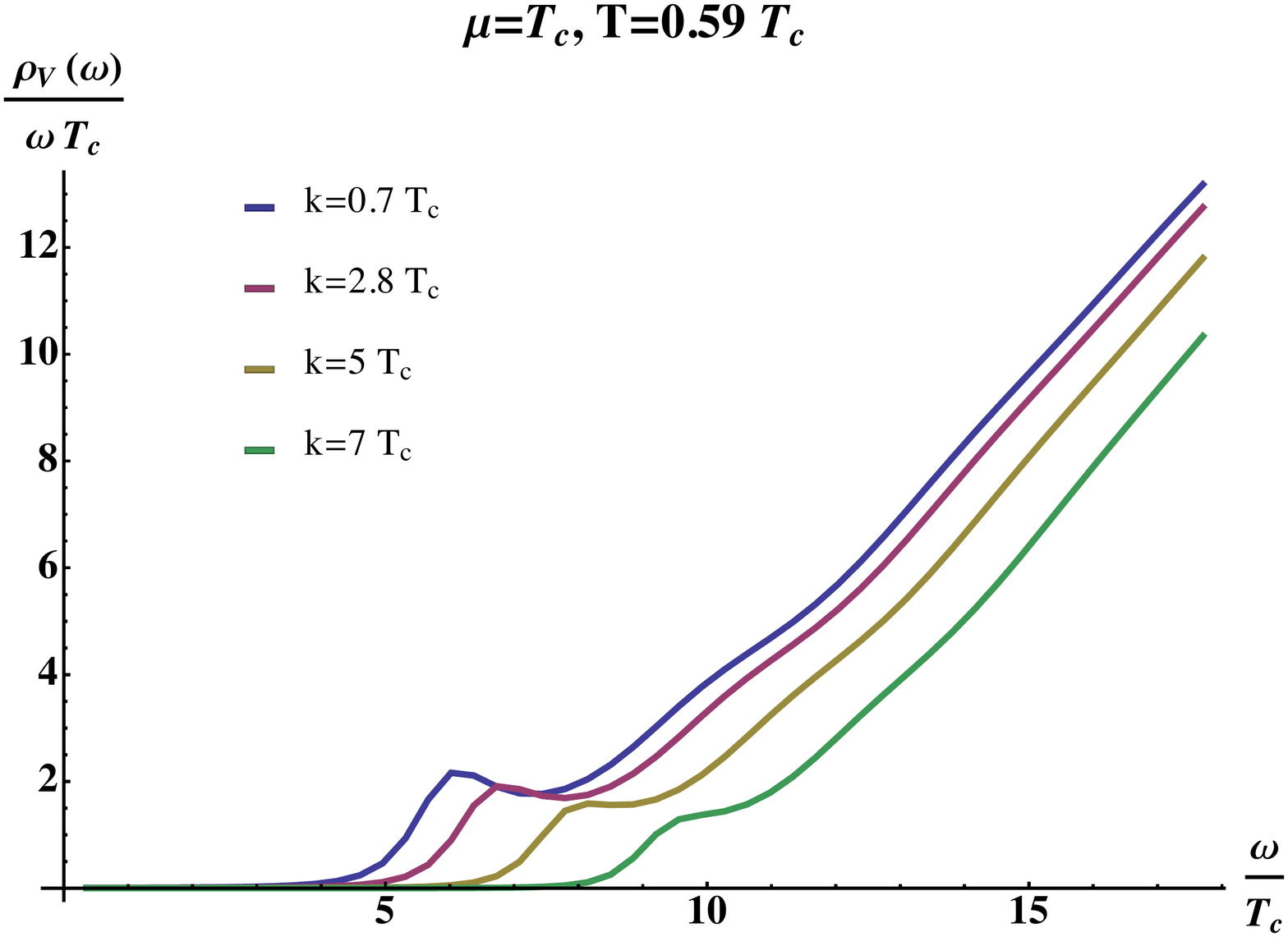}}
\subfigure[]{%
\includegraphics[width=0.49\textwidth]{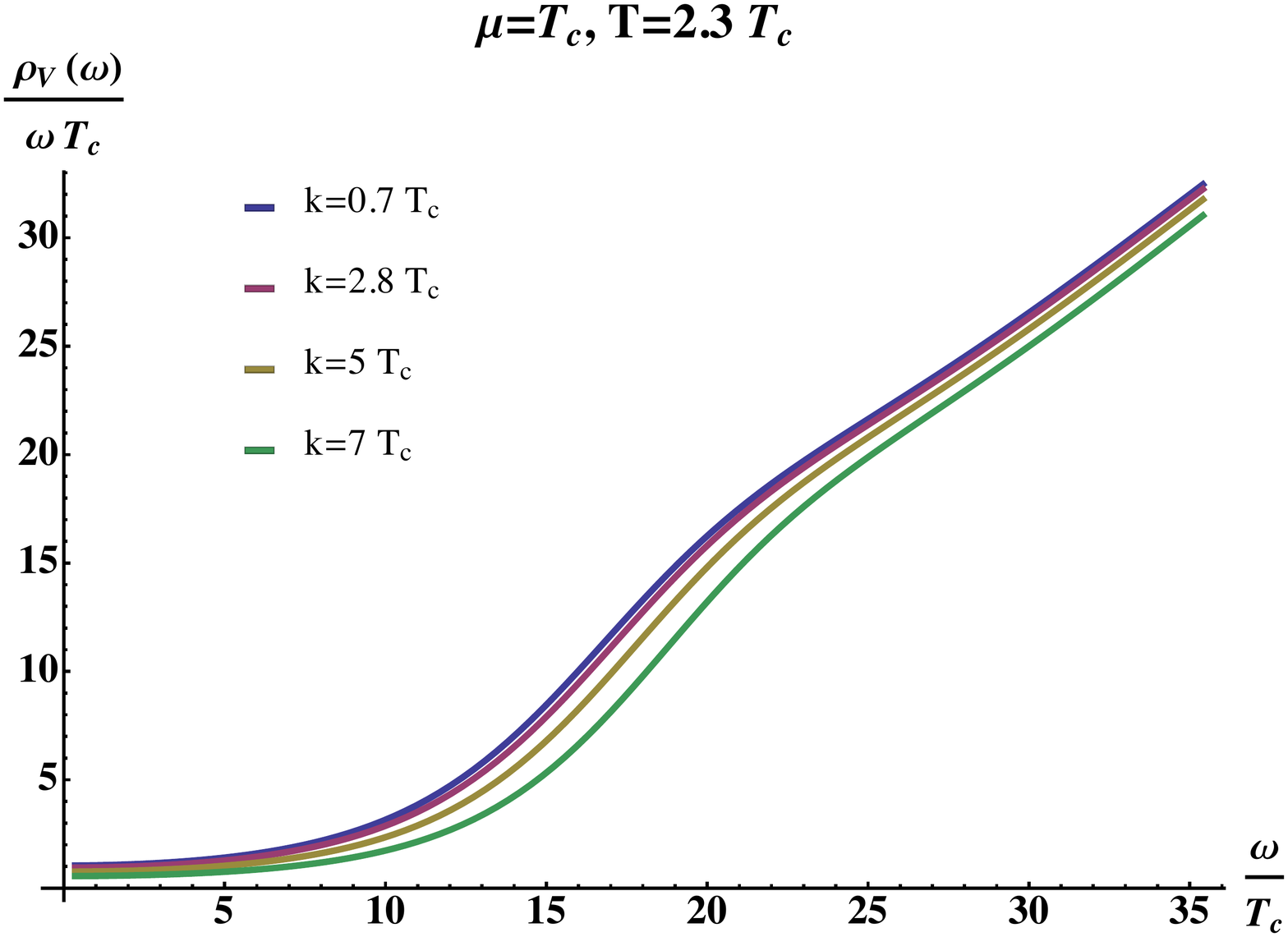}
} \quad
\subfigure[]{%
\includegraphics[width=0.49\textwidth]{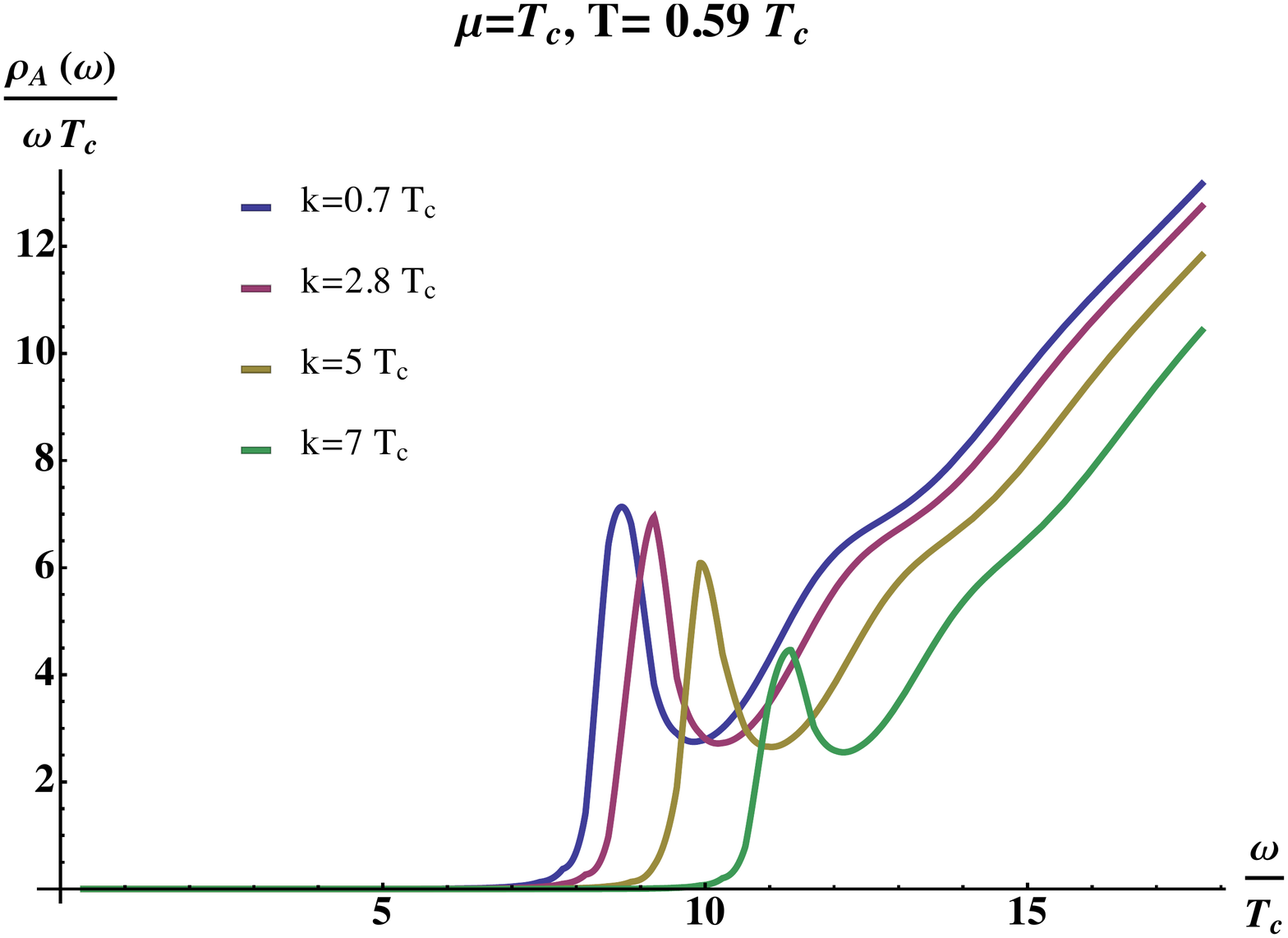}}
\subfigure[]{%
\includegraphics[width=0.49\textwidth]{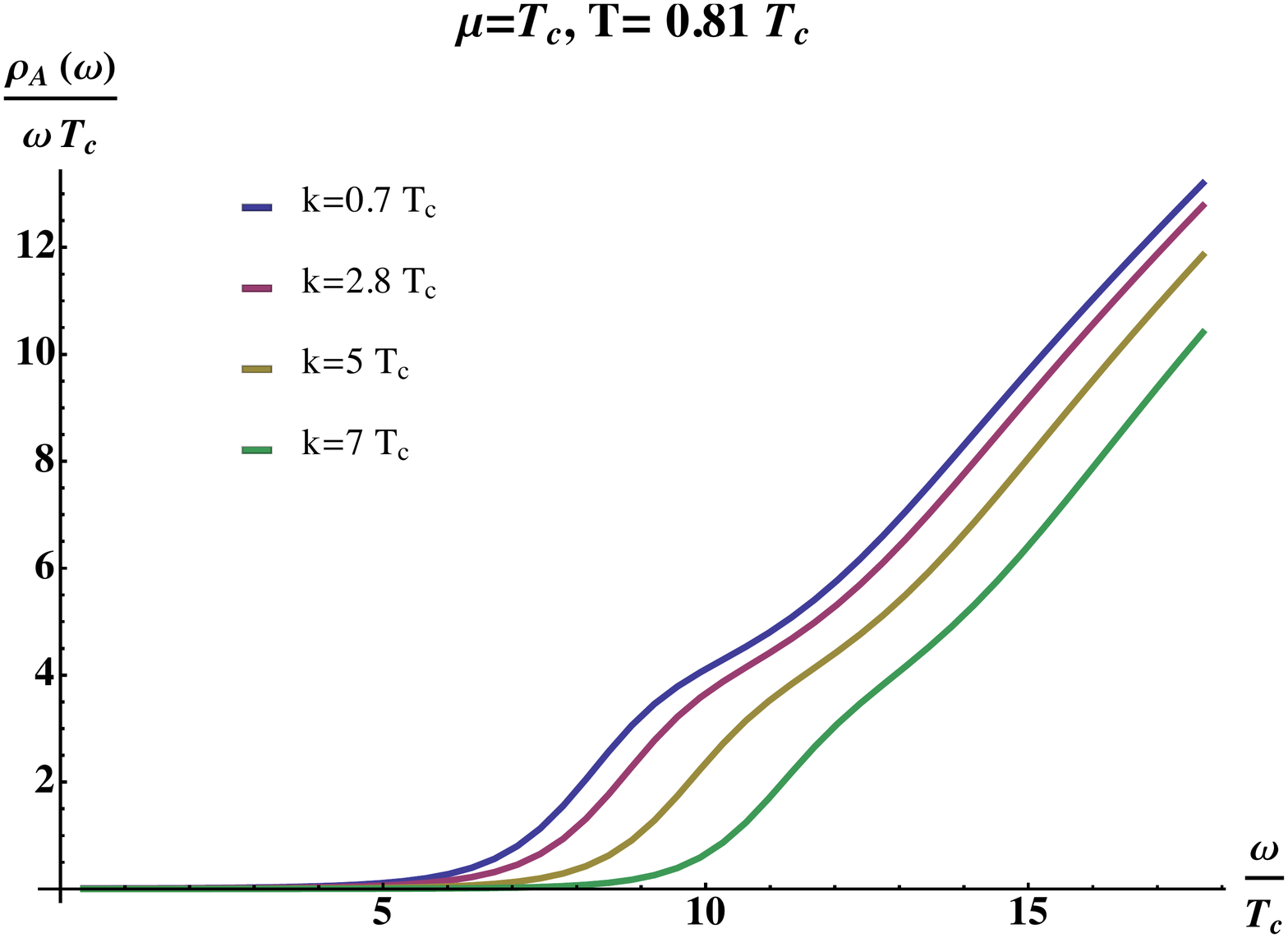}
}\quad
\end{center}
\caption{(a), (b): Vector and Axial-vector spectral densities, respectively, at $\mu=0.15$ for different temperatures as a function of $\omega$ for ${\bf k}=0$. 
(c), (d): Vector spectral density at $\mu=0.15$ for different spatial momenta ${\bf k}$ as a function of $\omega$ at $T=0.59 T_c$ and $T=2.3T_c$, respectively.
(e), (f):  Axial-vector spectral density at $\mu=0.15$ for different spatial momenta ${\bf k}$ as a function of $\omega$ at $T=0.59T_c$ and $T=0.81T_c$, respectively.The dimensional quantities are expressed in units of $T_c$.}
\label{spect3}
\end{figure}

\begin{figure}[H]
\begin{center}
\subfigure[]{%
\includegraphics[width=0.49\textwidth]{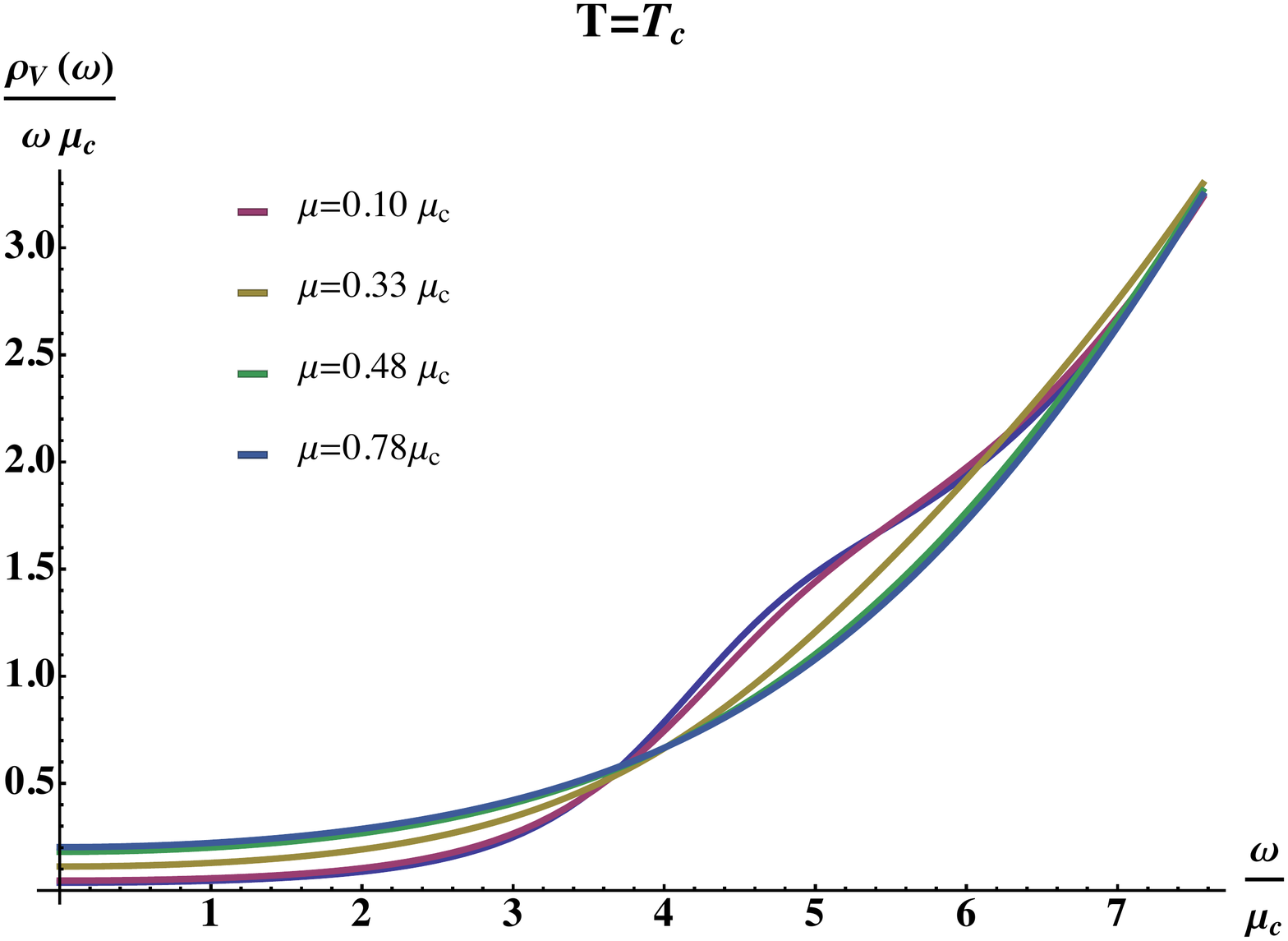}}
\subfigure[]{%
\includegraphics[width=0.49\textwidth]{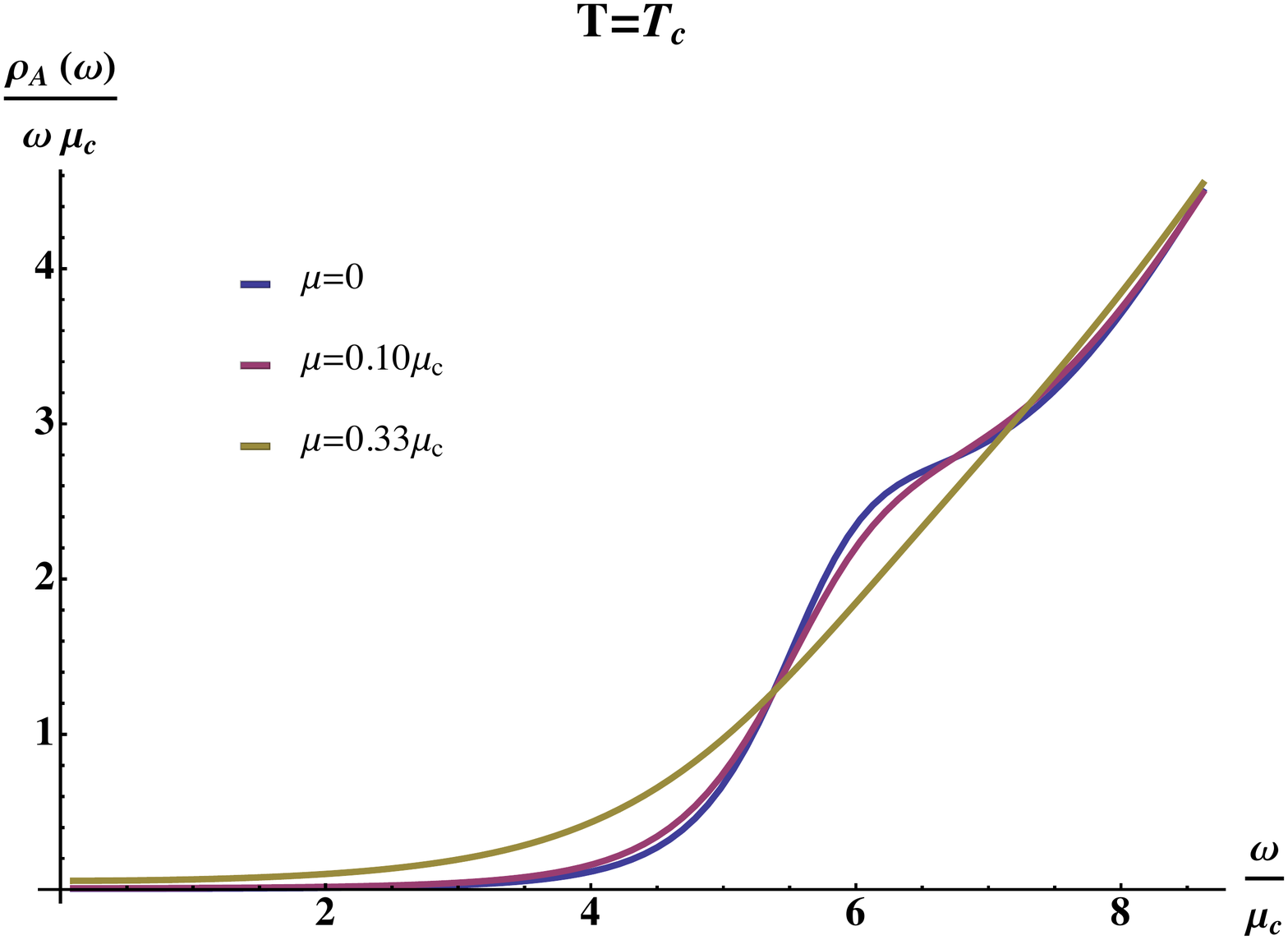}}\quad
\subfigure[]{%
\includegraphics[width=0.49\textwidth]{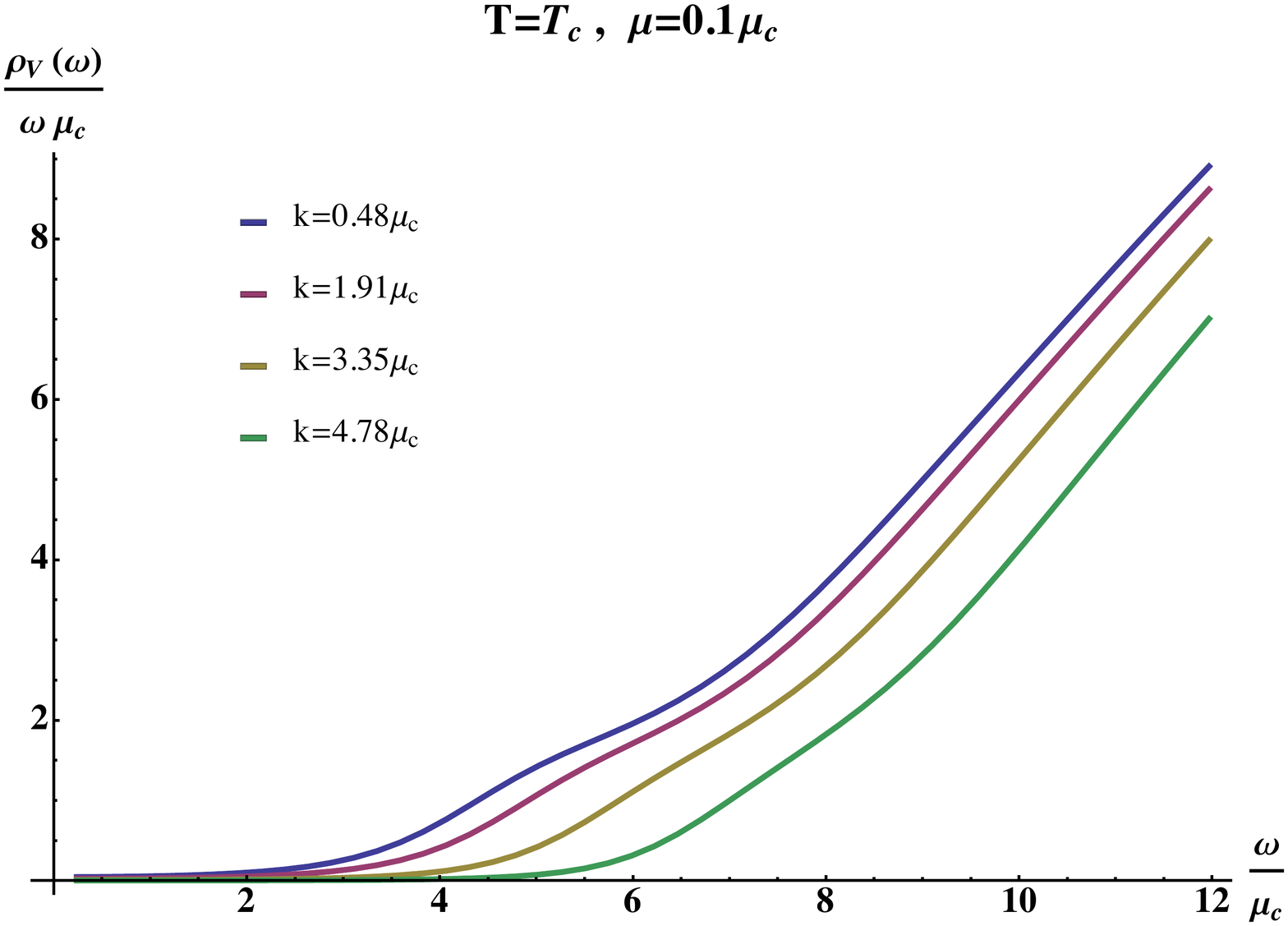}}
\subfigure[]{%
\includegraphics[width=0.49\textwidth]{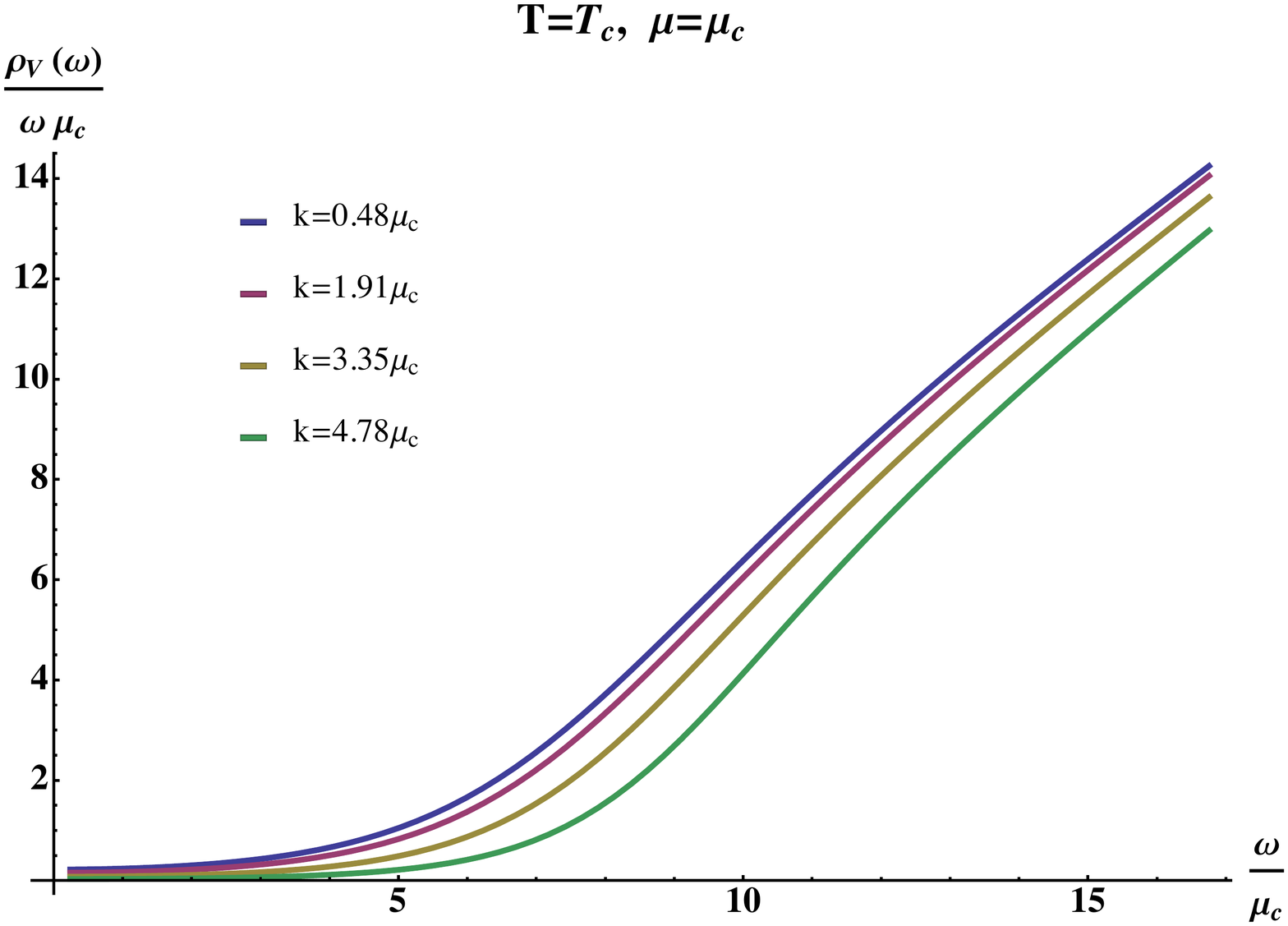}}\quad
\subfigure[]{%
\includegraphics[width=0.49\textwidth]{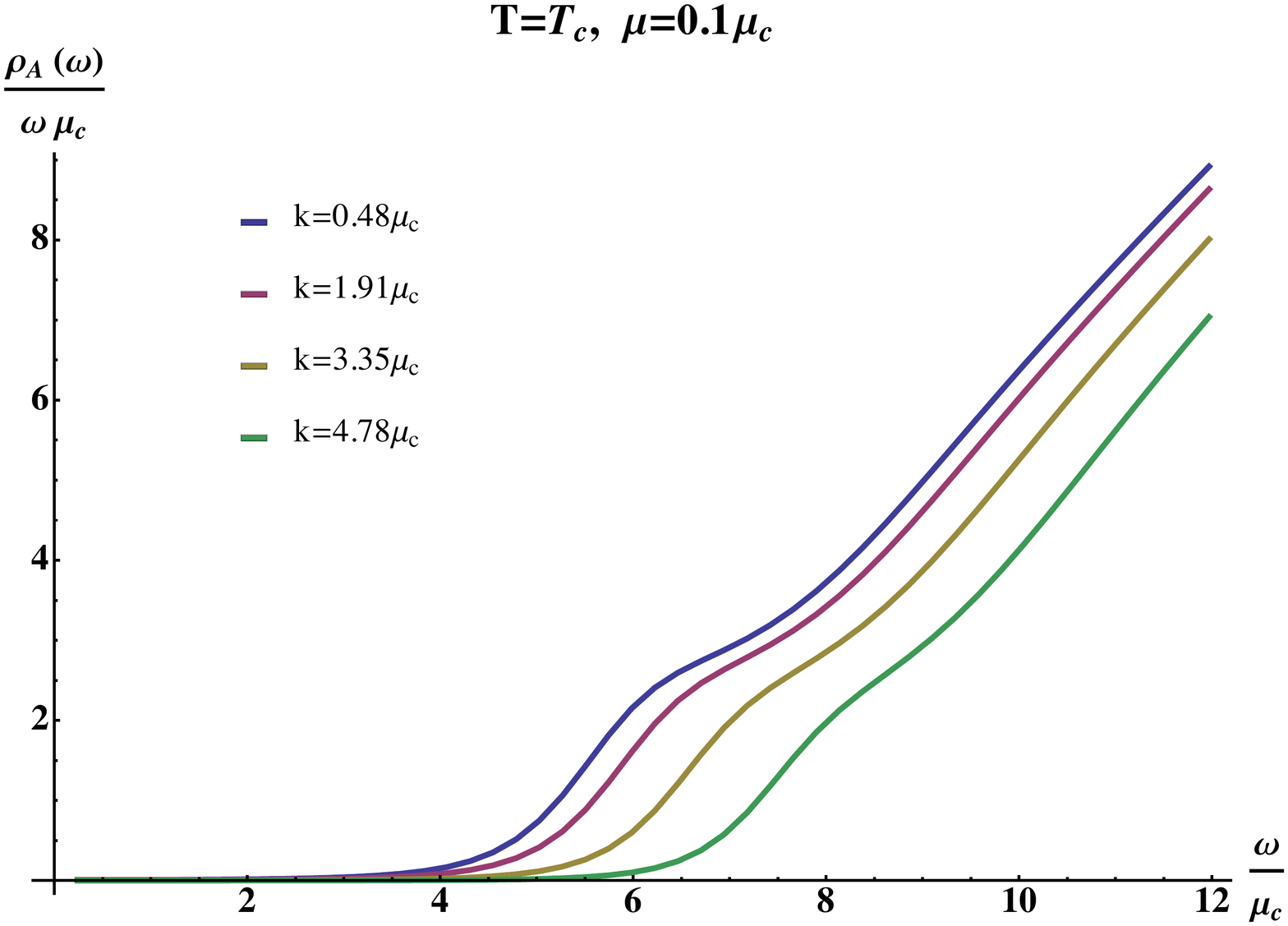}}
\subfigure[]{%
\includegraphics[width=0.49\textwidth]{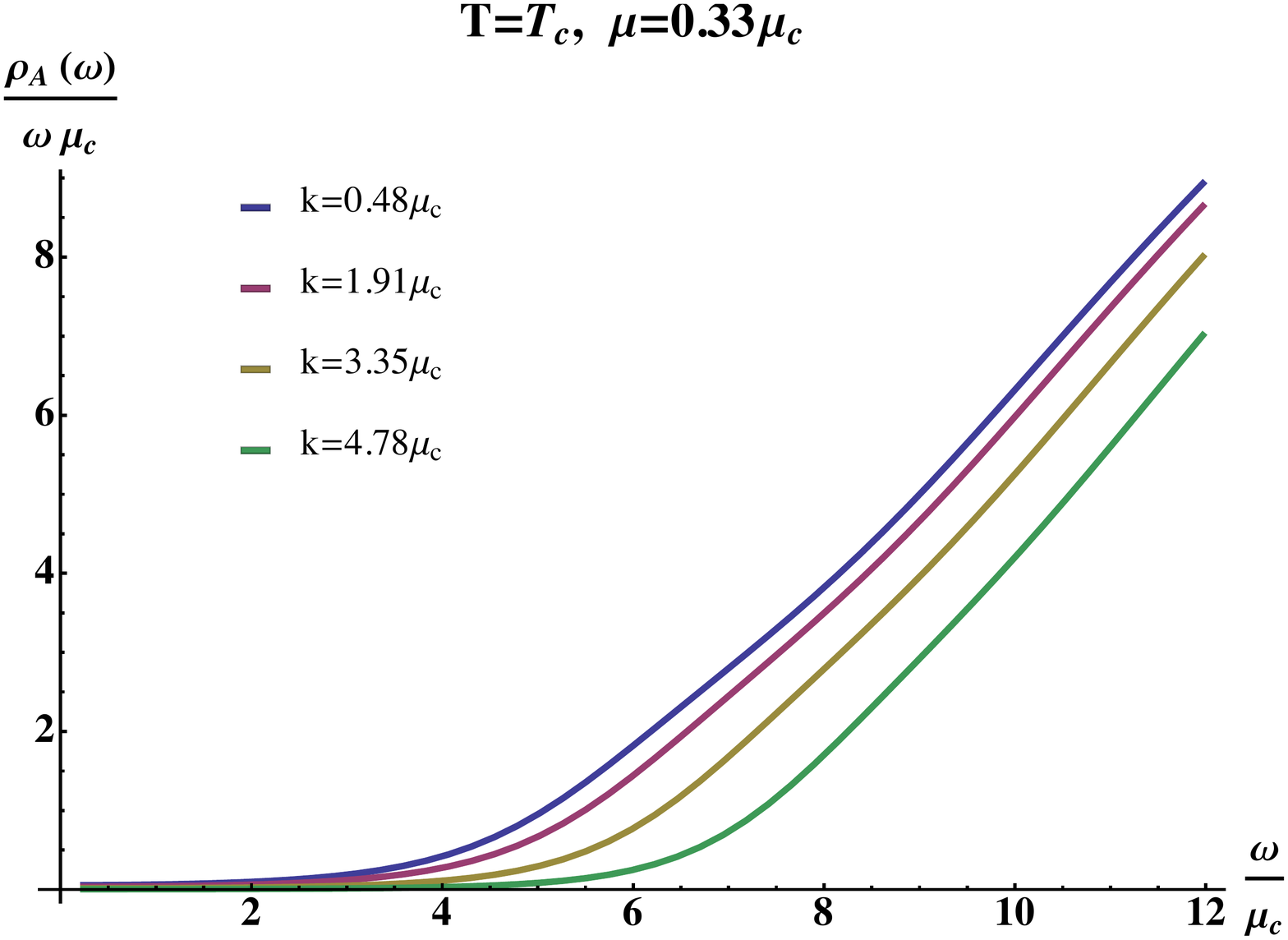}}\quad
\end{center}
\caption{(a), (b): Vector and Axial-vector spectral densities, respectively, at $T=T_c$ for different chemical potentials as a function of $\omega$ for ${\bf k}=0$. 
(c), (d): Vector spectral density at $T=T_c$ for different spatial momenta ${\bf k}$ as a function of $\omega$ at $\mu=0.1 \mu_c$ and $\mu=\mu_c$, respectively.
(e), (f):  Axial-vector spectral density at $T=T_c$ for different spatial momenta ${\bf k}$ as a function of $\omega$ at $\mu=0.1 \mu_c$ and $\mu=0.33 \mu_c$, respectively.The dimensional quantities are expressed in units of $\mu_c$.}
\label{spect4}
\end{figure}

\newpage

%%%%%%%%%%%%%%%%%%%%%%%%%%%%%%%%%%%%%%%%%%%%%%%%%%%%%%%%%%%%%%%%%%%%%%%%%%%%%%%%%%%%%
\section{Flavor Conductivity, Susceptibility and Diffusion}
\label{condifsus}
%%%%%%%%%%%%%%%%%%%%%%%%%%%%%%%%%%%%%%%%%%%%%%%%%%%%%%%%%%%%%%%%%%%%%%%%%%%%%%%%%%%%%

\subsection{Conductivity}
\begin{figure}[H]
\begin{center}
\includegraphics[width=0.49\textwidth]{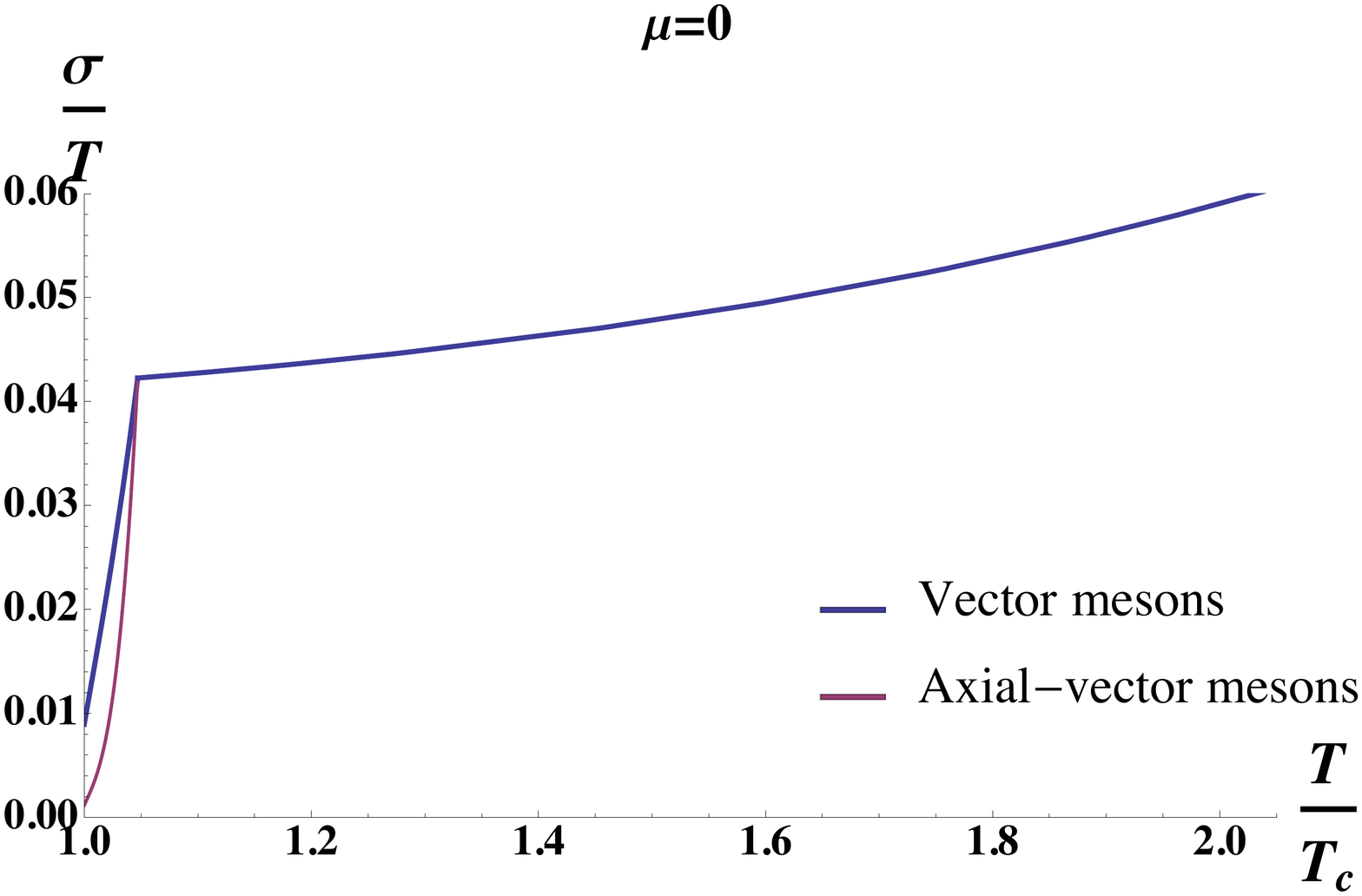}
\includegraphics[width=0.49\textwidth]{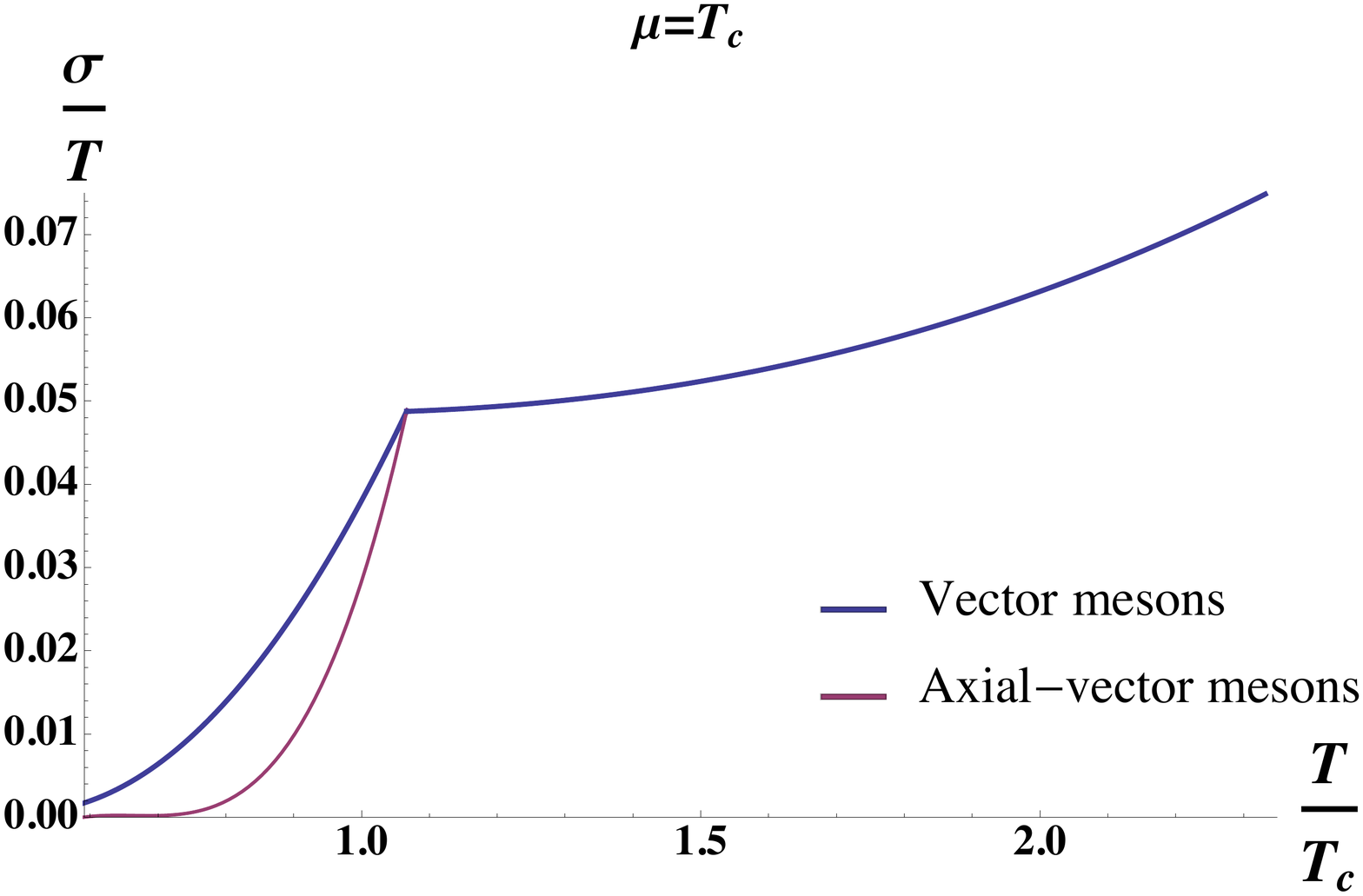}
\end{center}
\caption{  Transport coefficients of vector and axial-vector flavor currents divided by the temperature  along the three lines of the phase diagram in Fig. \protect\ref{phdialines}. The first plot corresponds to $\mu=0$ the second to $\mu=0.15$.}
\label{transp}
\end{figure}

The bulk flavor conductivity can be extracted from the time-like limit $\omega, k\rightarrow 0$ limit of the transverse
vector and axial spectral functions as $\sigma = \lim_{\omega \to 0} {\rho (\omega) \over 6 \, \omega}$. Solving \eqref{1ordeq} for ${\bf k}=0$ and $\omega=0$, with the boundary condition \eqref{bchor}, we  find 

\be
\sigma=
M^{3} N_c N_f V_{f}(\l_h,\t_h) \cla^4 \,  \kappa(\l_h,\t_h)^2\, \qt_h \,e^{\Awf_h} \,.
\label{trcoef}
\ee
The dependence on the chemical potential is explicit through the factor $\qt$ calculated on the boundary and implicit through the background fields. The numerical calculation of $\sigma$ is straightforward as soon as the background functions are known.
In Fig.~\ref{transp} we show the result both the vector and axial flavor conductivities versus $T/T_c$ with $T_c$ the deconfinement transition temperature at $\mu=0$. The numerical calculations are for $N_c=N_f=3$.  In Fig.~{transpmu} we show the behavior versus
$\mu/\mu_c$ with $\mu_c$  the deconfinement transition chemical potential at $T=0$. The electric vector and axial conductivities are
found to rise quickly across the transition region from a deconfined and strongly coupled phase to a strongly coupled chirally restored
phase, with both conductivities merging. At the deconfinement transition or $T=T_c$ the electric vector conductivity is amusingly comparable in magnitude to the electric conductivity derived from hadronic models, and rises quickly to agree with the electric 
conductivity reported on the lattice at about $T\approx 1.5T_c$. The rise is about an order of magnitude. The variation of the
flavor conductivities both vector and axisl at $T=T_c$ versus $\mu/\mu_c$ is slower.
 
The electric conductivity coincides with the flavor vector conductivity at $\mu=0$. This is the case since at $\mu=0$ the equation of motion for the singlet and non-singlet excitations coincide, see section (\ref{Svmesons}). Moreover, the definition of the flavor conductivity is such that it coincides with the singlet, section (\ref{specfun}). 

\begin{figure}[H]
\begin{center}
\includegraphics[width=0.49\textwidth]{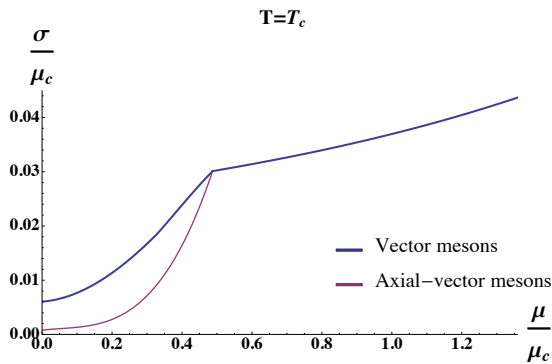}
\end{center}
\caption{  Transport coefficients of vector and axial-vector flavor currents  along the  $T=T_c$ line of the phase diagram in Fig. \protect\ref{phdialines}. We use $\mu_c$ as the energy unit in this plot.}
\label{transpmu}
\end{figure}

At non-zero chemical potential the electric conductivity (corresponding to the singlet vector field) develops a delta function at $\omega=0$, \cite{Hartnoll:2009sz}. The delta function cannot be seen from the numerical calculation of  $\sigma$ since it appears through the $\omega \to 0$ limit of the real part of the retarded Green function ($G_R$). In case of finite density, we have seen that the equation of the vector fluctuation acquires an effective mass term due to the coupling to the graviton, (\ref{totu1vecfl}). This mass term, in contrast to the massless case (\ref{vectoreom1m}),  generates a non-trivial solution of the equation of motion in the hydrodynamic limit $\omega \to 0$. This fact leads the real part of $G_R$ to be finite in the zero omega limit and the real part of the conductivity to diverge as $1/ \omega$, where we generalize the conductivity definition ${\tilde \sigma}=\lim_{\omega \to 0} G_R/\omega$. Then, the conductivity that we calculated above is $\sigma= {\rm Im} \, {\tilde \sigma}$.  Note that there is an $i$ difference in  our definition of ${\tilde \sigma}$ with  \cite{Hartnoll:2009sz}. Then, by the Kramer-Kr\"oning relation, ${\mathrm Im} \, G_R/\omega \sim \delta(\omega)$. At zero 
density ${\mathrm Re} \,G_R(\omega \to 0) \sim 0$, hence the conductivity is finite. The appearance of the delta function in $\sigma$ is a generic feature of charged black hole backgrounds and is related to charge  and momentum conservation of the vacuum state. The dual system to a homogeneous charged black hole is an infinite size homogeneously charged medium. Applying a constant electric field in such a system, an accelarated charge motion along the field will be generated and this contributes the $\delta(\omega)$ to the conductivity. It should be noticed that such systems are not natural since re-scattering should destroy such a state of an infinite homogeneously  charged medium.

\subsection{Diffusion}
\label{sec:dif}

Another important bulk flavor transport parameter is the the flavor diffusion constant. It follows from 
solving Eq.(\ref{longveceq1}) in the hydrodynamic limit $\omega,k \ll T $ and making use of the dispersion relation $\omega = -i D k^2$. The result for $D$ is in closed form in terms of the background field.  In order to expand around the hydrodynamic limit, we introduce a power counting parameter , $\omega \to \eta^2 \omega$, $k \to \eta k$. Then, the solution can be written in the form of

\be
E_L(r) = (r_h-r)^{- i{ \omega \over 4 \pi T}} F(r) \, ,
\ee
where $F(r)$ is regular at the horizon. $F(r)$ also admits a regular expansion in $\eta$, $F(r)=F_{0}(r)+\eta^2 F_{1}(r)+\dots$.Then, 
\be
E_L(r)= F_0(r)+\eta^2 \left(F_1(r) -i {\omega \over 4 \pi T} F_0(r)\right) +\ldots\, .
\ee
Expanding the \eqref{longveceq1} close to the horizon we find  $F_0(r)=1- i {k^2 \over \qt(r_h) \omega} (r_h-r) $. 
Solving  \eqref{longveceq1}  to zeroth order in $\eta$ it is found

\be
F_0(r) = 1+C \int_{r_h}^r dr {G(r) \over V_f(\l,\t) \, \kappa(\l,\t)^2 e^{A(r)} \qt(r)^3} \,.
\ee
Matching the derivatives, $F_0'(r_h)$, we determine $C=-i \left. {k^2 \over \omega} { V_f(\l,\t) \, \kappa(\l,\t)^2 e^{A} \qt^2 \over G} \right|_{r=r_h}$. Imposing the boundary condition on the boundary $F_0(\epsilon)=0$ gives

\be
D=  \left( \left.{ V_f(\l,\t) \, \kappa(\l,\t)^2 e^{A} \qt^2 \over G} \right|_{r=r_h}  \int_{r_h}^\epsilon dr {G(r) \over V_f(\l,\t) \, \kappa(\l,\t)^2 e^{A(r)} \qt(r)^3}  \right)\, ,
\ee
such that $\omega = - i D k^2$, for small $\omega$ and $k$. 

\begin{figure}[!tb]
\begin{center}
\subfigure[]{%
\includegraphics[width=0.49\textwidth]{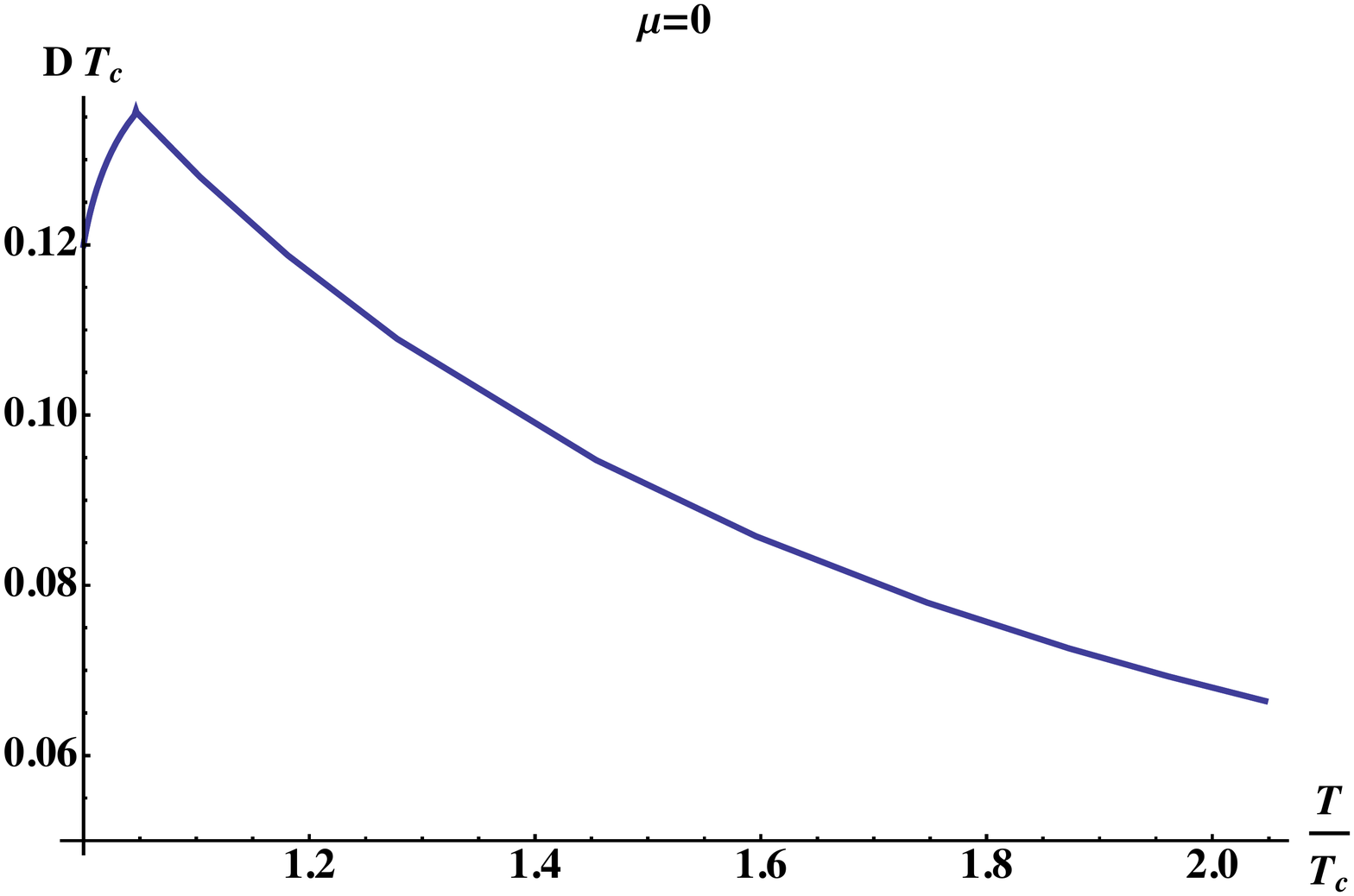}}
\subfigure[]{%
\includegraphics[width=0.49\textwidth]{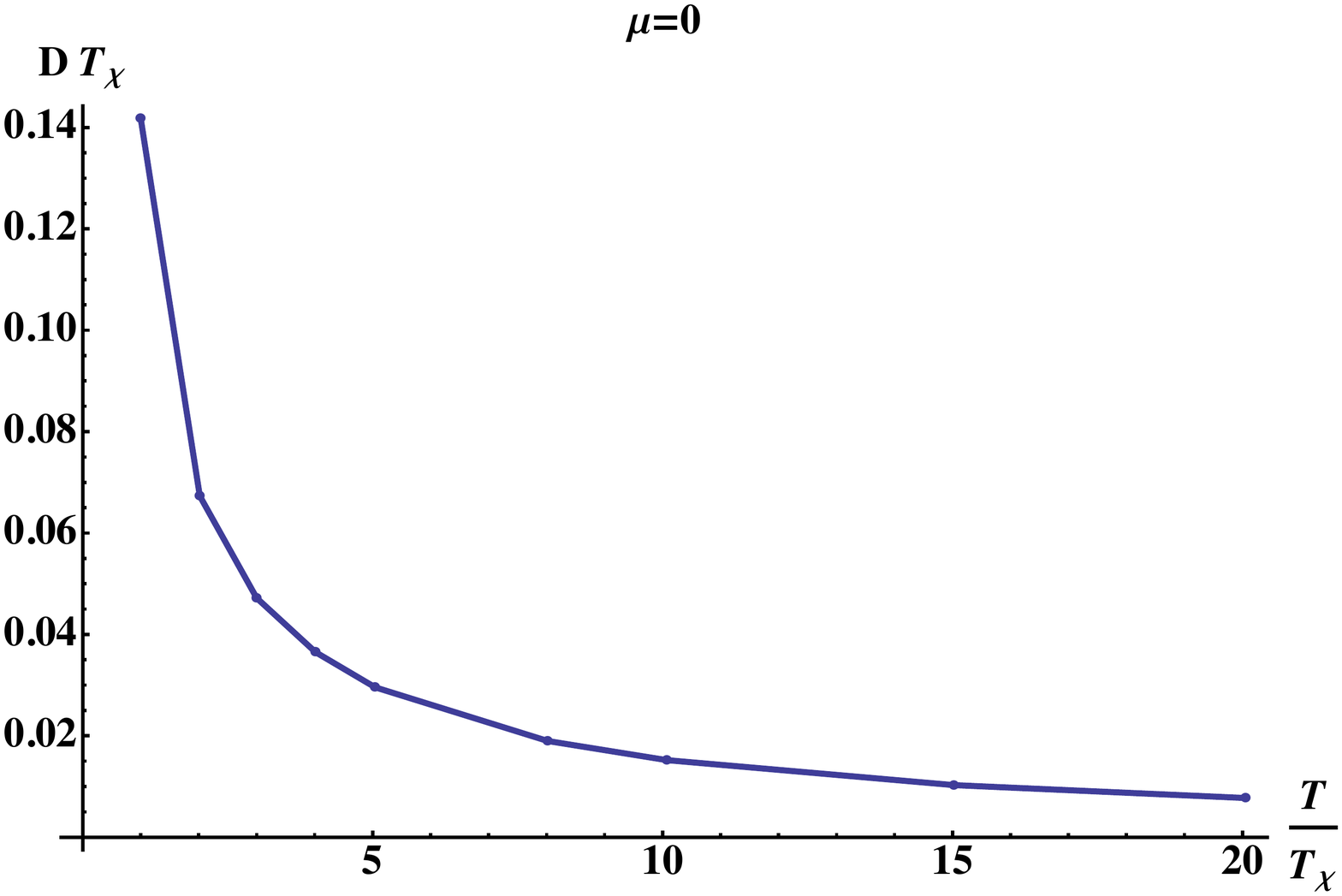}}\quad
\subfigure[]{%
\includegraphics[width=0.49\textwidth]{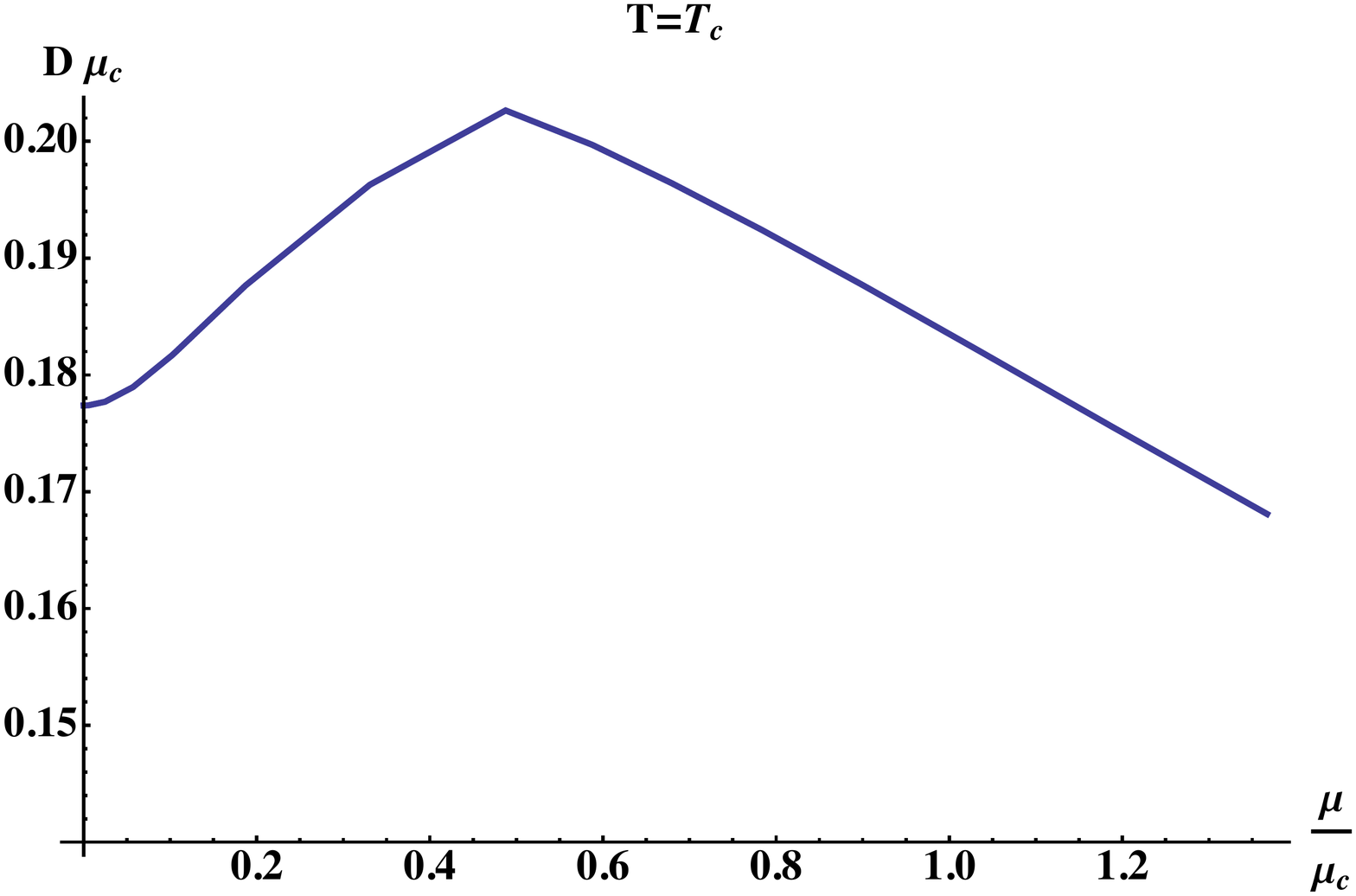}}
\subfigure[]{%
\includegraphics[width=0.49\textwidth]{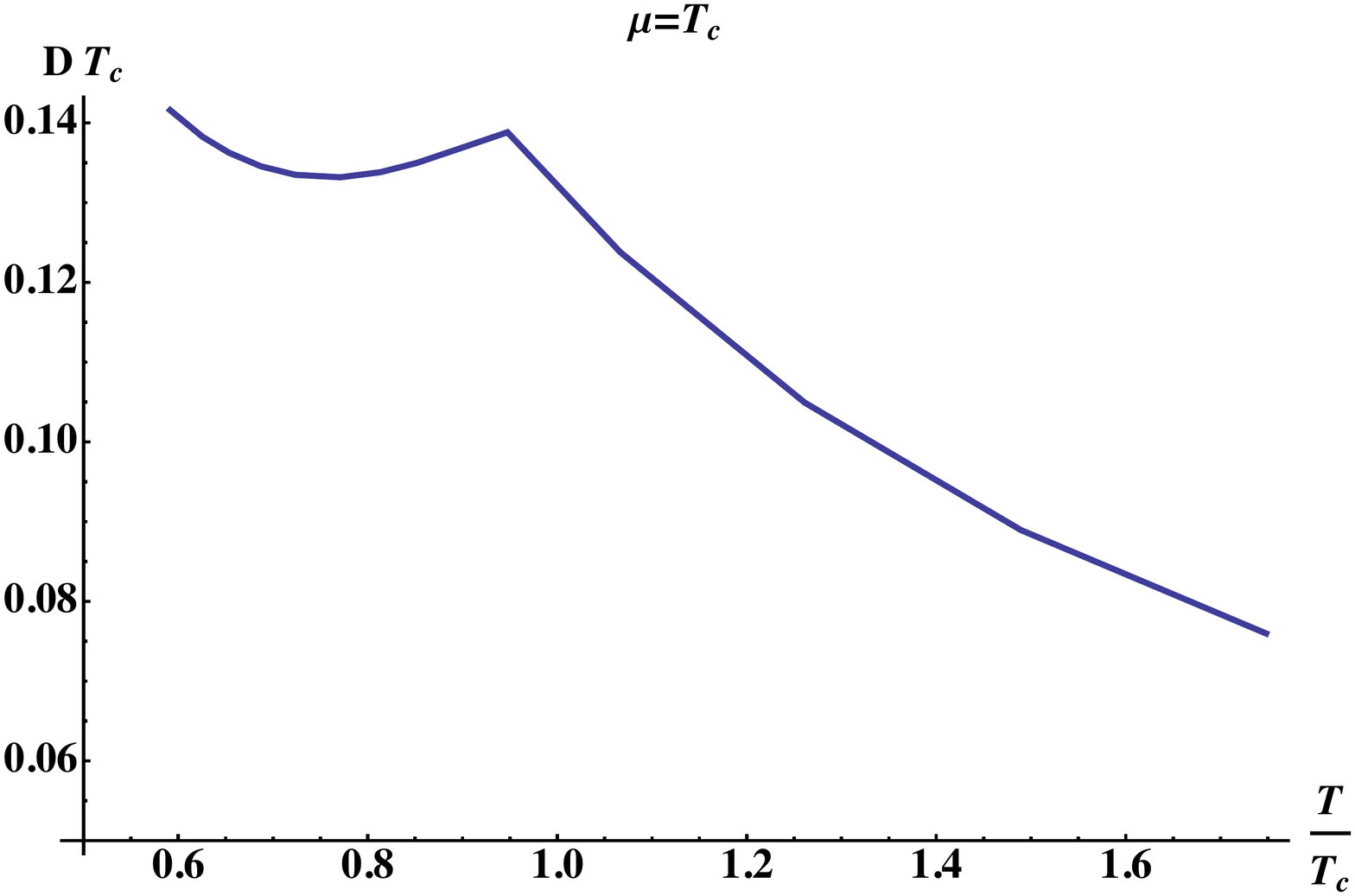}}
\end{center}
\caption{Diffusion constant $D$ as a function of temperature. (a): $D$ from $T_c$ to $2 T_c$ at $\mu=0$. (b): $D$ from $T_{\chi}$ to  $20 T_{\chi}$, where $T_{\chi}$ is the chiral transition temperature at $\mu=0$. Here $T_c= 0.95 T_{\chi}$.
 (c):$D$ versus $\mu/\mu_c$ at $T=T_c$. $D$ versus $T/T_c$ for fixed $\mu=T_c$.}
\label{difconfig}
\end{figure}

In Fig.~\ref{difconfig}a-b we show the flavor diffusion constant as a function of $T/T_c$ at zero chemical potential $\mu=0$. It rises
in the deconfined but chirally broken phase (a) and decreases in the chirally restored phase (b). The same rise and decrease is 
seen as a function of $\mu/\mu_c$ in Fig.~\ref{difconfig}c for fixed $T=T_c$. A slice of the flavor diffusion constant versus temperature
for fixed $\mu=T_c$ is shown in Fig.~\ref{difconfig}d. The fall of the flavor diffusion constant $D$ in the chirally restored phase at high temperature reflects on the de-corellation of the flavor momentum in the thermal but yet strongly coupled phase. The rise of $D$ in the
chirally broken but deconfined phase implies strong momentum correlations in the mixed phase prior to chiral restoration in this holographic model.

%\begin{figure}[!tb]
%\begin{center}
%\label{DC1}
%\includegraphics[width=0.49\textwidth]{plots/difusTmu0Tc.pdf}
%\includegraphics[width=0.49\textwidth]{plots/difusT1-20mu0.pdf}
%\end{center}
%\caption{Diffusion constant $D$ at $\mu=0$ as a function of temperature. Left: $D$ from $T_c$ to $2 T_c$. Right: $D$ from $T_{\chi}$ to  $20 T_{\chi}$, where $T_{\chi}$ is the chiral transition temperature at $\mu=0$. Here $T_c= 0.95 T_{\chi}$.}
%\label{difus}
%\end{figure}
%
%
%\begin{figure}[!tb]
%\label{DC2}
%\begin{center}
%\includegraphics[width=0.49\textwidth]{plots/difusmuTc.pdf}
%\includegraphics[width=0.49\textwidth]{plots/difusTmu015Tc.pdf}
%\end{center}
%\caption{Diffusion constant $D$ at $T=T_c$ versus $\mu/mu_c$ (left). $D$ versus $T/T_c$ for fixed $\mu=T_c$.}
%\label{difus}
%\end{figure}

\subsection{Susceptibility}

\begin{figure}[!tb]
\label{SUS}
\begin{center}
\includegraphics[width=0.69\textwidth]{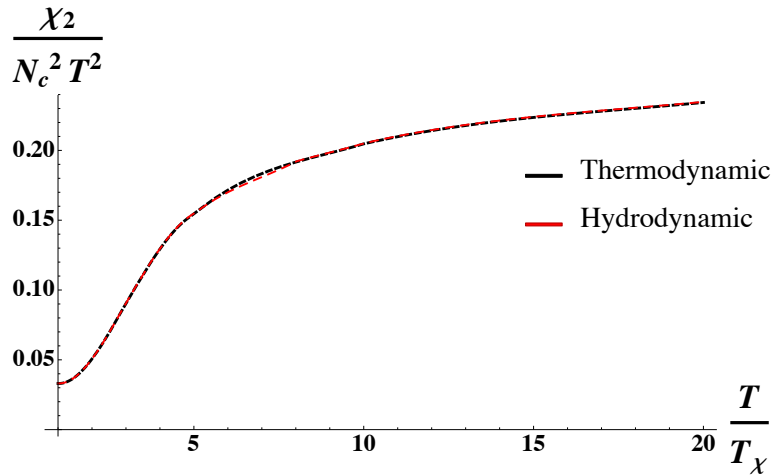}
\end{center}
\caption{The normalized flavor susceptibility $\chi_2$ versus $T/T_\chi$. The results from
thermodynamics~\protect{\cite{altemu}} (black dashed) agree with our results (red dashed).}
\label{difus}
\end{figure}

The electric flavor conductivity and diffusion constant are tied by the Einstein-Kubo relation, leading to the flavor susceptibility.
The latter was analyzed in this model using the bulk pressure in~\cite{altemu}. In large $N_c$, the non-singlet
flavor diffusion constant dwarfs the singlet one, leading to the general Einstein-Kubo relation for the flavor susceptibility
$\chi_2=\sigma/3D$. The free thermal flavor susceptibility is $\chi_{2F}= N_cT^2/3$ for each flavor species. In Fig.~\ref{SUS}
we show $\chi_2$ versus temperature $T/T_\chi$ for $\mu=0$  in the chirally restored phase. Our results in the hydrodynamical
limit using the Einstein-Kubo formula matches the flavor susceptibility obtained earlier in this model using the bulk pressure~{\cite{altemu}}. The agreement provides an overall consistency check on our numerical analysis. The use of the Einstein-Kubo relation together with our results for $\sigma$ and $D$ allow a determination of $\chi_2$ throughout the phase diagram in our model.

%%%%%%%%%%%%%%%%%%%%%%%%%%%%%%%%%%%%%%%%%%%%%%%%%%%%%%%%%%%%%%%%%%%%%%%%%%%%%%%%%%%%%
\section{Conclusions}
\label{concl}
%%%%%%%%%%%%%%%%%%%%%%%%%%%%%%%%%%%%%%%%%%%%%%%%%%%%%%%%%%%%%%%%%%%%%%%%%%%%%%%%%%%%%

A variant of improved holographic QCD in the Veneziano limit was used to probe the time-like structure of the QCD phase diagram
in the double limit of large $N_{c,f}$ but fixed $x=N_f/N_c$. The model exhibits a strongly coupled deconfined phase followed by a
chirally restoring phase. The transition region in the $\mu-T$ plane is narrow at $\mu=0$ but widens at $T=0$, see Fig.~\ref{phdia}.
The quasi-normal modes which are the analogue of the time-like masses survive the deconfinement transition to dissolve rapidly in 
the chirally broken phase with typically a real part $\omega_R \approx 2\pi T$ and and an imaginary part $ \omega_I \approx 2\pi T_c $.
The rise of the real part $\omega_R$ with high temperature is similar to the rise of the electric screening masses with temperature reported by lattice simulations above the deconfinement temperature
\cite{Bernard:1991ah,Petreczky:2003iz,Petreczky:2012rq}. They are readily understood from 
the dimensional reduction of QCD~\cite{Hansson:1991kb,Fayyazuddin:1993ua,Hansson:1994nb,Hansson:1997fz}

The detailed analysis of the  vector and axial spectral functions across 
the transition region reflects on these broad and low-lying quasi-normal modes showing that the strongly coupled phase of V-QCD
is about featureless across the transition region. Electromagnetic emissivities of dileptons at current collider energies are very 
sensitive to the vector and axial-vector spectral functions at and above the temperatures studied in this model. Our results should
help guide some of the model calculations in this strongly coupled phase of the QCD phase diagram.

The rapid rise of the electric conductivities across the transition region by almost an order of magnitude may explain the puzzling
discrepancy between current hadronic results in the confined phase of QCD and the reported lattice results in the the de-confined phase~\cite{Lee:2014pwa}. Indeed, our results for the vector electric conductivity $\sigma/T$ at $T=T_c$ is small and consistent in magnitude
with the hadronic electric conductivity at about the same temperature and zero $\mu$. At $T\approx 1.5T_c$ it is an order of magnitude larger and also consistent with currently reported (quenched) lattice results. We recall that the vector electric conductivity plays an
important role setting up the time-scale for the decor relation of the magnetic field in the chiral magnetic effect~\cite{Fukushima:2008xe} and its
variant~\cite{Qian:2012eq}. The bulk flavor susceptibility derived from our time-like analysis with the help of the Einstein relation is in total
agreement with the one derived from thermodynamics using the bulk pressure~\cite{altemu}.

The present framework and time-like analysis can be extended in a number of ways, to address the important issues of lighter
chiral excitations (scalar and pseudo-scalar) as well U(1) correlations across the transition region to cite a few. We hope to report on these issues next.

\section{Acknowledgements}

We are grateful to Timo Alho, Johanna Erdmenger, Matti J\"arvinen, Dima Kharzeev, 
Elias Kiritsis, Rene Meyer, Andy O' Bannon, Edward Shuryak and Dereck Teaney for useful discussions.
This work was supported by the U.S. Department of Energy under Contracts No.
DE-FG-88ER40388. I.I. would also like to thank the Mainz Institute for theoretical Physics for the hospitality and partial support during the last stage of this work.

%%%%%%%%%%%%%%%%%%%%%%%%%%%%%%%%%%%%%%%%%%%%%%%%%%%%%%%%%%%%%%%%%%%%%%%%%%%%%%%%%%%%%
\appendix
\renewcommand{\theequation}{\thesection.\arabic{equation}}
\addcontentsline{toc}{section}{Appendices}
%%%%%%%%%%%%%%%%%%%%%%%%%%%%%%%%%%%%%%%%%%%%%%%%%%%%%%%%%%%%%%%%%%%%%%%%%%%%%%%%%%%%%
\section*{APPENDIX}
%%%%%%%%%%%%%%%%%%%%%%%%%%%%%%%%%%%%%%%%%%%%%%%%%%%%%%%%%%%%%%%%%%%%%%%%%%%%%%%%%%%%%

%%%%%%%%%%%%%%%%%%%%%%%%%%%%%%%%%%%%%%%%%%%%%%%%%%%%%%%%%%%%%%%%%%%%%%%%%%%%%%%%%%%%%
\section{Schr\"odinger form}
%%%%%%%%%%%%%%%%%%%%%%%%%%%%%%%%%%%%%%%%%%%%%%%%%%%%%%%%%%%%%%%%%%%%%%%%%%%%%%%%%%%%%
\label{app:schro}

The Schr\"dingier formalism for a generic fluctuation in the bulk is briefly presented here.
The five-dimensional action for a field $\Psi(x^0,r)$ is of the form

\be
S = -\frac12 {\cal K}_\Psi \int d^4 x dr \left(
C_1(r) (\partial_r \Psi)^2 + C_2(r) \eta^{\mu\nu} \partial_\mu \Psi \partial_\nu \Psi +
C_3(r) \Psi \partial_r \Psi + M(r) \Psi^2
\right)\,\,,
\label{quadracti}
\ee
where an arbitrary constant is multiplying the action. Let us consider $\Psi = e^{-i \omega_n t} \psi(r)$, then  $\omega_n$ takes discrete complex values as soon as the appropriate boundary conditions are set. The boundary conditions are normalizability at the boundary $\Psi(r=\epsilon)=0$ and the incoming wave at the horizon $\Psi(r\to r_h)=(r_h-r)^{-i{\omega \over 4 \pi T}} \left(1+ c_h^{(1)} (r_h-r)+c_h^{(2)} (r_h-r)^2 \right. $  $ \left. + {\mathcal O} ((r_h-r)^3) \right)$.
  The excitation equation is derived readily from from (\ref{quadracti}):
\be
-\frac{1}{C_2(r)} \partial_r \left( C_1(r) \partial_r \psi_n (r)\right)
+ H(r) \psi_n(r) = \omega_n^2 \psi_n(r)\, ,
\label{eomschfa1}
\ee
where the bulk "mass" term is 
\be
H(r)\equiv \frac{1}{C_2(r)} \left( M(r) -\frac12 \partial_r C_3(r)\right)\,\,.
\label{defhofz}
\ee
%We can define the orthonormality condition:
%\be
% \int dr C_2(r) \psi_n(r) \psi_m (r) = \delta_{mn}\, .
%\label{normpsi}
%\ee
We now define a new radial variable $u$, and a rescaled field $\alpha$
in terms of a function $\Xi$ as
\be
du = \sqrt{\frac{C_2(r)}{C_1(r)}} dr\,\,,\qquad
\alpha = \Xi\, \psi \,\,,\qquad
\Xi(r) = (C_1(r)C_2(r))^\frac14\,\, .
\label{defsschrei}
\ee
The Sturm-Liouville problem now takes the Schr\"odinger form:
\be
- \frac{d^2 \alpha_n(u)}{du^2} + V(u) \alpha_n(u) = m_n^2 \alpha_n (u) \, ,
\label{schroeqap}
\ee
where the Schr\"odinger potential is
\be
V(u) = \frac{1}{\Xi(u)}\frac{d^2 \Xi(u)}{du^2} + H(u) \, .
\label{schrlike}
\ee

\section{Vector excitations}
\label{vecapp}
We expand the action to quadratic order in the vector fluctuation field. The quantities, ${\bf A}_{L/R}$, in the square roots of the tachyon DBI action in the vacuum are

\be
{\bf A}_{L\,MN} = {\bf A}_{R\,MN} = {\mathcal G}_{MN}= 
\left( \begin{array}{ccccc}
-e^{2 A(r)} f(r) & 0 & 0 & 0 & {e^{-A(r)}\, \nt \, G(r) \over \kappa(\lambda, \t) \, \qt(r) \, V_f(\l , \t)} \\
0 & e^{2 A(r)} & 0 & 0 & 0 \\
0 & 0 & e^{2 A(r)} & 0 & 0 \\
0 & 0 & 0 & e^{2 A(r)} & 0\\
-{e^{-A(r)} \nt G(r) \over \kappa(\lambda, \t) \, \qt(r) \, V_f(\l , \t)} & 0 & 0 & 0 & {e^{2 A(r)} \over f(r)} G(r)^2
\end{array}
\right) \, .
\ee
The inverse matrix is 

\be
 {\mathcal G}^{MN}= 
\left( \begin{array}{ccccc}
-{e^{-2 A(r)} \qt(r)^2 \over f(r)} & 0 & 0 & 0 & {e^{-5 A(r)} \, \nt \, \qt(r) \over \kappa(\lambda, \t) \, G(r) \, V_f(\l , \t)} \\
0 & e^{-2 A(r)} & 0 & 0 & 0 \\
0 & 0 & e^{-2 A(r)} & 0 & 0 \\
0 & 0 & 0 & e^{-2 A(r)} & 0\\
-{e^{-5 A(r)} \, \nt \, \qt(r) \over \kappa(\lambda, \t) \, G(r) \, V_f(\l , \t)} & 0 & 0 & 0 & {e^{- 2 A(r)} f(r) \qt(r)^2 \over G(r)^2}
\end{array}
\right) \,.
\ee
 Using the expansion

\be
\sqrt{\det(\mathbb{I}+X)}=1+{1\over2}\tr{X}-{1\over4}\tr{X^2}+{1\over8}(\tr{X})^2+\mathcal{O}(X^3)\,,
\ee
wheree $X$ is a matrix, the action becomes

\be
\begin{split}
S_V & =- M^3 \, N_c \,  {\mathbb Tr} \, \int d^4 x \, dr  V_f(\l,\t) \sqrt{-\mathrm{det}({\mathcal G})} \sqrt{ \mathrm{det} (\delta^M_N +\cla^2 \kappa(\lambda, \tau) {\mathcal G}^{MR}V_{RN} )} \\
 & = -{1 \over 2}M^3 \, N_c \, {\mathbb Tr} \, \int d^4 x \, dr  V_f(\l,\t) \sqrt{-\mathrm{det}({\mathcal G})} \biggl( -\cla^2 \kappa(\l,\t) \mathcal{G_A}^{MN} V_{MN} \\
 & - { \cla^4 \kappa(\l,\t)^2 \over 2} \left[\mgs^{MS} \, V_{ST} \, \mgs^{TN} \, V_{NM} + \mga^{MS} \, V_{ST} \, \mga^{TN} V_{NM} -{1 \over 2}\, (\mga^{MN} V_{MN})^2 \right] \biggr) \,. 
\end{split}
\ee
The linear term vanishes for the flavored excitations due to the trace over flavor indices. The action is written in the $V_r=0$ gauge as

\be
\begin{split}
S_V &= -  {1\over2}\, M^3 N_c\,  {\mathbb Tr} \, \int d^4x\, dr
V_f(\l,\t) \, \cla^4 \,  \kappa(\l,\t)^2\, \qt \, \G(r)^{-1}\,e^{A(r)} \\
& \left[ \frac12\, {\G(r)^2 \over \qt(r)^2}\, V_{ij}V^{ij} -\frac12 \, {\G(r)^2 \over f(r)^2}\, V_{i0}V^{i0}  
 +f(r)  \partial_r V_i \partial_r V^{i} -\qt(r)^2 \partial_r V_0 \partial_r V^{0}
\right]\, ,
\end{split}
\label{vectoracti}
\ee
where   $V_{ij}=\partial_i V_j-\partial_j V_i$, and the trace is over the flavor indices.
The fluctuation equation is written as 

\be
\partial_N \biggl[ V_f(\l,\t) \sqrt{-\mathrm{det}({\mathcal G})} \kappa(\l,\t)^2 \biggl(2 \mgs^{MS} \mgs^{TN} V_{ST} +2 \mga^{MS} \mga^{TN} V_{ST} +\mga^{ST} \mga^{MN} V_{ST} \biggr) \biggr] =0
\ee
The bulk field is Fourier expanded as  $V_\mu (t,{\bf x},r) =\int {d^4 k \over (2 \pi)^4} e^{-i \omega t+i {\bf k x}}  V_\mu (r,\omega,k)$.
In the $V_r=0$ gauge, the fluctuation equations for the different components of the vector field read

%\begin{eqnarray}
%&& e^{-2 A(r)}\, \partial_r (k^i V_i)- \omega \, \mgs^{00} \partial_r V_0=0 \nn \\
%&& -\partial_r \biggl( {e^{5 A(r)} G(r) \over \qt(r) } V_f(\l,\t) \, \kappa(\l,\t)^2 \, \mgs^{00} \, \mgs^{rr} \, \partial_r \,V_0 \biggr) \nn \\
%&&+  {e^{3 A(r)} G(r) \over \qt(r) } V_f(\l,\t) \, \kappa(\l,\t)^2 \, \mgs^{00} \, \left( {\bf k}^2 V_0 + \omega (k^i V_i) \right)=0 \nn \\
%&& -\partial_r \biggl( {e^{3 A(r)} G(r) \over \qt(r) } V_f(\l,\t) \, \kappa(\l,\t)^2 \, \mgs^{rr} \, \partial_r (k^iV_i) \biggr)  \\ 
%&&+  {e^{3 A(r)} G(r) \over \qt(r) } V_f(\l,\t) \, \kappa(\l,\t)^2 \, \mgs^{00} \omega \, \left( {\bf k}^2 V_0 + \omega (k^i V_i \right)=0 \nn \\
%&& -\partial_r \biggl( {e^{3 A(r)} G(r) \over \qt(r) }  V_f(\l,\t) \, \kappa(\l,\t)^2 \, \mgs^{rr} \, \partial_r ( V^\bot_i) \biggr) +{e^{A(r)} G(r) \over \qt(r) }  V_f(\l,\t) \, \kappa(\l,\t)^2\,  {\bf k}^2 V^\bot_i \nn \\
%&& +  {e^{3 A(r)} G(r) \over \qt(r) }  V_f(\l,\t) \, \kappa(\l,\t)^2 \, \mgs^{00} \omega^2 \,  V^\bot_i \,=0 \, ,\nn
%\end{eqnarray}
%where we have used that $\mgs^{ij}=e^{-2 A(r)} \delta^{ij}$. Replacing $\mgs$ we have

{\small 
\begin{eqnarray}
&&  \partial_r \, (k^i V_i)+ \, {\qt(r)^2 \over f(r)} \omega \, \partial_r V_0=0 \nn \\
&& -\partial_r \biggl( {e^{A(r)} \qt(r)^3 \over G(r) } V_f(\l,\t) \, \kappa(\l,\t)^2 \, \partial_r V_0 \biggr)
 +  {e^{A(r)} \, \qt(r) \, G(r) \over f(r) } V_f(\l,\t) \, \kappa(\l,\t)^2 \, \left( {\bf k}^2 V_0 + \omega (k^i V_i) \right)=0 \nn \\
&& \partial_r \biggl( {e^{A(r)} \, f(r) \, \qt(r) \over G(r) } V_f(\l,\t) \, \kappa(\l,\t)^2  \, \partial_r (k^iV_i) \biggr) 
+{e^{A(r)} \, \qt(r) \, G(r) \over f(r) } V_f(\l,\t) \, \kappa(\l,\t)^2 \,  \omega \, \left( {\bf k}^2 V_0 + \omega (k^i V_i \right)=0 \nn \\
&& -\partial_r \biggl(  {e^{A(r)} \, f(r) \, \qt(r) \over G(r) }   V_f(\l,\t) \, \kappa(\l,\t)^2  \, \partial_r \, V^\bot_i  \biggr) +{e^{A(r)} G(r) \over \qt(r) }  V_f(\l,\t) \, \kappa(\l,\t)^2\, \biggl( {\bf k}^2
-  {\qt(r)^2  \over f(r) } \, \omega^2 \biggr)   V^\bot_i \,=0 \, , \nn
\end{eqnarray}
}
The first equation follows from the two equations of motion of $V_0$ and $k^i V_i$. Without  loss of generality we may take only $k_3=k$ as non zero. Then the two equations can be diagonalized  by the gauge invariant variable ${E_L}= k V_0 +\omega V_3$. The equation of motion for the longitudinal field $E_L$ is

\begin{align}
&E_L(r)'' - \left( \partial_r \log  {V_f(\l,\t) \, \kappa(\l,\t)^2 e^{A(r)} \qt(r)^3 \over G(r)}  +{\omega^2 \over \omega^2 - k^2 {f \over \qt(r)}} \partial_r  \log {\qt(r)^2 \over f}   \right) E_L(r)' \\
&+{G(r)^2 \over f(r)^2} \left( \omega^2 - k^2 {f(r) \over \qt(r)^2} \right) E_L(r)=0 \, .
\label{longveceq}
\end{align}
The equation for the transverse components  is the last one and is decoupled from the rest.
For ${\bf k}=0$, the equations of $V^\bot(r)$ and $E_L(r)$ reduce to the same equation

\be
\frac{1}{V_f(\l,\t)\, \h(\l,\t)^2\, e^{\Awf}\,\G \, \qt \, f^{-1}}
\partial_r \left( V_f(\l,\t)\, \h(\l,\t)^2 \,e^{\Awf}\,
\G^{-1}\, \qt \, f \, \partial_r \psi_V \right)
+\, \omega^2  \,\psi_V = 0 \, ,
\label{vectoreomap}
\ee
the vector field in the bulk is written as $V^{\bot}=E_L=\psi_V$.
Eq. \eqref{vectoreomap} can be transferred to Schr\"odinger form as shown in Appendix~\ref{app:schro}.
The Schr\"odinger functions for the vector meson equation are
$C_3(r)=M(r)=0$ and
\begin{equation}
\begin{split}
C_1(r)&=V_f(\l,\t)\,  \cla^4 \, \kappa(\l,\t)^2 \, e^{\Awf(r)}\, f(r) \, \G(r)^{-1} \, \qt(r)\,, \\
C_2(r)&=V_f(\l,\t) \, \cla^4 \, \kappa(\l,\t)^2 \, e^{\Awf(r)}\, f(r)^{-1}  \,\G(r)\, \qt(r)\,.
\end{split}
\label{ABdefs}
\end{equation}
Further defining
\be \label{XiHV}
 \Xi_V(r) = \left(C_1(r)C_2(r)\right)^{1/4} = \,  \cla^2 \, \kappa(\l,\t) \,  \sqrt{V_f(\l,\t)\, \qt(r) \, e^{\Awf(r)}}\,,\qquad H_V(r) = \frac{M(r)}{C_2(r)} = 0\,,
\ee
the Schr\"odinger potential for the flavor non-singlet vectors reads
\be
 V_V(u) = \frac{1}{\Xi_V(u)}\frac{d^2\Xi_V(u)}{du^2} + H_V(u)\,.
\ee
Here the Schr\"odinger coordinate $u$ is defined by
\be \label{udefAppA}
 \frac{du}{dr} = \sqrt{\frac{C_2(r)}{C_1(r)}} = {G(r) \over f(r)}\,
\ee
and the boundary condition that $u \to 0$ in the UV.  $G$ is defined in~\eqref{Gdef}. Taking into account the expansion of the background fields in the UV, Eqs. (\ref{uvasal}) (\ref{tauuv}), and close to the horizon, \cite{altemu}, we find the $u$ coordinate asymptotically 

\be
{\rm UV:} \,\,  u_{UV}(r)\sim r\,\,\,\,,\,\,\,\,\,  {\rm Horizon:} \,\, u_{hor.}(r)=-{1 \over 4 \pi T} \ln (r_h-r) \, .
\ee
 The definition of the coordinate $u$ will be the same for all non-singlet meson towers, but the potential will 
change, as we shall see below.  The leading UV asymptotics of the potential follow from the AdS form of the asymptotics close to the boundary

\be
V_{UV}(u) \simeq {3 \over 4 u^2} \, ,
\ee
where higher order logarithmic corrections have been neglected. Close to the horizon $\Xi(r)$ is analytic and the potential reads

\be
V_{hor.}(u) \simeq  - (4 \pi T)^2 {\Xi'(r_h)\over  \Xi(r_h)}e^{-4 \pi T \, u} \, ,
\ee
where prime denotes the derivative of $\Xi$ with to $r$. $\Xi'(r_h) / \Xi(r_h)$ is a negative finite number.

%%%%%%%%%%%%%%%%%%%%%%%%%%%%%%%%%%%%%%%%%%%%%%%%%%%%%%%%%%%%%%%%%%%%%%%%%%%%%%%

\section{Axial-Vector \& Pseudoscalar Mesons}
\label{axvecapp}
Axial-vector and pseudoscalar mesons  mix at finite temperature. The quadratic action for those non-singlet mesons is

\be
\begin{split}
S_A & =- {1 \over 2}M^3 \, N_c \int d^4 x \, dr  V_f(\l,\t) \sqrt{-\mathrm{det}({\mathcal G})} \biggl( \kappa(\l,\t) \, \tau(r)^2 \, \mathcal{G_S}^{MN} \, (\partial_M \theta +2 A_M) \, (\partial_N \theta +2 A_N)\\ 
&-\cla^2 \kappa(\l,\t) \mathcal{G_A}^{MN} A_{MN} - { \cla^4 \kappa(\l,\t)^2 \over 2} \biggl[ \mgs^{MS} \, A_{ST} \, \mgs^{TN} \, A_{NM} + \mga^{MS} \, A_{ST} \, \mga^{TN} A_{NM}  \\
& -  {1 \over 2}\, (\mga^{MN} A_{MN})^2 \biggr] \biggr)
\end{split}
\ee
The fluctuation equations are written as 

\begin{eqnarray}
&& \partial_M\biggl[\kappa(\l, \t) \,\tau^2 \,\mgs^{MN}\,(\partial_N \theta + 2 A_N) \biggr]=0 \nn \\
&& \partial_N \biggl[ V_f(\l,\t) \, \sqrt{-\mathrm{det}({\mathcal G})} \, \cla^4 \, \kappa(\l,\t)^2 \biggl(2\,  \mgs^{MS} \mgs^{TN} A_{ST} +2 \, \mga^{MS} \mga^{TN} A_{ST} \nn\\
&&  + \, \mga^{ST} \mga^{MN} A_{ST} \biggr) \biggr] +4 \, V_f(\l,\t) \, \sqrt{-\mathrm{det}({\mathcal G})} \,\kappa(\l,\t) \,\tau^2 \,\mgs^{MN}\,(\partial_N \theta + 2 A_N) =0
\end{eqnarray}
In the $A_r=0$ gauge, the fluctuation equations in terms of the components of the gauge field

{\small 
\begin{eqnarray}
&& \partial_r \biggl( V_f(\l,\t) \, \kappa(\lambda, \tau) \, \tau(r)^2  \, {e^{3 A(r)} \, f(r) \, \qt(r) \over G(r) } \partial_r \, \theta \biggr) - V_f(\l,\t) \, \kappa(\l, \t) \, \t(r)^2 \, {e^{3 A(r)} \, \qt(r) \, G(r) \over f(r) }   \nn \\
&& \left( 2 \, {\mathrm i}\, \omega\,  A_0 - \omega^2 \, \theta \right) + V_f(\l, \t) \, \kappa(\l, \t) \, \tau(r)^2 \, {e^{3 A(r)} \, f(r) \, \qt(r) \over G(r)} \, \left( 2 {\mathrm i} (k^i A_i) - k^2 \theta \right)  = 0 \\
&& {\mathrm i}\, \cla^4 \, \kappa(\l, \t)^2 \, e^{-2 A(r)} \, \biggl( \partial_r (k^i A_i)+ \omega \, {\qt(r)^2 \over f(r) }  \, \partial_r A_0 \biggr) +2 \,  \kappa(\l,\t) \, \t(r)^2 \, \partial_r \theta=0  \\
&& -\partial_r \biggl( {e^{A(r)} \qt(r)^3 \over G(r) } V_f(\l,\t)\, \cla^4 \, \kappa(\l,\t)^2 \, \partial_r A_0 \biggr)
+ V_f(\l,\t) \, \cla^4 \, \kappa(\l,\t)^2 \,  {e^{A(r)} \, \qt(r) \, G(r) \over f(r) }  \nn \\
&&   \left( {\bf k}^2 A_0 + \omega (k^i A_i) \right)  +2\, V_f(\l,\t) \, \kappa(\l,\t) \, \t(r)^2 \, {e^{3 A(r)} \, \qt(r) \, G(r) \over f(r) } \left( 2 A_0 - {\mathrm i} \, \omega \, \theta \right) =0 \\
&&  -\partial_r \biggl( {e^{A(r)} \, f(r) \, \qt(r) \over G(r) } V_f(\l,\t) \, \cla^4 \, \kappa(\l,\t)^2  \, \partial_r (k^iA_i) \biggr)- V_f(\l,\t) \, \cla^4 \,\kappa(\l,\t)^2 \,  {e^{A(r)} \, \qt(r) \, G(r) \over f(r) } \, \omega \,  \nn \\
&&  \left( {\bf k}^2 A_0 + \omega (k^i A_i) \right)+ 2\, V_f(\l,\t) \, \kappa(\l,\t) \, \t(r)^2 \, {e^{3 A(r)} \, G(r) \over \qt(r) } \left( {\mathrm i} \, {\bf k^2} \, \theta +2 (k^i A_i) \right) =0 \\
&& -\partial_r \biggl(  {e^{A(r)} \, f(r) \, \qt(r) \over G(r) }   V_f(\l,\t) \, \cla^4 \, \kappa(\l,\t)^2  \, \partial_r \, A^\bot_i  \biggr) +{e^{A(r)} G(r) \over \qt(r) }  V_f(\l,\t) \, \cla^4 \, \kappa(\l,\t)^2\, 
\nn \\
&&\biggl( {\bf k}^2-  {\qt(r)^2  \over f(r) } \, \omega^2 \biggr)   A^\bot_i   + \, 4 \, V_f (\l,\t) \, \kappa(\l,\t) \, \t^2 \,  {e^{3 A(r)} G(r) \over \qt(r) }   A^\bot_i  =0 \, ,
\end{eqnarray}
}
%The quadratic action for the $SU(N_f)$ sector of the axial vector meson excitations reads
%\be
%\begin{split}
%S_A = & - {1\over2}\, M^3 N_c\, {\mathbb Tr} \int d^4x\, dr
%V_f(\l,\t)\, e^{\Awf}\, f^{-1}\, \qt \, \G^{-1}
%\bigg[ {1 \over 2}\,  \cla^4 \, \kappa(\l,\t)^2 \, \G^2 \, A_{\mu\nu}A^{\mu\nu} +
% \\
%&+
% \cla^4 \, \kappa(\l,\t)^2 \, f^2 \, \partial_r A^{\bot F}_\mu \partial_r A^{\bot F\,\mu}+4\h(\l,\t)\, \tau^2\, f\, e^{2 \Awf}\, \G^2\,  \, \qt^{-2} \,
%A^{\bot F}_\mu A^{\bot F\,\mu}
%\bigg]
%\label{axacti}
%\end{split}
%\ee
%where  $A_{\m\n}=\partial_\m A_\n^{\bot F}-\partial_\n A_\m^{\bot F}$.
where we have Fourier expanded the fields $A_{\mu} (t,{\bf x},r) =\int {d^4 k \over (2 \pi)^4} e^{-i \omega t+i {\bf k x}}  A_{\mu}(r)$. In case of  $k_i=0$, the excitation equation of the transverse modes reads

\begin{eqnarray}
&&\frac{1}{V_f(\l,\t)\, \h(\l,\t)^2\, e^{\Awf}\,\G \, \qt \, f^{-1}}
\partial_r \left( V_f(\l,\t)\, \h(\l,\t)^2 \,e^{\Awf}\,
\G^{-1}\, \qt \, f \, \partial_r \psi_A \right)-
{4 \t(r)^2\, e^{2 \Awf(r)} \, f(r) \over \qt(r)^2 \cla^4 \kappa(\l,\t)}\, \psi_A  \nn \\
&&+\omega^2 \, \psi_A= 0 \, ,
\label{axvectoreom}
\end{eqnarray}
where $A^{\bot}(r)=\psi_A(r)$. The Schr\"odinger functions are otherwise the same
as for vectors
but now
\be \label{MdefA}
M(r)=4 \, V_f(\l,\t) \, \h(\l,\t) \, \t(r)^2\, e^{3 \Awf(r)}  \, G(r) \, \qt(r)^{-1} 
\ee
 is nonzero. Therefore we find that
\be \label{XiHA}
 \Xi_A(r) = \Xi_V(r)\,,\qquad H_A(r) =  {4 \t(r)^2\, e^{2 \Awf(r)} \, f(r) \over \qt(r)^2 \cla^4 \kappa(\l,\t)} \,,
\ee
and the definition of $u$ is as in~\eqref{udefAppA}. For $k_i=0$, $\theta$ mixes with $A_0$. The coupled equations are 

\begin{eqnarray}
&& {\mathrm i}\, \cla^4 \, \kappa(\l, \t)^2 \, e^{-2 A(r)}   \, {\qt(r)^2 \over f(r) } \, \omega \, \partial_r A_0  +2 \,  \kappa(\l,\t) \, \t(r)^2 \, \partial_r \theta=0  
\label{k0eq1} \\
&& -\partial_r \biggl( {e^{A(r)} \qt(r)^3 \over G(r) } V_f(\l,\t)\, \cla^4 \, \kappa(\l,\t)^2 \, \partial_r A_0 \biggr) \nn \\ 
&& +2\, V_f(\l,\t) \, \kappa(\l,\t) \, \t(r)^2 \, {e^{3 A(r)} \, \qt(r) \, G(r) \over f(r) } \left( 2 A_0 - {\mathrm i} \, \omega \, \theta \right) =0 \,.
\label{k0eq2}
\end{eqnarray} 
We define the new variable in order to combine the two equations to one

\be
\hat \psi_P=- {e^{A(r)} \qt(r)^3 \over G(r) } V_f(\l,\t)\, \cla^4 \, \kappa(\l,\t)^2 \, \partial_r A_0 \, .
\ee
Then the two equations \eqref{k0eq1} and \eqref{k0eq2} are combined to

\begin{eqnarray}
&&V_f(\l,\t) \, \kappa(\l,\t) \, \t(r)^2 \, e^{3A(r)}\, \qt(r)^{2} \, G(r)^{-1} \, f(r) \, \partial_r \left[  { 1 \over V_f(\l,\t)\, \t(r)^2\, \kappa(\l,\t)\, e^{3 \Awf} \, \qt(r)\,G(r) \, f(r)^{-1} }\partial_{r} \hat \psi_{P} \right] \nn \\
&& - 4 \tau(r)^2  {e^{2 A(r)} \, f(r) \over \cla^4 \, \kappa(\l,\t) \, \qt(r)^2}  \hat \psi_P +\omega^2 \,\hat \psi_P=0 \, ,
\end{eqnarray}
and the corresponding Schr\"odinger functions read
\be
\begin{split}
C_1(r)&= V_f(\l,\t)^{-1}\, \t(r)^{- 2}\, \kappa(\l,\t)^{-1}\, e^{-3 \Awf} \, \qt(r)^{-1}\,G(r)^{-1} \, f(r) 
\,,\\
C_2(r)&=V_f(\l,\t)^{-1}  \, \kappa(\l,\t)^{-1} \, \t(r)^{-2} \, e^{-3A(r)}\, \qt(r)^{-1} \, G(r) \, f(r)^{-1}
\,, \\
M(r)&= C_2(r){ 4 \tau(r)^2  e^{2 A(r)} \, f(r) \over \cla^4 \, \kappa(\l,\t) \, \qt(r)^2} \,,
\end{split}
\ee
and $C_3(r)=0$. Therefore,
\be \label{XiHP}
 \Xi_P(r) = \frac{1}{\t(r)\sqrt{V_f(\l,\t)\,\h(\l,\t)\, e^{3 \Awf(r)} \, \qt(r)}}\,,\qquad H_P(r) = { 4 \tau(r)^2  e^{2 A(r)} \, f(r) \over \cla^4 \, \kappa(\l,\t) \, \qt(r)^2}\,.
\ee

%%%%%%%%%%%%%%%%%%%%%%%%%%%%%%%%%%%%%%%%%%%%%%%%%%%%%%%%%%
\section{Scalar Mesons}
\label{scapp}
The quadratic action is
\begin{eqnarray}
S&&= -{1\over2}\, M^3 N_c^2\,  {\mathbb Tr} \int d^4 x\, dr\,  e^{3\Awf} \, G^{-1} 
\bigg\{ V_{f}(\l,\t) \,\h(\l,\t) \,G^{-2}  \, \qt \, f  \left( G^2  +\qt^2 -G^2 \, \qt^2 \right)
  (\partial_r \mathfrak{s})^2  \nn \\
&&+ \qt \, f\, \tau' \, \left( 2\, \h(\l,\t)  \, \partial_\t V_f(\l,\t) + (1+G^{-2}) \qt^2 \, \,V_{f}(\l,\t)\, \partial_{\t}\h(\l,\t)
\right) \mathfrak{s} \partial_r \mathfrak{s} \nn \\
&&+{e^{-2 A} \over 4  \, G^2 \, \qt \, \kappa(\l,\t)^2 }
\Big[ 4  \, e^{4 \Awf}  \, G^4 \kappa (\l,\t)^2 \, \partial_\t^2 V_f(\l,\t)   \nn \\
&&+4 \, e^{4 A}\, G^2 \, \kappa (\l , \t) \,\left(  \qt^2+\qt^2 \, \G^2-2 G^2  \right) \partial_\t V_f(\l,\t)\,  \partial_\t \h(\l,\t) \nn \\
&&  +   V_f(\l,\t) \left(\kappa(\l,\t)^2 \qt^4 \,f^2 \, \t'^4+ 4 \, e^{4 A} \, G^2 (\qt^2 - 1) (1+G^2\, (\qt^2-1))  \right. \nn \\
&&  -   \left. 4 \, e^{2 A} \, f \, G^2 \, \kappa(\l,\t) (\qt^2-1)^2 \tau^{\prime \,\, 2}  \right) \, (\partial_\t\h(\l,\t) )^2  \nn \\
&&  +     2 e^{4 A} G^2 V_f(\l,\t) \kappa(\l,\t) \left( \qt^2 + \qt^2 G^2-2 G^2  \right) \,    \partial^2_\t \h(\l,\t)    \Big]  \mathfrak{s}^2 
  - V_{f}(\l,\t) \,\h(\l,\t)\, \qt \, f^{-1} \,(\partial_{0}  \mathfrak{s})^2 \nn \\
&& +  V_{f}(\l,\t)\,\h(\l,\t)\,  {G^2 + \qt^2-G^2 \, \qt^2 \over \qt}\,(\partial_{i}  \mathfrak{s})^2 \bigg\} \,,
\end{eqnarray}
where $\mathfrak{s}=\mathfrak{s}^a t^a= \int {d^4 k \over (2 \pi)^4} e^{- i \omega t + i k^i x_i} \psi_S(r)$. The fluctuation equation therefore becomes

\be \label{flscalflucts}
\psi_S''+ \partial_r (\log C_1(r)) \psi_S' - {M-{1\over 2} \partial_r C_3
  \over C_1} \psi_S + \left( {C_2^{(0)} \over C_1} \omega^2 - {C_2^{(i)} \over C_1} k^2  \right)\psi_S=0\,,
\ee
where the Schr\"odinger functions read
\begin{eqnarray}
 C_1(r) && =  V_{f}(\l,\t) \,\h(\l,\t) \,e^{3\Awf}  \,G^{-3}  \, \qt \, f \,  \left( G^2  +\qt^2 -G^2 \, \qt^2 \right) \\
 C_2^{(0)}(r) && = V_{f}(\l,\t) \,\h(\l,\t)\,  e^{3\Awf} \, G^{-1} \,\qt  \, f^{-1}  \\
 C_2^{(i)}(r) && =V_{f}(\l,\t)\,\h(\l,\t)\, e^{3\Awf} \, G^{-1} \,  {G^2 \, \qt^2 -G^2 - \qt^2 \over \qt}  \\
 C_3(r) && = e^{3\Awf} \, G^{-1} \, \qt \, f\, \tau' \, \left( 2\, \h(\l,\t)  \, \partial_\t V_f(\l,\t) + (1+G^{-2}) \qt^2 \, \,V_{f}(\l,\t)\, \partial_{\t}\h(\l,\t)
\right)  \\
 M(r)&& = {e^{ A} \over 4  \, G^3 \, \qt \, \kappa(\l,\t)^2 }
\Big[ 4  \, e^{4 \Awf}  \, G^4 \kappa (\l,\t)^2 \, \partial_\t^2 V_f(\l,\t) \nn \\
&&  +4 \, e^{4 A}\, G^2 \, \kappa (\l , \t) \,\left(  \qt^2+\qt^2 \, \G^2-2 G^2  \right) \partial_\t V_f(\l,\t)\,  \partial_\t \h(\l,\t) \nn \\
&&  +   V_f(\l,\t) \left(\kappa(\l,\t)^2 \qt^4 \,f^2 \, \t'^4+ 4 \, e^{4 A} \, G^2 (\qt^2 - 1) (1+G^2\, (\qt^2-1))  \right. \nn \\
&&  -   \left. 4 \, e^{2 A} \, f \, G^2 \, \kappa(\l,\t) (\qt^2-1)^2 \tau^{\prime \,\, 2}  \right) \, (\partial_\t\h(\l,\t) )^2  \nn \\
&&  +     2 e^{4 A} G^2 V_f(\l,\t) \kappa(\l,\t) \left( \qt^2 + \qt^2 G^2-2 G^2  \right) \,    \partial^2_\t \h(\l,\t)  \Big]  \,.
\end{eqnarray}

%%%%%%%%%%%%%%%%%%%%%%%%%%%%%%%%%%%%%%%%%%%%%%%%%%%%%%%%%%%%%%%%%%%%%%%%%%%%%%%%%

\end{document}